\documentclass[]{aa}

\usepackage{graphicx}
\usepackage{natbib}
\usepackage{txfonts}

\usepackage{subfigure}
\usepackage{color}

\newcommand{\dphi}{\delta \phi}

\title{Optimized fringe sensors for the VLTI next generation instruments}
\author{N. Blind \inst{1} \and O. Absil \inst{2}\fnmsep\thanks{Postdoctoral Researcher F.R.S.-FNRS (Belgium).} \and J.-B. Le Bouquin \inst{1} \and J.-P. Berger \inst{3} \and A. Chelli \inst{1}}

\institute{UJF-Grenoble 1/CNRS-INSU, Institut de Plan\'etologie et d'Astrophysique de Grenoble (IPAG) UMR 5274, Grenoble, France  \and  Institut d'Astrophysique et de G\'eophysique de Li\`ege (IAGL), University of Li\` ege, B-4000 Sart Tilman, Belgium \and European Southern Observatory, Casilla 19001, Santiago 19, Chile}

\date{Received xxx / Accepted xxx}

\begin{document}

\abstract
	{With the arrival of the next generation of ground-based imaging interferometers combining from 4 to possibly 6 telescopes simultaneously, there is also a strong need for a new generation of fringe trackers able to cophase such arrays. These instruments have to be very sensitive and to provide robust operations in quickly varying observational conditions.}
	{We aim at defining the optimal characteristics of fringe sensor concepts operating with 4 or 6 telescopes. The current detector limitations impose us to consider solutions based on co-axial pairwise combination schemes.}
	{We independently study several aspects of the fringe sensing process: 1) how to measure the phase and the group delay, and 2) how to combine the telescopes in order to ensure a precise and robust fringe tracking in real conditions. Thanks to analytical developments and numerical simulations, we define the optimal fringe-sensor concepts and compute the expected performance of the 4-telescope one with our dedicated end-to-end simulation tool sim2GFT.}
	{We first show that measuring the phase and the group delay by obtaining the data in several steps (i.e. by temporally modulating the optical path difference) is extremely sensitive to atmospheric turbulence and therefore conclude that it is better to obtain the fringe position with a set of data obtained simultaneously. Subsequently, we show that among all co-axial pairwise schemes, moderately redundant concepts increase the sensitivity as well as the robustness in various atmospheric or observing conditions. Merging all these results, end-to-end simulations show that our 4-telescope fringe sensor concept is able to track fringes at least 90\% of the time up to limiting magnitudes of 7.5 and 9.5 for the 1.8- and 8.2-meter VLTI telescopes respectively.}
	{}

\keywords{Techniques: high angular resolution - Techniques: interferometric - Instrumentation: high angular resolution - Instrumentation: interferometers - Methods: analytical - Methods: numerical}

\maketitle

			%%%%%%%%%%%%%%%%%%%%%%%%%%%%%%%%%%%%%%%%%%%%%%%%%%%%%%%
			%%%%%%%%%%%%%%%%%%%%%%%%%%%%%%%%%%%%%%%%%%%%%%%%%%%%%%%
			%%%%%%%%%%%%%%%%%%%%%%%%%%%%%%%%%%%%%%%%%%%%%%%%%%%%%%%
			\section{Introduction}                                                         \label{part:intro}

The sensitivity of ground-based interferometers is highly limited by the atmospheric turbulence and in particular by the random optical path difference (OPD) between the telescopes, the so-called piston. By making the fringes randomly move on the detector, the piston blurs the interferometric signal and prevents from using integration times longer than the coherence time of the atmosphere $\tau_0$ (typically a few 10\,ms in the near infrared). To reach their ultimate performance and increase their number of potential targets, interferometers need fringe trackers, i.e. instruments dedicated to measuring and compensating in real-time the random piston. By keeping the fringes locked with a precision better than $\lambda/10$, they ensure a fringe visibility loss lower than 20\% with integration times of a few seconds. Up to now, fringe trackers had to cophase array up to 3 telescopes by combining 2 baselines \citep[e.g., the FINITO fringe tracker at VLTI;][]{gai_2003b, lebouquin_2009}. The new generation of interferometric instruments, such as MIRC at CHARA \citep{monnier_2004}, MROI \citep{jurgenson_2008} or GRAVITY \citep{gillessen_2010}, MATISSE \citep{lopez_2008} and VSI \citep{malbet_2008} at the VLTI, requires to cophase arrays of 4 and possibly 6 telescopes, raising new fringe tracking challenges. This paper aims at defining the optimal concept of fringe sensor for such arrays.

This study is focused on solutions based on co-axial pairwise combination of the light beams, as currently used in existing and planed fringe-tracker such as FINITO, CHAMP and GRAVITY. The reason is that fringe sensing is generally carried out in the detector-noise limited regime and that multi-axial combination requires a larger number of pixels than pairwise co-axial combination. Additionally, we consider only the concepts providing measurements of both the phase delay (phase of the interferometric fringes) and the group delay (position of the white-light fringe). Indeed, the group delay resolves the $2\pi$ ambiguity on the phase and is mandatory to ensure an efficient and robust fringe tracking.

To define the optimal 4- and 6-telescope fringe sensor concepts based on the co-axial pairwise combination, we study 3 independent points. In Section \ref{part:phase} we study the phase estimator. We compare two different implementations of the ABCD fringe coding depending on whether the ABCD samples are obtained simultaneously or sequentially. In Section \ref{part:groupdelay} we study the two possible ways to measure the group delay, either by temporally modulating the OPD or by spectrally dispersing the fringes. In Section \ref{part:optschemes} we compare the efficiency of beam combination schemes with various degrees of redundancy (that is forming all the possible baselines of the array or not). We show that the result is a tradeoff between precision and operational robustness. Finally in Section \ref{part:finalperf} we merge the results of the 3 previous sections to define the optimal concept in the 4-telescope case. We perform a detailled estimate of its performance in the VLTI environnement.

			%%%%%%%%%%%%%%%%%%%%%%%%%%%%%%%%%%%%%%%%%%%%%%%%%%%%%%%
			%%%%%%%%%%%%%%%%%%%%%%%%%%%%%%%%%%%%%%%%%%%%%%%%%%%%%%%
			%%%%%%%%%%%%%%%%%%%%%%%%%%%%%%%%%%%%%%%%%%%%%%%%%%%%%%%
			\section{Phase estimation}                                       \label{part:phase}

%
%
%%%%%%%%%%%%%%%%%%%%%%%%%%%%%%%%%%%%%%%%%%%%%%%%%%%%%%%
\begin{figure}[t]
 \begin{minipage}[t]{\linewidth}
   \centering
   \subfigure{ \includegraphics[angle=0, width=0.35\textwidth]{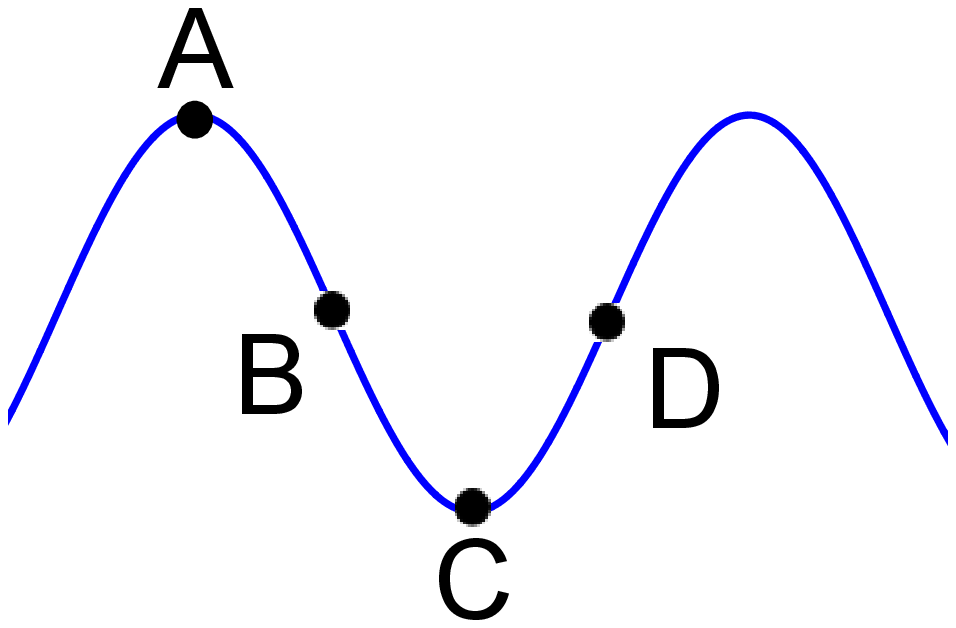}}
   \subfigure{ \includegraphics[angle=0, width=0.6\textwidth]{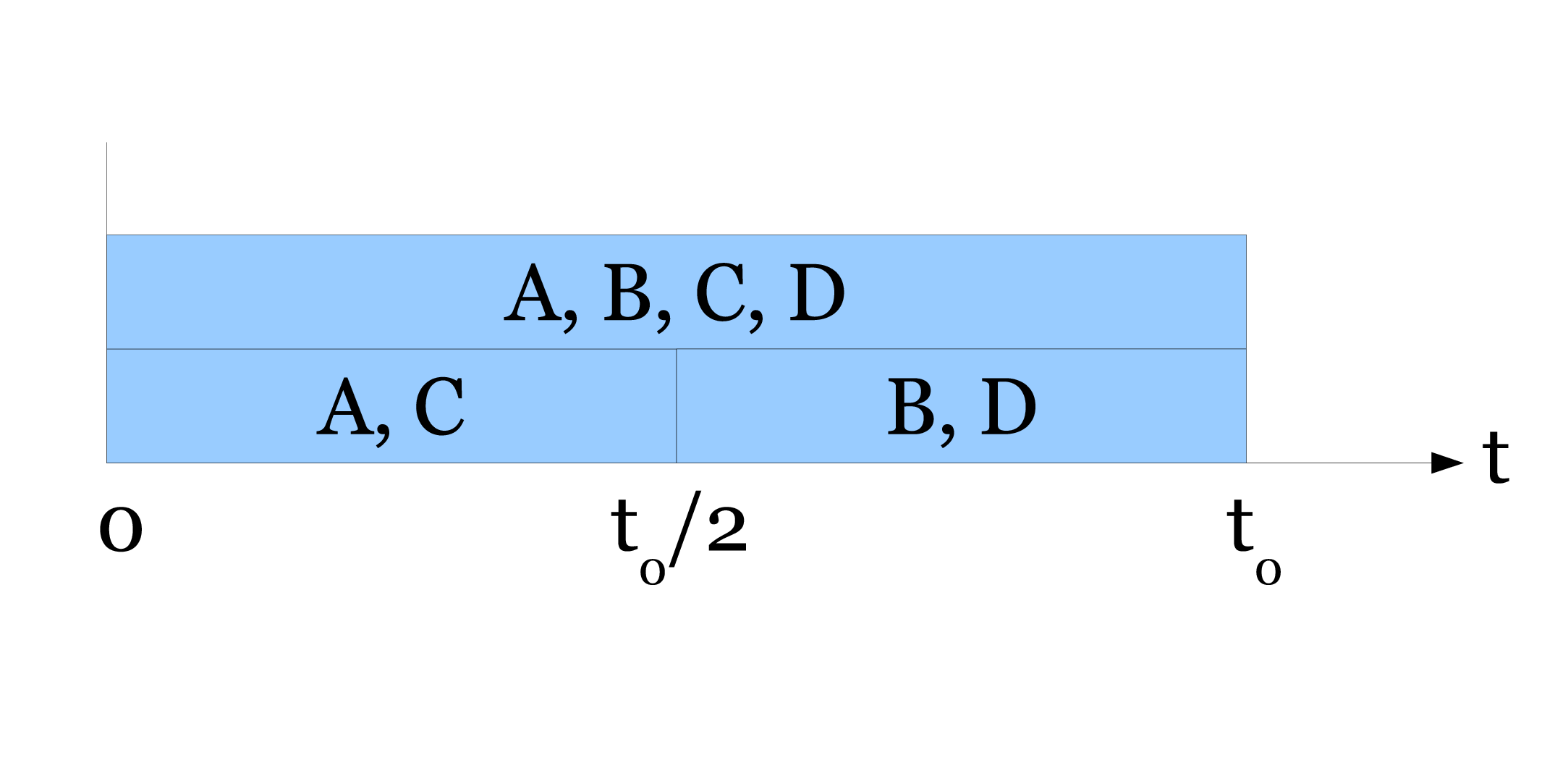}}
  \caption{The ABCD estimator. Left: conceptual representation of the 4 phase states sampling the fringes. Right: the measured phase states functions of the time for static (top) and temporally modulated (bottom) ABCD. The total integration time is $t_0$.}
  \label{fig:ABCD}
\end{minipage}
\end{figure}
%%%%%%%%%%%%%%%%%%%%%%%%%%%%%%%%%%%%%%%%%%%%%%%%%%%%%%%
%
%

Measuring the phase is essential for a fringe tracker in order to stabilize the fringes and to cophase the array within a fraction of wavelength. In this section, we therefore consider we are in a cophasing/phase tracking regime in which the group delay is known. We compare the precision of two different implementations of a phase estimator depending on whether the required measurements are simultaneous or not. The simplest and most efficient way to measure the fringe phase is the so-called ABCD estimator \citep{shao_1988}. It consists in sampling 4 points in quadrature in the same fringe (see Fig. \ref{fig:ABCD}, left), so that the real and imaginary parts of the coherent signal are extracted:
\begin{equation} \label{eq:idealABCD}
\left\{
	\begin{array}{ll}
	 A-C &\varpropto V \cos\,\phi \\
	 D-B &\varpropto V \sin\,\phi
	\end{array}
\right.
\end{equation}
where $V$ and $\phi$ are the fringe visibility and phase respectively, the cotangent of the latter being then estimated by:
\begin{equation}	\label{eq:ABCDphase}
\mathrm{tan}\, \hat{\phi} = \frac{D-B}{A-C}
\end{equation}

Considering a total integration time $t_0$ to obtain a phase estimation, there are two possible ways to perform the ABCD measurements (Fig. \ref{fig:ABCD}, right):
\begin{itemize}
 \item \textbf{Temporal ABCD}: it consists in temporally modulating the OPD like in the cases of FINITO at VLTI \citep{gai_2004}, CHAMP at CHARA \citep{bergerDh_2006} or the Keck Interferometer fringe tracker \citep{colavita_2010b}. We will consider in the following an implementation using a sampling of both outputs of a beam-splitter (in phase opposition) simultaneously. This allows the recording of two phase states A and C (in phase opposition) from $t = 0$ to $t_0/2$, and the B and D phase states by adding a temporal $\pi/2$ phase and recording between $t = t_0/2$ and $t_0$. This way one can generate an  ABCD fringe coding (see Fig. \ref{fig:ABCD}, right and bottom). There is consequently a $t_0/2$ time delay between the (A,C) and (B,D) samples. Other possible implementations (for instance at the Keck Interferometer fringe tracker) consider a continuous modulation over 1 fringe and only use one of the two interferometric outputs to measure the phase. Providing an exhaustive comparison between possible temporal algorithms is out of the scope of the paper but might lead to select a different implementation.
 \item[]
 \item \textbf{Static ABCD}: with this method we simultaneously  measure the four phase states from $t=0$ to $t_0$. This method is implemented in the PRIMA FSU at the VLTI \citep{sahlmann_2009} and is expected to be used on future instruments such as GRAVITY. In this case, there is no time delay between the ABCD samples.
\end{itemize}
In both cases the same signal-to-noise ratio (SNR) is achieved since the same number of photons is collected. The static ABCD requires to make twice more measurements simultaneously, so that the output flux is divided by 2, but each pixel integrates the signal twice as long. However the temporal and static ABCDs are not fully equivalent in real conditions because of atmospheric and/or instrumental disturbances. We now compare them by taking into account such effects.

		%%%%%%%%%%%%%%%%%%%%%%%%%%%%%%%%%%%%%%%%%%%%%%%%%%%%%%%
		%%%%%%%%%%%%%%%%%%%%%%%%%%%%%%%%%%%%%%%%%%%%%%%%%%%%%%%
		\subsection{Phase measurement errors}

When considering piston or photometric disturbances, the phase quadratic error $\sigma_\phi^2$ decomposes into the sum of two terms:
\begin{equation}
	\sigma_\phi^2 = \sigma_{\rm sig}^2 + \sigma_{\rm del}^2
\end{equation}
The first one is the noise due to the interferometric signal detection $\sigma_{\rm sig}$ which includes detector and photon noises \citep{shao_1988}. The second one, the so-called delay noise $\sigma_{\rm del}$, is due to external disturbances (piston or photometric variations) that combine with a delay between the ABCD measurements. By definition, the temporal ABCD is affected by such a noise, but not the static ABCD, since the four measurements are simultaneous. As this noise is an additionnal term, independent of the source brightness, we can already anticipate that it limits the phase measurements precision at high flux.

	\subsubsection{Detection noise}  \label{part:detection_noise}
	
While integrating the signal, the fringes move slightly because of the atmospheric piston. Their contrast is attenuated by a factor $\mathrm{exp} \left( \sigma^2(\phi_p, t_1)/2 \right)$, where $\phi_p$ is the piston phase and $\sigma^2(\phi_p, t_1)$ its variance for an integration time $t_1$. The integration time per phase state is twice larger in the static case than in the temporal case (see Fig. \ref{fig:ABCD}, right) implying a more important contrast loss. Combining this effect with the expression of the detection noise for an ABCD estimator derived from \citet{shao_1988}, we obtain:
\begin{equation}
 \label{eq:det_noise}
 	\sigma^2_{\rm sig}  = 2\, \displaystyle\frac{4\,\sigma_e^2 + K} {V^2 \, K^2} \times
 	\left\{
		\begin{array}{l}
		 	\mathrm{exp}\left(\,0.5 \, \sigma^2(\phi_p, t_0/2)\right) \\
		 	\qquad \mathrm{in \,\, the \,\, temporal \,\, case} \\
		 	\\
			\mathrm{exp}\left(\,0.5\,\sigma^2(\phi_p, t_0)\right) \\
			\qquad \mathrm{in \,\, the \,\, static \,\, case}
		\end{array}
\right.
\end{equation}
where $\sigma_e$ is the read-out noise in electrons per pixel, $V$ is the fringe contrast and $K$ is the number of photo-events for a total integration time $t_0$. The left term corresponds to the sum of the detector and photon noises respectively.

%
%\begin{eqnarray} \label{eq:contrast_loss}
%  V \rightarrow & V \: \mathrm{exp}\left(-\,\sigma_{t_0/2}^2(\phi_p)/2\right) &\mathrm{in \,\, the \,\, temporal \,\, case.} \\
%  V \rightarrow & V \: \mathrm{exp}\left(-\,\sigma_{t_0}^2(\phi_p)/2\right)  &\mathrm{in \,\, the \,\, static \,\, case.}
%\end{eqnarray}
%
%\begin{equation} \label{eq:contrast_loss}
%  V \rightarrow V \: \mathrm{exp}\left(-\,\sigma_{t_1}^2(\phi_p)/2\right)
%\end{equation}
%

%Secondly, in the case of the modulated ABCD, there is a temporal delay between the (A,C) and (B,D) measurements. Because of possible important piston or photometric variations, the phase estimation can be highly biased. As a result an additionnal delay noise $\sigma_{\rm del}$ appears (see appendix \ref{app:atmnoise} for details) in the phase error $\epsilon$ when fringes are scanned. On the contrary, in the static case, the four ABCD measurements see and integrate exactly the same disturbances so that $\sigma_{\rm del}$ is null.

	\subsubsection{Delay noise}  \label{part:delay_noise}

Delay noise is the consequence of piston and photometric variations between the (A,C) and (B,D) measurements, and therefore only affects the temporal method. These disturbances can be induced by the atmosphere (piston and scintillation) or by the instruments (vibrations). Since atmospheric piston and scintillation are independent \citep{fried_1966a}, we can decompose the delay noise in two terms due to the piston and the scintillation respectively:
\begin{equation} \label{eq:del_noise}
 \sigma_{\rm del}^2 = \sigma_{\rm pist}^2 + \sigma_{\rm sci}^2
\end{equation}
To compute these noises we assume the disturbances are constant while integrating the (A,C) signal, and suddenly change while integrating (B,D).
%justification: A and C see the same disturbances, and basically measure an average of the phase during this integration. SO a static ABCD measures the mean phase during a time t_0 whereas a temporal one measures two different phase.

\paragraph{{\bf Piston noise}} --  Because of the piston variation between the (A,C) and (B,D) measurements, the phase difference between them is not $\pi/2$ as it should. Taking the point in the middle of the interval $t_0$ as the reference, the measured signal is therefore:
\begin{equation}
\left\{
	\begin{array}{ll}
 		A-C &\varpropto V\,\cos(\phi + \phi_p(t - t_0/4)) \\
		D-B  &\varpropto V\,\sin(\phi + \phi_p(t + t_0/4))
	\end{array}
\right.
\end{equation}
The comparison to the ideal signal in Eq. \ref{eq:idealABCD} shows that the estimated phase $\hat{\phi}$ is biased if $\phi_p$ has varied between measurements. When we take into account the piston statistics, this bias results in the following piston noise of variance:
\begin{eqnarray} \label{eq:piston_noise}
 &\sigma_{\rm pist}^2 &= 0.125 \, \sigma^2(\delta\phi_p, t_0/2) %\; ,
\end{eqnarray}
where $\sigma^2(\delta\phi_p, t_0/2)$ is the variance of the difference of piston separated by $t_0/2$. Details of the computation can be found in Appendix~\ref{app:atmnoise}.

%
%%%%%%%%%%%%%%%%%%%%%%%%%%%%%%%%%%%%%%%%%%%%%%%%%%%%%%
\begin{table}
\begin{center}
	\begin{tabular}{lcccc}
	\hline \hline
	Condition & Excellent & Good & Medium & Bad \\
   \hline
	Seeing [arcsec] & 0.46 & 0.55 & 0.64 & 1.10 \\
	$\tau_0$ [ms] & 8.7    & 3.1   & 2.7   & 2.0 \\
   \hline
	\end{tabular}
	\caption{Typical seeing and atmospheric coherence time $\tau_0$ for the different observing conditions considered. \label{tab:atm_conditions}}
	\end{center}
\end{table}
%%%%%%%%%%%%%%%%%%%%%%%%%%%%%%%%%%%%%%%%%%%%%%%%%%%%%%
%
%
%%%%%%%%%%%%%%%%%%%%%%%%%%%%%%%%%%%%%%%%%%%%%%%%%%%%%%
\begin{table}[t]
 \begin{center}
  \begin{tabular}{ccccccccc}
\hline
\hline
 & \multicolumn{3}{c}{Piston noise $\sigma_{\rm pist}$} && \multicolumn{3}{c}{Scintillation noise $\sigma_{\rm sci}$} \\
%\\
 ATs & \multicolumn{7}{c}{} \\
 \hline
 $t_0$ [ms]                  & 2 & 4 & 8 && 2 & 4 & 8 \\
 Good & $\lambda/$92 & $\lambda/$60  & $\lambda/$35 && $\lambda/$499  & $\lambda/$369 & $\lambda/$290 \\
 Bad & $ \lambda/$29 & $ \lambda/$19 & $ \lambda/$12  && $\lambda/$101   & $\lambda/$67   & $\lambda/$37    \\
 \\
 UTs & \multicolumn{7}{c}{} \\
\hline
 $t_0$ [ms]                  & 1 & 2 & 4 && 1 & 2 & 4 \\
 Good & $\lambda/$33 & $\lambda/$21  & $\lambda/$12  && $\lambda/$162 & $\lambda/$122 & $\lambda/$59  \\
 Bad & $ \lambda/$21 & $  \lambda/$14 & $ \lambda/$9 && $\lambda/$101 & $\lambda/$52 & $\lambda/$21\\
%\\
  \end{tabular}
  \caption{Piston and scintillation noises computed from Eq. \ref{eq:piston_noise} and \ref{eq:photom_noise}. They are expressed as a function of the wavelength (in the H band), for three different integration times. Atmospheric conditions are Good (G) and Bad (B). For more details, see Tab. \ref{tab:piston_noise} and \ref{tab:photom_noise} in Appendix \ref{app:atmnoise}.}
 \label{tab:atmnoise}
 \end{center}
\end{table}
%%%%%%%%%%%%%%%%%%%%%%%%%%%%%%%%%%%%%%%%%%%%%%%%%%%%%%
%

\paragraph{{\bf Scintillation noise}} -- The fringe visibility depends on the flux imbalance between the two beams $I_1$ and $I_2$ of the interferometer. These unequal fluxes reduce the fringe visibility by a factor:
\begin{equation}
   V_{\rm sci} = \frac{2\,\sqrt{I_1 I_2} }{I_1 + I_2} %\; .
\end{equation}
Because of scintillation, $I_1$ and $I_2$, and therefore $V_{\rm sci}$, change between the (A,C) and (B,D) measurements. Still considering the middle of the interval $t_0$ as the reference, the measured signal is:
\begin{equation}
\left\{
	\begin{array}{ll}
	A-C &\varpropto V_{\rm sci}(t - t_0/4) \, \cos\,\phi \\
	D-B  &\varpropto V_{\rm sci}(t + t_0/4) \, \sin\,\phi
	\end{array}
\right.
\end{equation}
By comparing this equation to the ideal signal (Eq. \ref{eq:idealABCD}), we see that a single phase estimation is biased if $V_{\rm sci}$ varies, that is if $I_1$ and/or $I_2$ vary. Assuming the beams $I_1$ and $I_2$ to be independent and of same statistics, the scintillation noise is:
\begin{equation} \label{eq:photom_noise}
 \sigma_{\rm sci}^2 \sim 0.04 \, \sigma^2(x, t_0/2)
\end{equation}
where $x = (I_1(t+t_0/4)-I_1(t-t_0/4))/I_1(t)$ is the relative flux variation between the (A,C) and (B,D) exposures, $\langle x \rangle$ its mean and $\sigma^2(x, t_0/2)$ its variance during a time $t_0/2$. Note that to compute this noise, we consider the extreme case of a mean unbalance between the interferometric inputs equal to 10. Details of the calculations can be found in Appendix~\ref{app:atmnoise}.

		%%%%%%%%%%%%%%%%%%%%%%%%%%%%%%%%%%%%%%%%%%%%%%%%%%%%%%%
		%%%%%%%%%%%%%%%%%%%%%%%%%%%%%%%%%%%%%%%%%%%%%%%%%%%%%%%
		\subsection{Performance comparison}

%
%
%%%%%%%%%%%%%%%%%%%%%%%%%%%%%%%%%%%%%%%%%%%%%%%%%%%%%%%
\begin{figure}[]
   \centering
   \includegraphics[angle=0, width=0.4\textwidth]{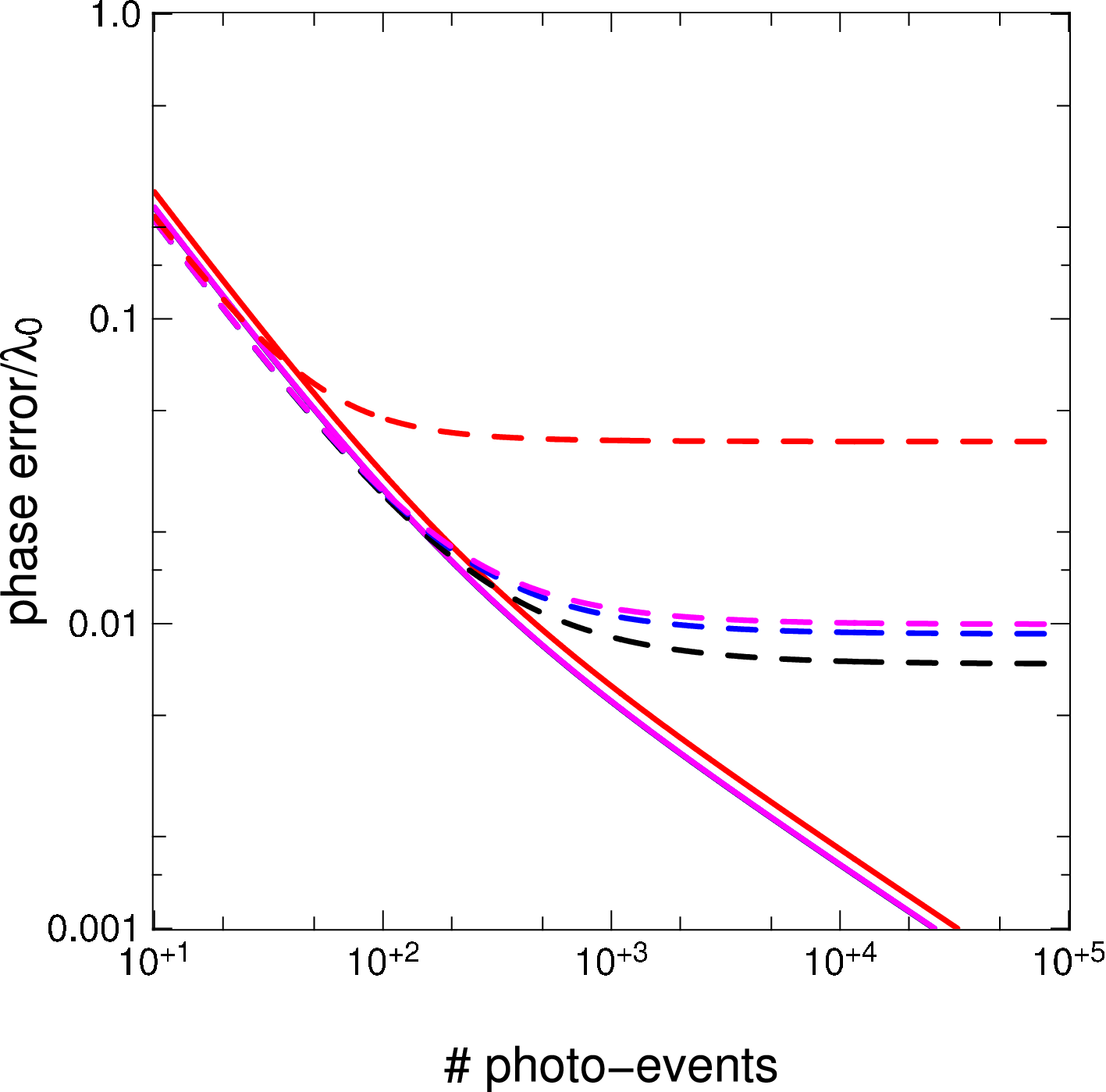}
   \caption{Relative errors $\sigma_\phi/\lambda$ of temporal (dash) and static (solid) ABCD phase estimators in H band as a function of the number of detected photo-events $K$. Black, blue, magenta and red curves represent Excellent, Good, Medium and Bad conditions respectively as defined in Tab.\ \ref{tab:atm_conditions}. The plots are done in the case of the ATs for an integration time of 2\,ms for the specific ABCD implementation considered here. Note that for the static ABCD, the black, blue and magenta curves are superimposed because of close performances.}
   \label{fig:ABCDvsACBD}
\end{figure}
%%%%%%%%%%%%%%%%%%%%%%%%%%%%%%%%%%%%%%%%%%%%%%%%%%%%%%%
%
%

In order to put quantitative numbers on the previous results, we used data provided by ESO and collected at the Paranal Observatory in 2008. The FITS files contain the photometric flux and the fringe phase as measured by the FINITO fringe-tracker in the H-band. Data were collected at a frequency of 1\,kHz for ATs and 2\,kHz for UTs, and for various atmospheric conditions (see Tab. \ref{tab:atm_conditions}). We have computed the variance of the difference of piston and photometries separated by $t_0/2$, for different values of $t_0$. We have finally injected the results in Eq.~\ref{eq:piston_noise} and~\ref{eq:photom_noise} to evaluate piston and scintillation noises, in atmospheric conditions ranging from Excellent to Bad (see Tab.~\ref{tab:atmnoise}). We note that whatever the conditions and the integration time, $\sigma_{\rm pist}$ is always at least twice larger than $\sigma_{\rm sci}$: when measuring the phase, the piston is therefore far more harmful than the relative variations of flux -- this is all the more true than we consider an extremely unfavorable case for scintillation noise, as explained in the previous section.

We now compute the phase error $\sigma_\phi$ in realistic conditions for the temporal and static ABCD methods.
Fig.~\ref{fig:ABCDvsACBD} represents the phase error relative to the wavelength (i.e.\ $\sigma_\phi/\lambda$) in H-band with both methods. It clearly shows that the static ABCD outperforms the temporal one in almost all regimes. It is only in the photon poor regime and in bad conditions that modulating the fringes is a little more efficient, that is when the fringe contrast attenuation on the static ABCD becomes important. Yet regarding the large phase error ($\sigma_\phi > \lambda/10$, see Tab.~\ref{tab:atmnoise}), phase tracking would be very poor -- if possible -- in such conditions.

In the photon rich regime, the plateau for the temporal method is due to the delay noise. For the 1.8-m Auxiliary Telescopes (ATs) at the VLTI, it has an almost null impact on phase tracking in good conditions ($\sigma_\phi<\lambda/35$) even for integration times as long as 8\,ms. In bad conditions with integration times  longer than 2\,ms there could be some limitations ($\sigma_\phi>\lambda/20$) depending on the actual implementation of the temporal ABCD.

Observations on the 8-m Unit Telescopes (UTs) show a higher piston noise, partly due to instrumental vibrations \citep{dilieto_2008}: in good conditions, the noise level is similar to the one of ATs in bad conditions. Passing from good to bad conditions, the integration time has to be divided by 2 to maintain the performance in a photon rich regime. In particular, in bad conditions and $t_0 > 4$\,ms, the noise level is higher than $\lambda/10$ whatever the source, and phase tracking could be hardly possible with a temporally modulated ABCD. This probably explains the difficulty of the FINITO fringe-tracker to close the loop on the UTs for faint objects.

In conclusion, with a temporal phase estimator, the fringe tracking capabilities are compromised in bad atmospheric conditions and on faint sources requiring long integration times. Therefore, from a performance point of view, a static method should be preferred thanks to its lower sensitivity to disturbances.

			%%%%%%%%%%%%%%%%%%%%%%%%%%%%%%%%%%%%%%%%%%%%%%%%%%%%%%%
			%%%%%%%%%%%%%%%%%%%%%%%%%%%%%%%%%%%%%%%%%%%%%%%%%%%%%%%
			%%%%%%%%%%%%%%%%%%%%%%%%%%%%%%%%%%%%%%%%%%%%%%%%%%%%%%%
			\section{Group delay estimation methods}                     \label{part:groupdelay}

\begin{figure}[t]
\begin{center}
	\label{fig:envco}
 	\includegraphics[angle=0, width=.45\textwidth]{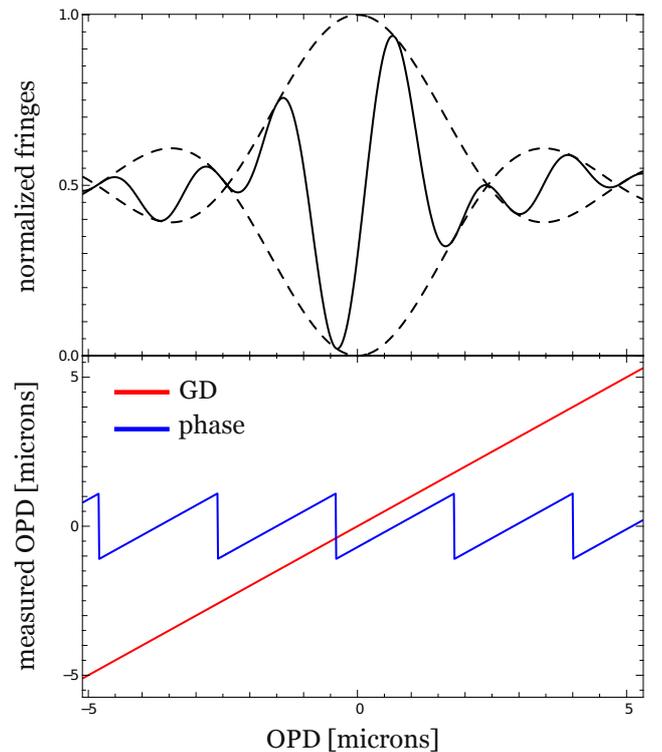}
 	\caption{Top: example of polychromatic fringes (solid line) with longitudinal dispersion, modulated by the coherence envelope (dashed lines). Bottom: corresponding phase and group delay measurements (in blue and red respectively) presented in microns.}
\end{center}
\end{figure}

The group delay (GD) is a measurement complementary to the phase and is mandatory to ensure an efficient fringe tracking. Indeed, a phase estimator only determines the fringe position modulo $2 \pi$. The GD lifts this ambiguity (see Fig.~\ref{fig:envco}). It allows to find and recover the position of maximum contrast, therefore providing the highest SNR. This is of particular interest when the fringe-tracking is unstable and/or when unseen fringe jumps occur regularly. Moreover, monitoring both the GD and the phase allows to determine the amount of dispersion induced by atmospheric water vapor \citep{meisner_2003}. This is done routinely at the Keck Interferometer for cophasing in N-band while measuring the phase and group delay in K-band \citep{colavita_2010a}. 

%
%
%%%%%%%%%%%%%%%%%%%%%%%%%%%%%%%%%%%%%%%%%%%%%%%%%%%%%%%
%%%%%%%%%%%%%%%%%%%%%%%%%%%%%%%%%%%%%%%%%%%%%%%%%%%%%%%
%
%
%

$I(\lambda)$ and $V(\lambda)$ being the flux and the complex visibility of the interferometric signal, the coherence envelope is linked to the complex coherent flux $I(\lambda) V(\lambda)$ through a Fourier transform:
\begin{equation} \label{eq:def-envco}
   E(x) \varpropto \left|\,  \int_0^\infty I(\lambda) V(\lambda) e^{i 2 \pi x_{\rm GD} /\lambda}\, e^{-i 2 \pi x /\lambda} \,d\lambda \,\right|
\end{equation}
where $x$ is the OPD. Consequently it is possible to estimate the group delay
%(noted $\hat{x}_{\rm GD}$)
with two different methods:
\begin{itemize}
 \item The \textbf{temporal method} estimates the GD by measuring the envelope amplitude (in other words the fringe contrast) $E(x)$ at several points around its maximum by modulating the OPD. Since the phase needs to be measured at the same time to ensure fringe tracking, the OPD is modulated near the envelope center to keep a high SNR. This method is currently used in FINITO and CHAMP.
 \item[]
 \item The \textbf{spectral method} uses the Fourier relation between the coherent spectrum $I(\lambda) V(\lambda)$ and the coherence envelope $E(x)$. The coherence envelope is recovered by measuring the coherent spectrum over few spectral channels. This method has been successfully implemented at PTI \citep{colavita_1999a}, and more recently in PRIMA \citep{sahlmann_2009} and in the KI fringe tracker \citep{colavita_2010b}.
\end{itemize}
We could not obtain a realistic analytical description of these group delay estimators. Therefore we decided to compare them with Monte-Carlo simulations taking into account atmospheric disturbances.

%
%
%%%%%%%%%%%%%%%%%%%%%%%%%%%%%%%%%%%%%%%%%%%%%%%%%%%%%%%
\begin{figure*}[t]
 \begin{minipage}[]{\linewidth}
   \centering
   \includegraphics[angle=0, width=0.65\textwidth]{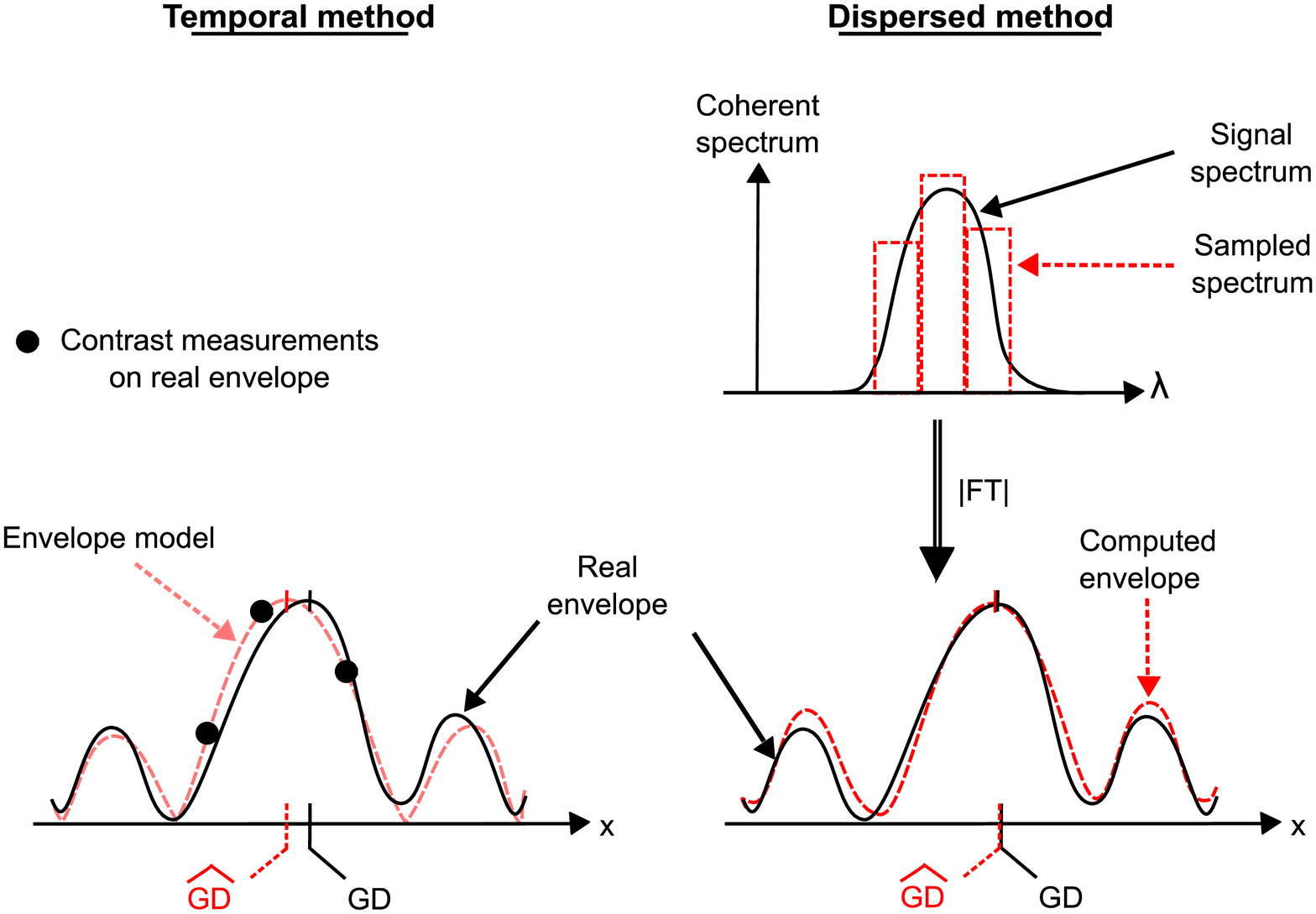}
    \caption{Conceptual representation of the signal processing for group delay estimation. Temporal method (left): an envelope model is fitted on the 3 envelope amplitude measurements to determine the group delay. Dispersed method (right): from the spectral sampling of the complex coherent signal, an approximated envelope is computed with a Fourier transform operation. The envelope position is determined by fitting an envelope model.}
    \label{fig:sigprocess}
 \end{minipage}
\end{figure*}
%%%%%%%%%%%%%%%%%%%%%%%%%%%%%%%%%%%%%%%%%%%%%%%%%%%%%%%
%
%

		%%%%%%%%%%%%%%%%%%%%%%%%%%%%%%%%%%%%%%%%%%%%%%%%%%%%%%%
		%%%%%%%%%%%%%%%%%%%%%%%%%%%%%%%%%%%%%%%%%%%%%%%%%%%%%%%
		\subsection{Description of the simulations}

We want to fairly compare both methods, so that:
\begin{itemize}
 \item We use the same fringe coding, i.e.\ a static ABCD because of its lower sensitivity to disturbances (see the previous section).
 \item The signal is integrated during the same amount of time so that each method collects the same amount of photons and is prone to the same disturbances.
 \item In both cases, the group delay is estimated in the same way by fitting an envelope model to the processed data. This allows a comparison of the intrinsic quality of the data for both methods. There are obviously many other ways to estimate $x_{\rm GD}$ from a set of data, but we assume that this is a second order problem. Indeed, \citet{pedretti_2004} compared three different algorithms to estimate the group delay with a temporal method and noted only little differences on the performance, even with an algorithm as sophisticated as the one proposed by \citet{wilson_2004}.
\end{itemize}
These choices made, temporal and dispersed methods can also be optimized in order to improve their performances. Here below, we describe the characteristics of each method.

	%%%%%%%%%%%%%%%%%%%%%%%%%%%%%%%%%%%%%%%%%%%%%%%%%%%%%%%
	\subsubsection{Temporally modulated interferogram}

Simulations have shown that the temporal estimator is strongly affected by atmospheric and instrumental disturbances. Their effect is all the more minimized than the envelope is quickly scanned. Our study shows that the optimal way to proceed is to successively measure the fringe contrast in three different points over a 5-fringe range (OPD equal to -2.5$\lambda$, 0 and 2.5$\lambda$ near the envelope maximum). This result is in agreement with the CHAMP choice \citep{bergerDh_2006}. Once the three contrasts are measured, they are fitted with an envelope model to determine the group delay. A schematic overview of this method is displayed in Fig.~\ref{fig:sigprocess}, left.

The input fluxes have to be monitored to compensate in real time for the photometric/contrast variations that occur between the 3 measurements. For sake of simplicity, we consider these photometric estimations to be noise-free. The simulated performance for the temporal method will thus be optimistic.

%
%
%%%%%%%%%%%%%%%%%%%%%%%%%%%%%%%%%%%%%%%%%%%%%%%%%%%%%%%
\begin{figure*}[t]
   \centering
   \begin{tabular}{ccc}
   \includegraphics[angle=0, width=0.35\textwidth]{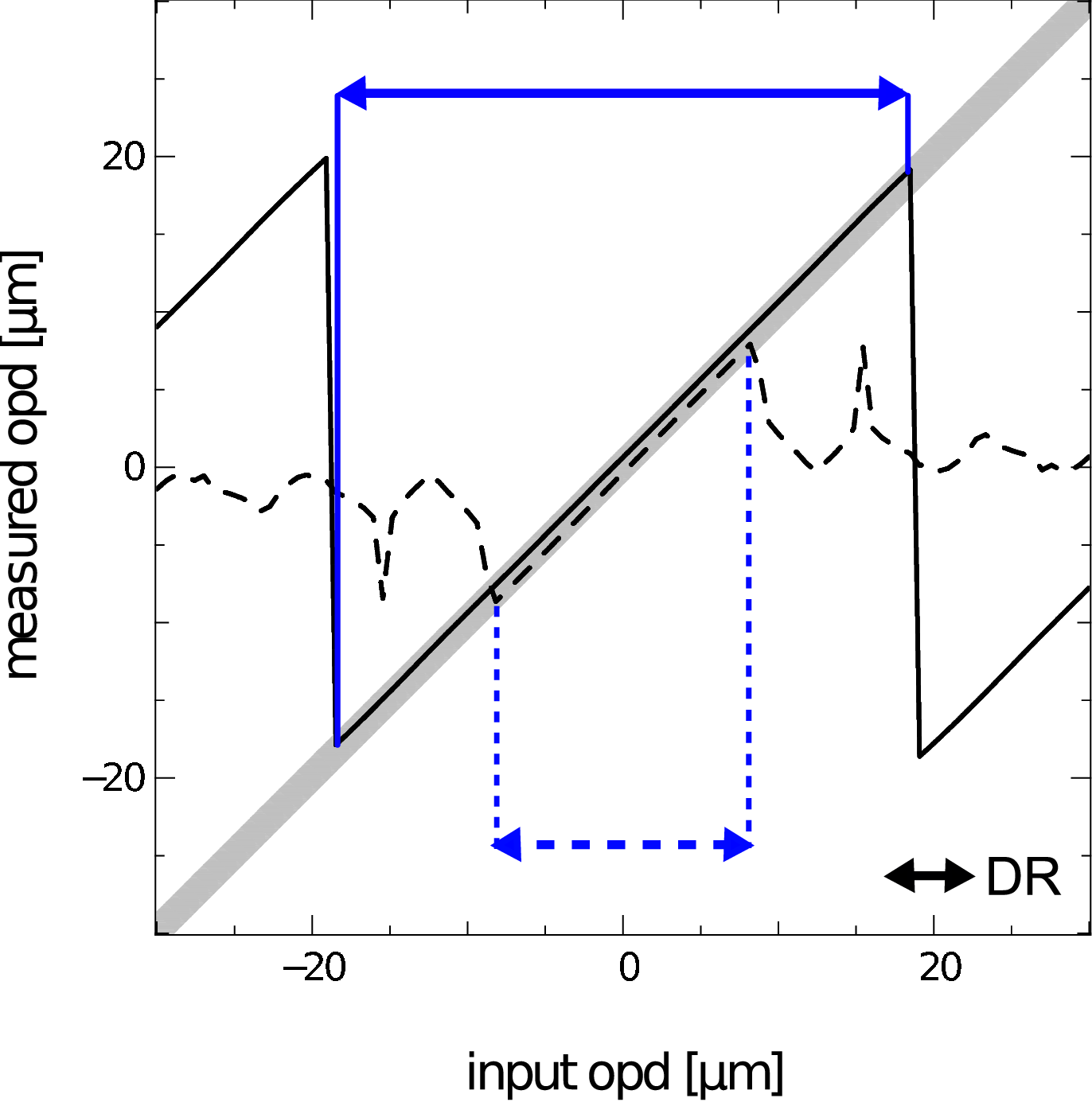} &
   \includegraphics[angle=0, width=0.35\textwidth]{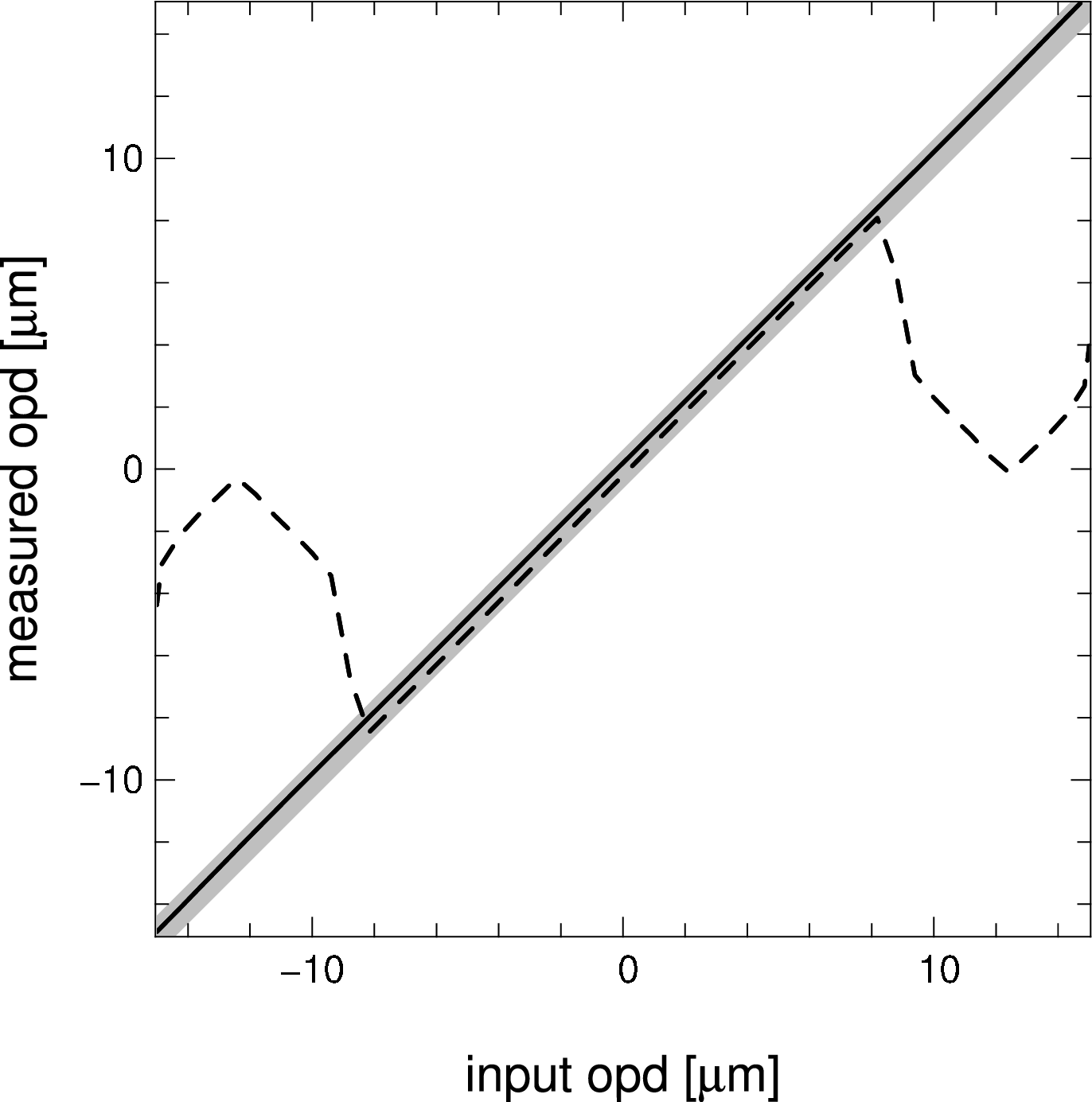} \\
   \includegraphics[angle=0, width=0.35\textwidth]{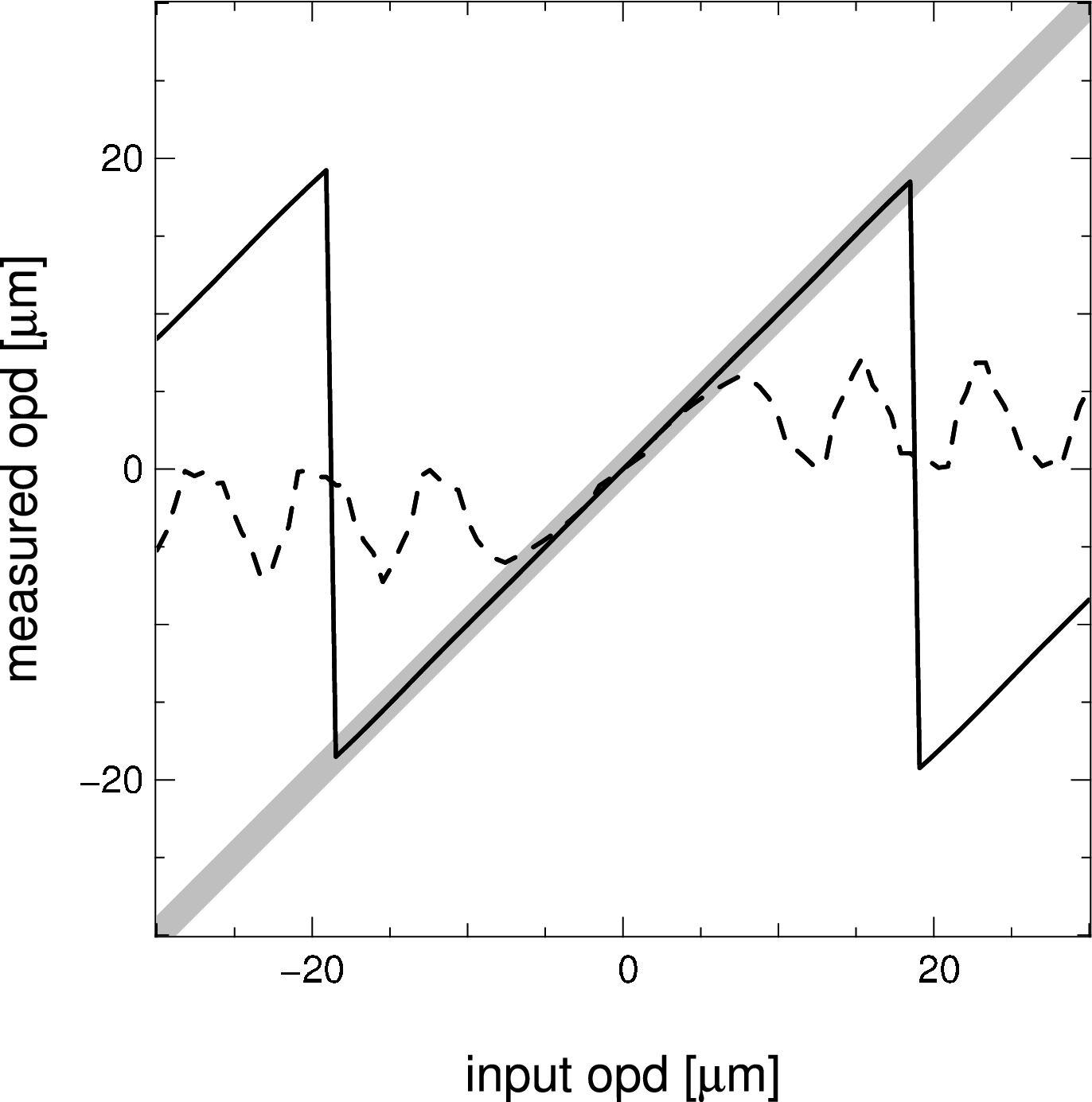} &
   \includegraphics[angle=0, width=0.35\textwidth]{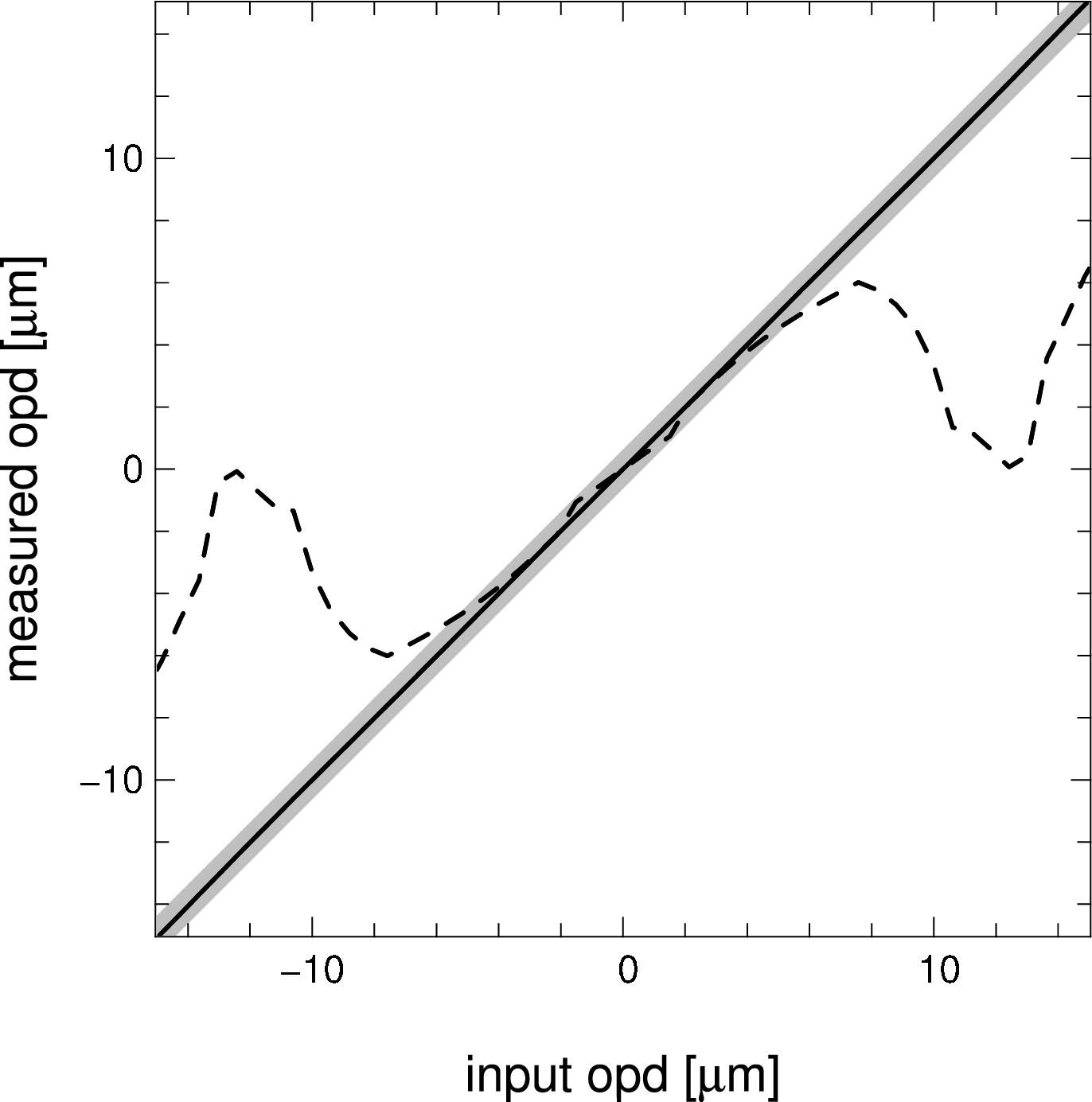}
   \end{tabular}
   \caption{Response of the dispersed and temporal group delay estimators (solid and dashed lines respectively) to an OPD ramp $x_{\rm GD}$ in H band. The ideal response is represented by the large grey line. In all cases, the coherence envelope has a sinc shape. Top: the envelope model is a sinc function.  Bottom: the envelope model is a parabola. Figures on the right are zoom on the central part of the left-hand side figures. The DR limits are represented with blue arrows on the top-left plot in the ideal case for both estimators.}
   \label{fig:linGD}
\end{figure*}
%%%%%%%%%%%%%%%%%%%%%%%%%%%%%%%%%%%%%%%%%%%%%%%%%%%%%%%
%
%

	%%%%%%%%%%%%%%%%%%%%%%%%%%%%%%%%%%%%%%%%%%%%%%%%%%%%%%%
	\subsubsection{Spectrally dispersed interferogram}

Thanks to the ABCD measurements, we can compute the chromatic complex visibility $I(\lambda) V(\lambda)$ on each spectral channel. An approximated coherence envelope is then computed by taking their discrete Fourier transform \citep{colavita_1999a}. It is possible to disperse the fringes over three channels to optimize the sensitivity, but we decide to use five channels to enhance the spectral sampling and thus the robustness of the estimator (see Section~\ref{part:GDlinearity}). For each exposure, a set of dispersed ABCD data is obtained, which enables a new GD estimation.

For a fair comparison between the temporal and the spectral method, they are both fed with the same disturbances and number of photons: therefore we make three GD estimations with the dispersed estimator, introducing disturbances between each estimation, and finally average them. \\

		%%%%%%%%%%%%%%%%%%%%%%%%%%%%%%%%%%%%%%%%%%%%%%%%%%%%%%%
		%%%%%%%%%%%%%%%%%%%%%%%%%%%%%%%%%%%%%%%%%%%%%%%%%%%%%%%
		\subsection{Linearity and dynamic range}                      \label{part:GDlinearity}

A reliable estimation of the group delay is of prime importance since it ensures the measurements to be made in the highest SNR area. We study in this part two quantities, the linearity and the dynamic range, by looking at the response $\hat{x}_{\rm GD}$ of both methods to a given OPD ramp $x_{\rm GD}$. We define the linearity $\eta$ as the local slope of $\hat{x}_{\rm GD}$ versus $x_{\rm GD}$:
\begin{equation}
	\eta = \frac{\partial \hat{x}_{\rm GD}}{\partial x_{\rm GD}}
\end{equation}
A perfectly linear estimator is such that $\eta=1$. Otherwise the estimator is biased and the envelope is not perfectly stabilized.
%$\eta$ therefore quantifies the bias between the group delay estimation and its real value.

The group delay is extremely important for the fringe tracking robustness, that is the ability of the estimator to keep the fringes locked in the highest SNR area, in particular after a strong piston stroke ($\geq 15 \,\mu m$). In practice, there are limits outside which the group delay estimation is highly biased and makes the fringe tracker diverge from its operating point. The interval between these limits corresponds to the so-called dynamic range (DR), which is used here to characterize the robustness of the estimators. In practice, the limits of the DR are reached when the slope of $\hat{x}_{\rm GD}$ versus $x_{\rm GD}$ changes sign (in other words when $\eta$ becomes negative) or when we observe a strong wrapping effect.

In the following paragraphs, we simulate noise-free ideal interferograms in the H-band with a sinc-shaped coherence envelope. We fit the results with two different envelope models (a parabola and a sinc function) in order to study its impact on the GD estimation. The results of this study are presented in Fig.~\ref{fig:linGD}.

\subsubsection{Temporally modulated interferogram}

In the temporal method, the envelope model is critical to ensure a good linearity.
% since we only sample a limited number of points inside the envelope.
Using the most appropriate sinc model with ideal interferograms (Fig.\ \ref{fig:linGD}, top, dashed line), the linearity is excellent ($\eta=1$), but the DR is limited to 10 fringes ($\pm 8\, \mu m$), i.e.\ to the width of the central lobe.
%Note it is more important than the scan length thanks to the fit of the envelope model.
Outside this range the GD estimation is totally non-linear but never cross the y-axis: the fringe tracking loop should not diverge but it should recover the envelope center with difficulty, or even could risk to lock the fringes far away from the envelope center.

Using a wrong envelope model (e.g., a parabola; see Fig.~\ref{fig:linGD}, bottom, dashed line) leads to a relative bias higher than $10\%$ ($\eta\sim 0.9$) whatever the OPD within the dynamic range. Increasing the number of samples or the scan length does not improve the results, emphasizing that the problem comes from the wrong envelope model. Because of the number of chromatic variables (particularly the longitudinal dispersion) which continually vary during a night and slightly modify the envelope shape, the envelope model cannot be perfect and the temporal estimator will therefore be consistently non linear by a few percents. Interestingly the DR is still equal to the width of the main lobe\footnote{Simulations show that the DR can be increased with a higher number of contrast samples and a higher scan length. But in real operations it also increases the influence of atmospheric disturbances, which is not suitable for precision purposes (see Section \ref{part:GDrms}).} and seems weakly affected by the model quality. \\

\subsubsection{Spectrally dispersed interferogram}

On the contrary, the dispersed method is not affected by the envelope model (see Fig. \ref{fig:linGD}, solid lines): since we sample the complex coherent spectrum, we can directly compute a realistic coherence envelope and the fitting model has therefore a weak influence. Dispersing fringes on 5 spectral channels in H-band, the linearity is excellent ($\eta\sim 1$) over an OPD range of $\pm 20\mu m$. Beyond these points a sharp wrapping effect is observed (Fig.~\ref{fig:linGD}, left), marking the DR limits: the discrete sampling of the spectrum induces aliasing effects on the computed envelope (obtained from a discrete Fourier transform of the complex coherent signal, see Eq. \ref{eq:def-envco}), so that outside the DR the GD is estimated on a replica of the true envelope. In practice, if the GD is measured after such a wrap, the fringe tracker will correct the OPD in the wrong direction and finally lock the fringes on a point even more distant from the envelope center than previously. However, since we have chosen to use 5 spectral channels, the DR ($\pm 20\mu m$) is larger than the strongest piston fluctuations typically observed on a few milliseconds ($\sim 15\mu m$). Note that working in K-band increases the dynamic range up to $\pm 40\mu m$, almost cancelling such issues. It is actually possible to infer an expression for the DR with dispersed fringes. Let us assume a spectral band with an effective wavelength $\lambda_0$ and a width $\Delta \lambda$, and that the fringes are dispersed over $N_\lambda$ channels. The dynamic range is then ideally (see Appendix \ref{app:DRdisp}):
\begin{equation}
DR = N_\lambda \frac{\lambda_0^2}{\Delta \lambda}
\end{equation}
The larger the number of spectral channels, the lower the aliasing and therefore the larger the DR. This relation is in excellent agreement with the simulation results.

%
%
%%%%%%%%%%%%%%%%%%%%%%%%%%%%%%%%%%%%%%%%%%%%%%%%%%%%%%%
\begin{figure}[t]
   \centering
   \includegraphics[angle=0, width=0.4\textwidth]{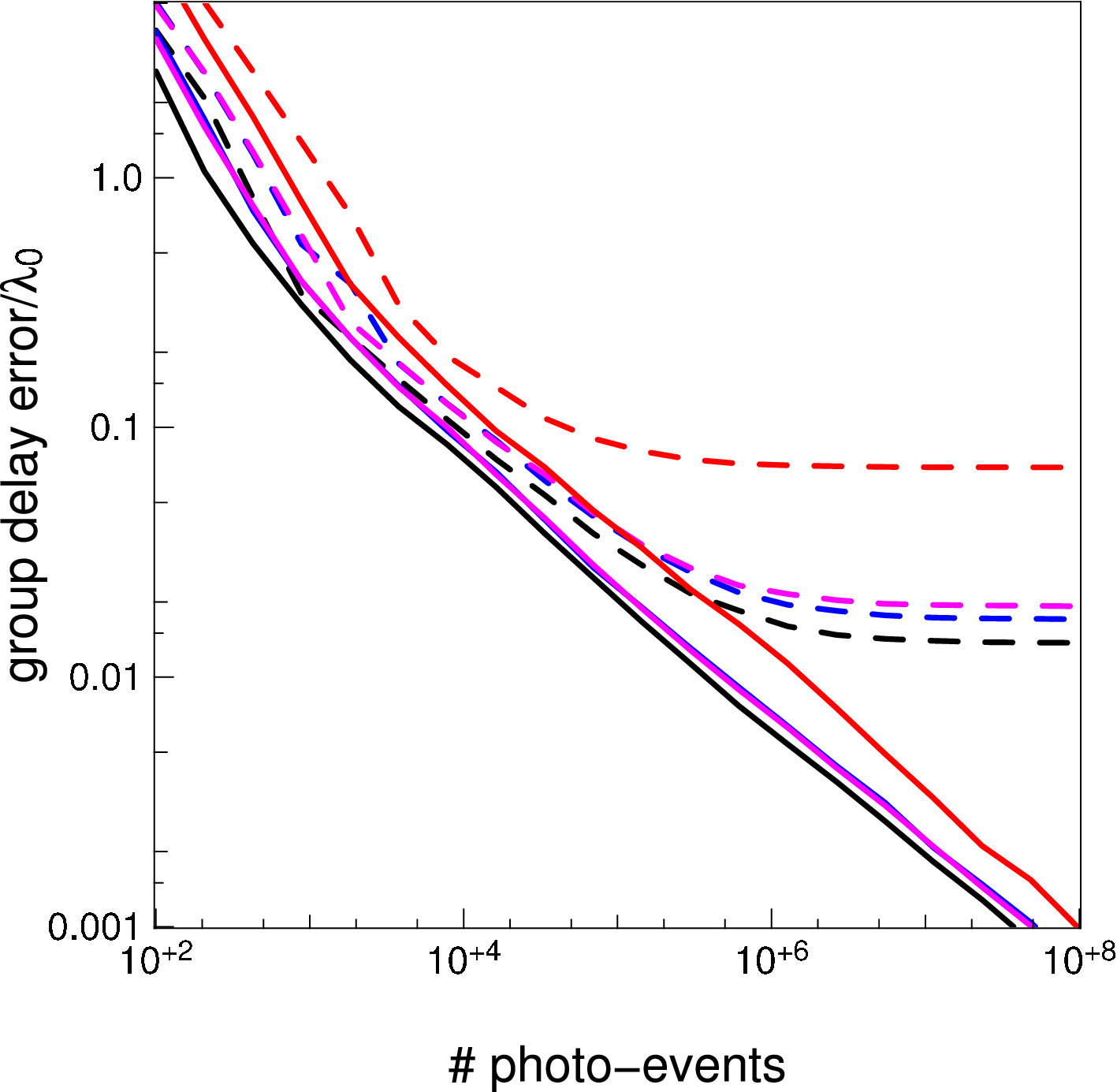}
   \caption{Relative errors $\sigma_{\rm GD}/\lambda$ of temporal (dash) and dispersed (solid) GD estimators in different atmospheric conditions. Black, blue, magenta and red curves represent Excellent, Good, Medium and Bad conditions respectively as defined in Tab.\ \ref{tab:atm_conditions}. The plots are done in the case of the ATs, for a total integration time of 3\,ms, constituted of 3 single exposures of 1\,ms.}
   \label{fig:GDrmsA1}
\end{figure}
%%%%%%%%%%%%%%%%%%%%%%%%%%%%%%%%%%%%%%%%%%%%%%%%%%%%%%%
%
%

When longitudinal dispersion is taken into account, the linearity and DR are slightly reduced because the undersampling of the coherent spectrum leads to a less precise envelope computation. Refining the spectral sampling with more channels improves both linearity and DR as shown by the simulations.

In conclusion, spectrally dispersing the fringes appears to be the most robust method to measure the group delay. It provides an estimator with:
\begin{itemize}
\item a good linearity without the need of a good envelope model, as it inherently computes a realistic envelope;
\item a large DR allowing robust operations and quick recovery of the fringes over an OPD range larger than typical piston variations.
\end{itemize}
%

		%%%%%%%%%%%%%%%%%%%%%%%%%%%%%%%%%%%%%%%%%%%%%%%%%%%%%%%
		%%%%%%%%%%%%%%%%%%%%%%%%%%%%%%%%%%%%%%%%%%%%%%%%%%%%%%%	
		\subsection{Group delay measurements precision}               \label{part:GDrms}

We now compare the precision of the GD estimators as a function of the incoming flux and of the disturbances strength. The simulations consist in computing noisy interferograms in H-band, introducing detector
%noise (we consider a read-out noise equal to 10$e^-$ per pixel)
and photon noises as well as piston and photometric disturbances, which are taken from actual FINITO data. For each simulation, we estimate a noisy GD ($\hat{x}_{\rm GD}$). Its statistics over several thousands of iterations gives the statistical error $\sigma_{\rm GD}$ for both estimators.

The results for ATs and an integration time of 1\,ms are presented in Fig.~\ref{fig:GDrmsA1}. It shows the relative error $\sigma_{\rm GD}/\lambda$ on the group delay measurements as a function of the number of photo-events for various atmospheric conditions. The limitation of the temporal estimator is obvious, with a plateau due to atmospheric disturbances (piston mainly) which acts like an independent, additional noise at high flux, increasing when atmospheric conditions get worse. On the contrary the dispersed estimator appears weakly sensitive to these disturbances. Although we have used favorable hypothesis for the temporal method (the required photometric monitoring is considered noise-free), there is no regime in which this concept is better than the dispersed one. For UTs, results are similar but with stronger limitations: it appears that the statistical error of the temporal estimator never goes below $\lambda/4$ with integration time as low as 1\,ms whatever the conditions.

Additionally, all the simulations show the same dependency of the statistical error of both GD estimators with respect to the incoming flux $K$ and the visibility $V$:\footnote{These empirical relations are only valid when there is no disturbance for the temporal method.}
\begin{eqnarray}
\label{eq:GDnoisemodel1}
&\sigma_{\rm GD}^2 \varpropto \displaystyle\frac{1}{K \, V^2}     &\quad \mathrm{in \, photon \, noise \, regime} \\
%,} \\
\label{eq:GDnoisemodel2}
&\sigma_{\rm GD}^2 \varpropto \displaystyle\frac{1}{K^2 \, V^2} &\quad \mathrm{in \, detector \, noise \, regime}
%.}
\end{eqnarray}
Interestingly, we find the same kind of dependency than for the phase (Eq.~\ref{eq:det_noise}) in the equivalent regimes. 

In conclusion, temporally modulating the OPD to estimate the group delay is not competitive with the spectrally dispersed fringe method, both in terms of robustness and precision. This is in line with the conclusion of Section~\ref{part:phase}, which showed the sensitivity of temporal fringe coding to external disturbances. We therefore strongly conclude that a static fringe coding scheme dispersed across a few spectral channels should be used to measure the fringe phase and group delay.

			%%%%%%%%%%%%%%%%%%%%%%%%%%%%%%%%%%%%%%%%%%%%%%%%%%%%%%%
			%%%%%%%%%%%%%%%%%%%%%%%%%%%%%%%%%%%%%%%%%%%%%%%%%%%%%%%
			%%%%%%%%%%%%%%%%%%%%%%%%%%%%%%%%%%%%%%%%%%%%%%%%%%%%%%%
			\section{Optimal co-axial pairwise combination schemes}                                \label{part:optschemes}

%
%
%%%%%%%%%%%%%%%%%%%%%%%%%%%%%%%%%%%%%%%%%%%%%%%%%%%%%%%
\begin{figure*}[t]
	\centering
		\begin{tabular}{cccccc}
			\includegraphics[angle=0, width=0.14\textwidth]{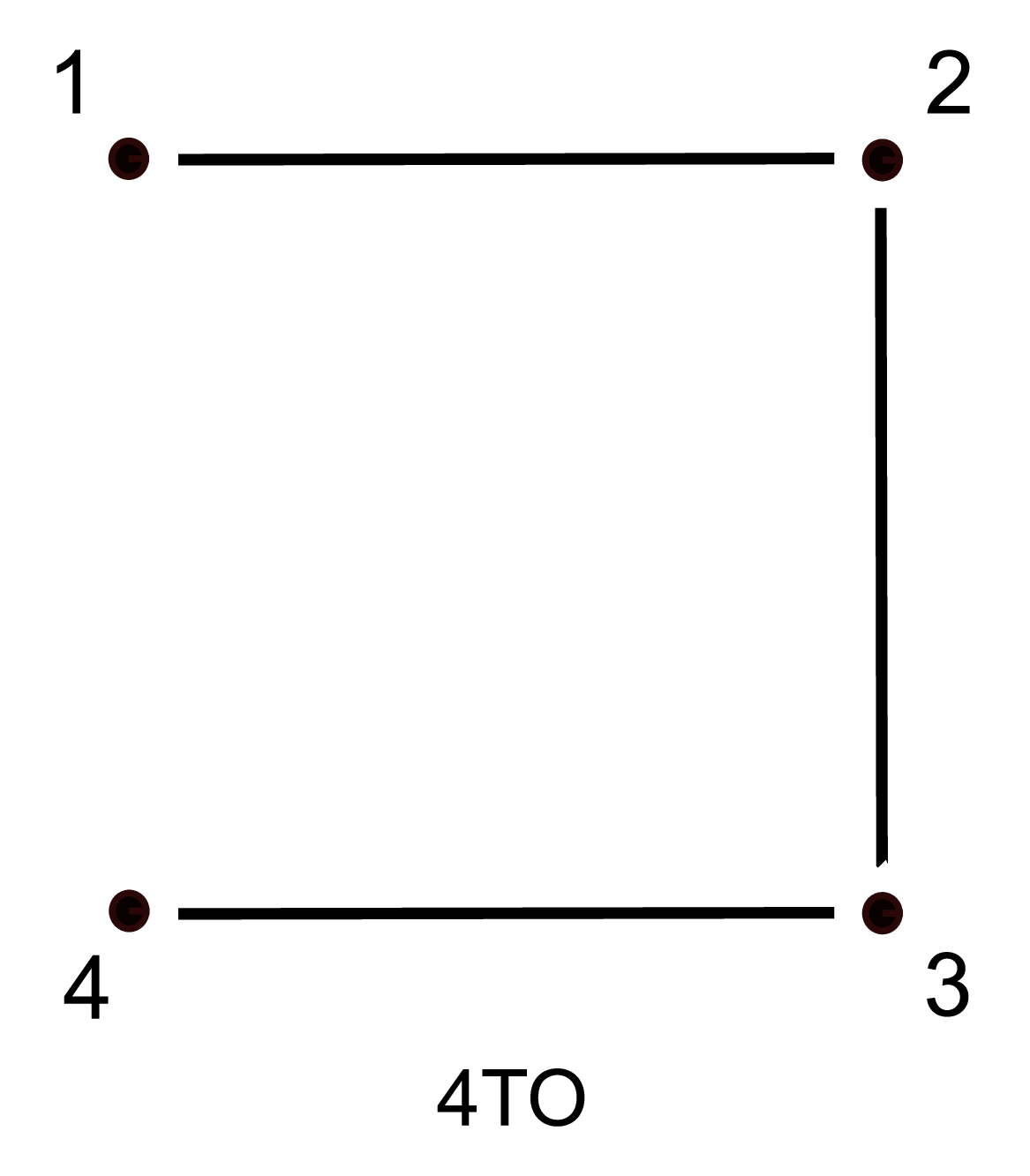}     &
			\includegraphics[angle=0, width=0.14\textwidth]{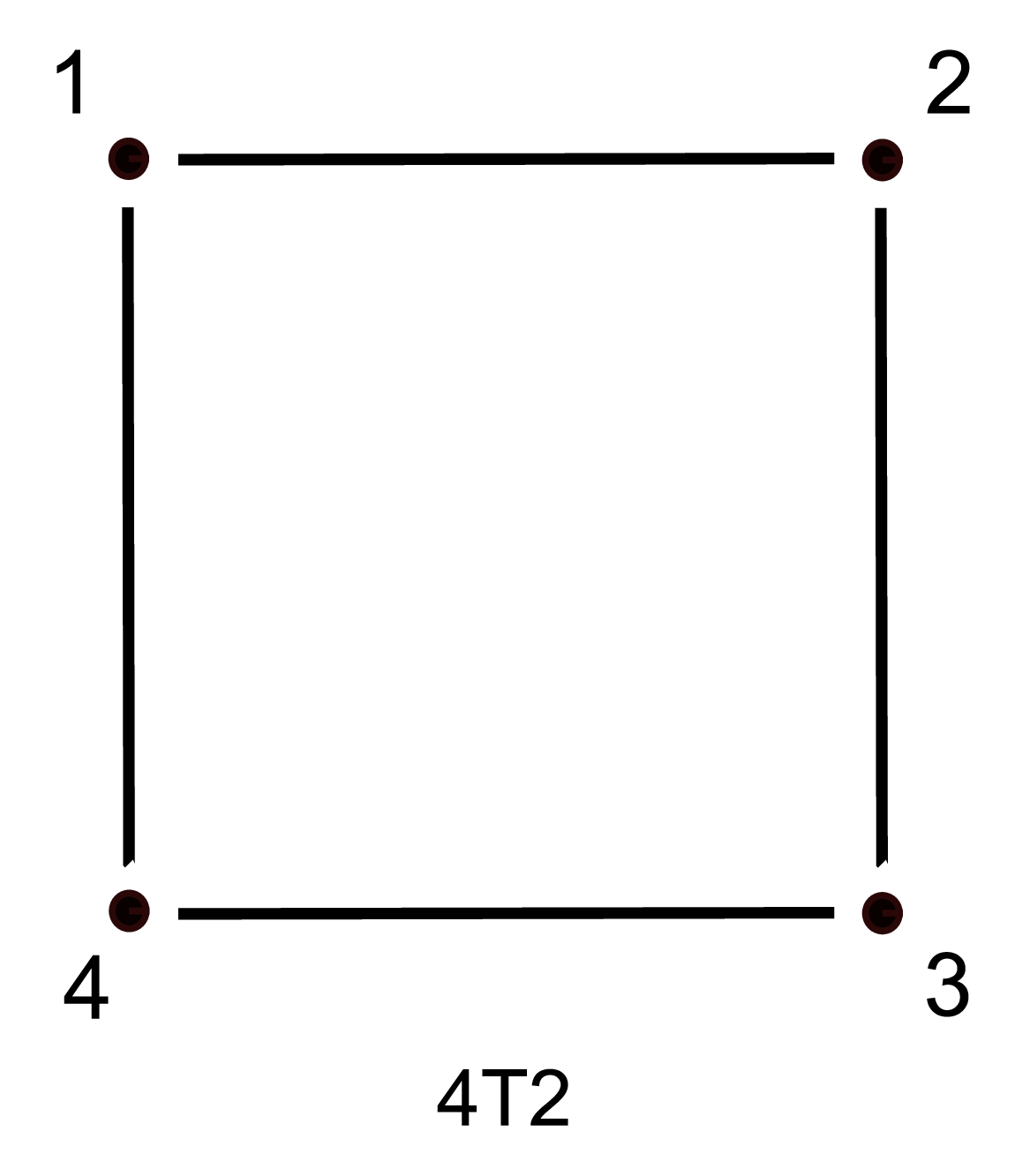}   &
			\includegraphics[angle=0, width=0.14\textwidth]{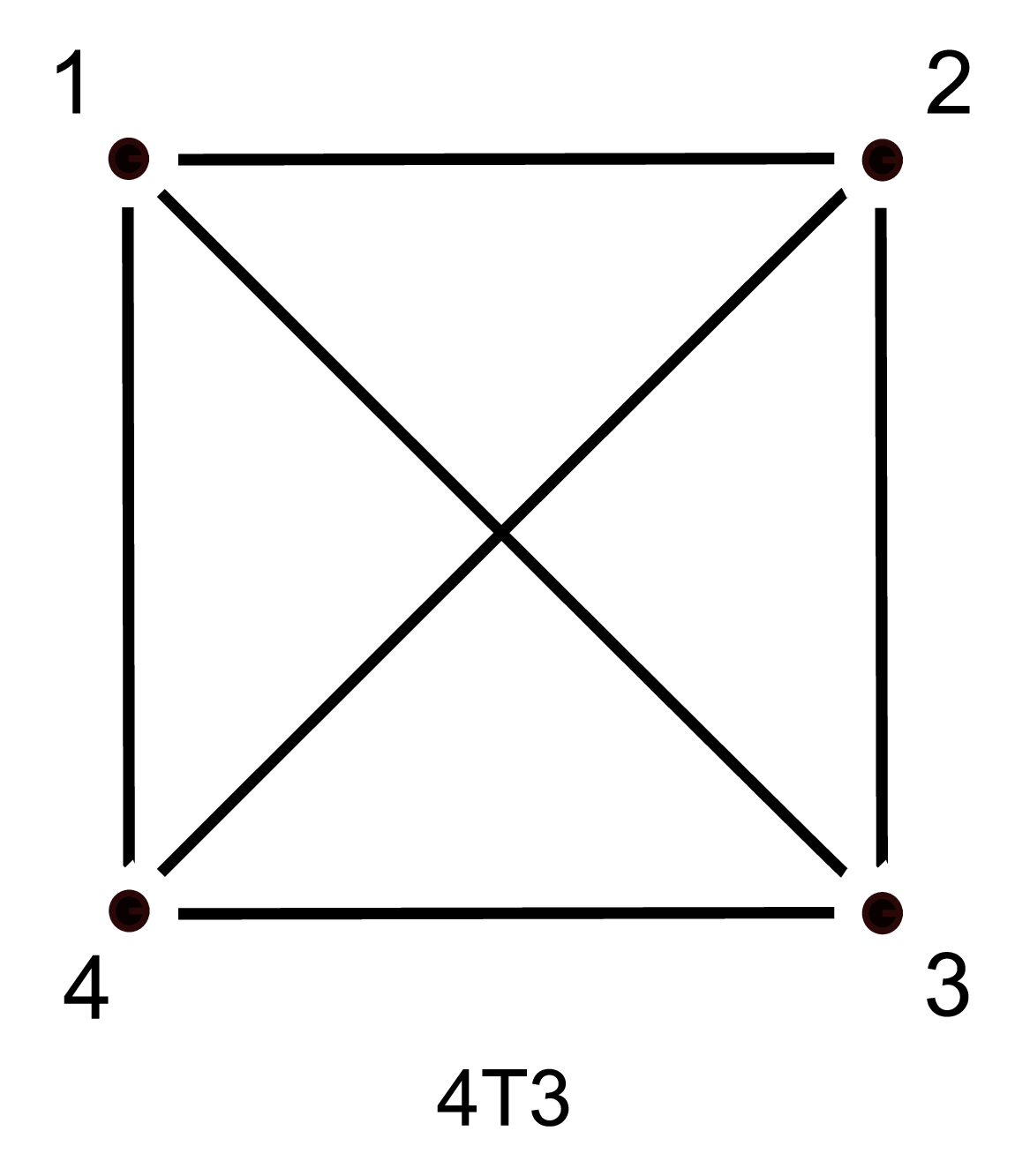}   &
		                                                                                      &
		                                                                                      &
		                                                                                      \\ \\
			\includegraphics[angle=0, width=0.14\textwidth]{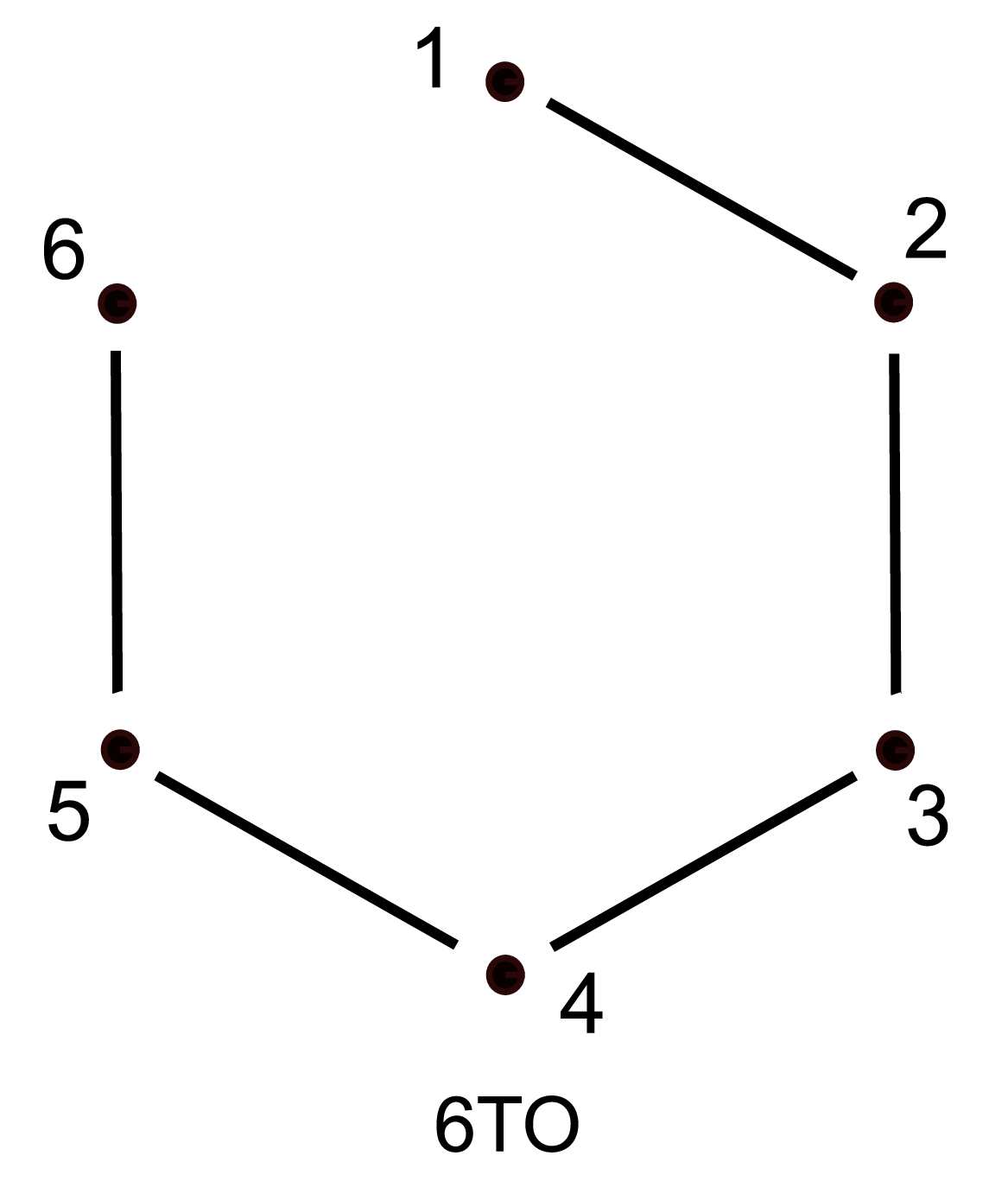}     &
			\includegraphics[angle=0, width=0.14\textwidth]{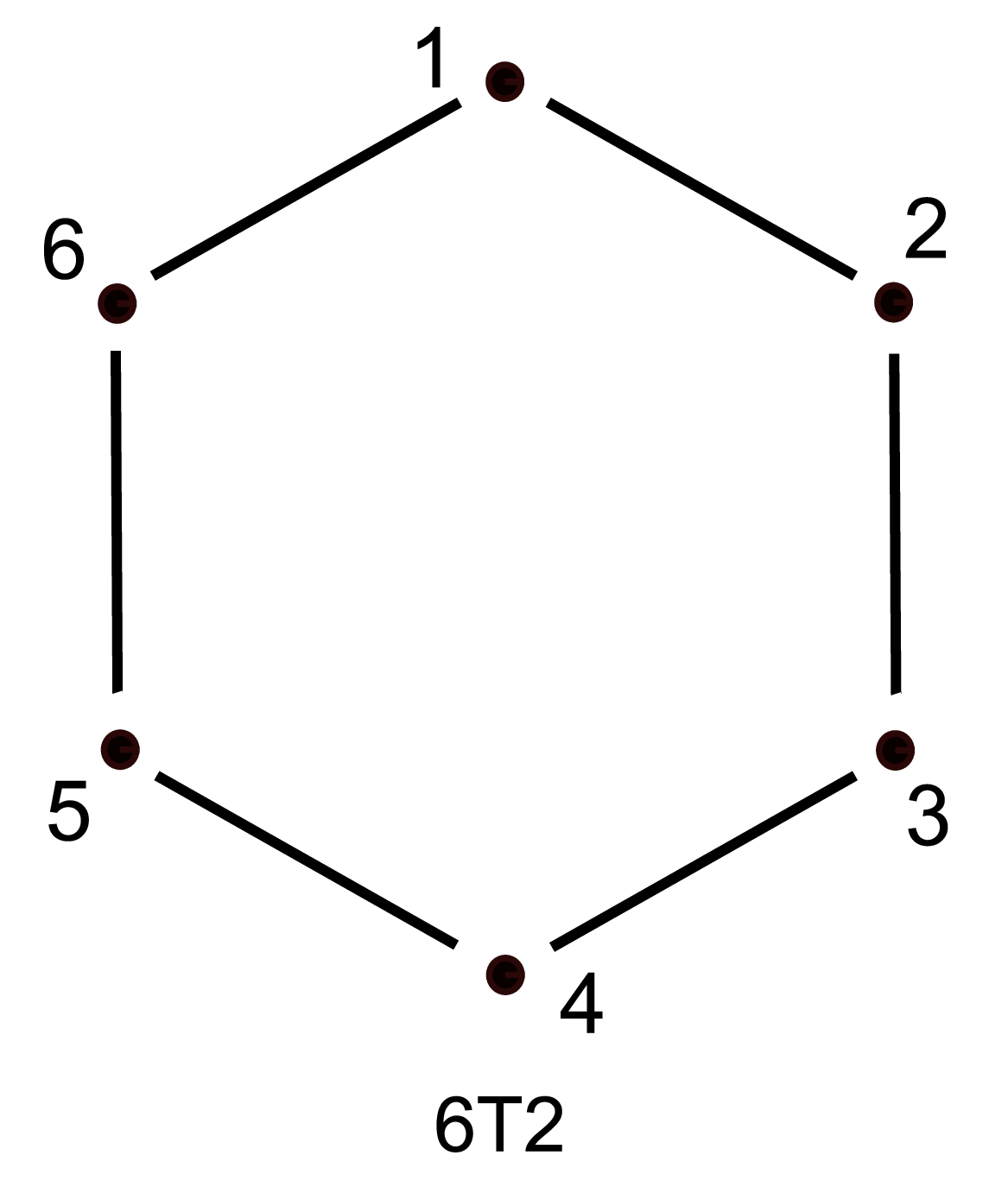}     &
			\includegraphics[angle=0, width=0.14\textwidth]{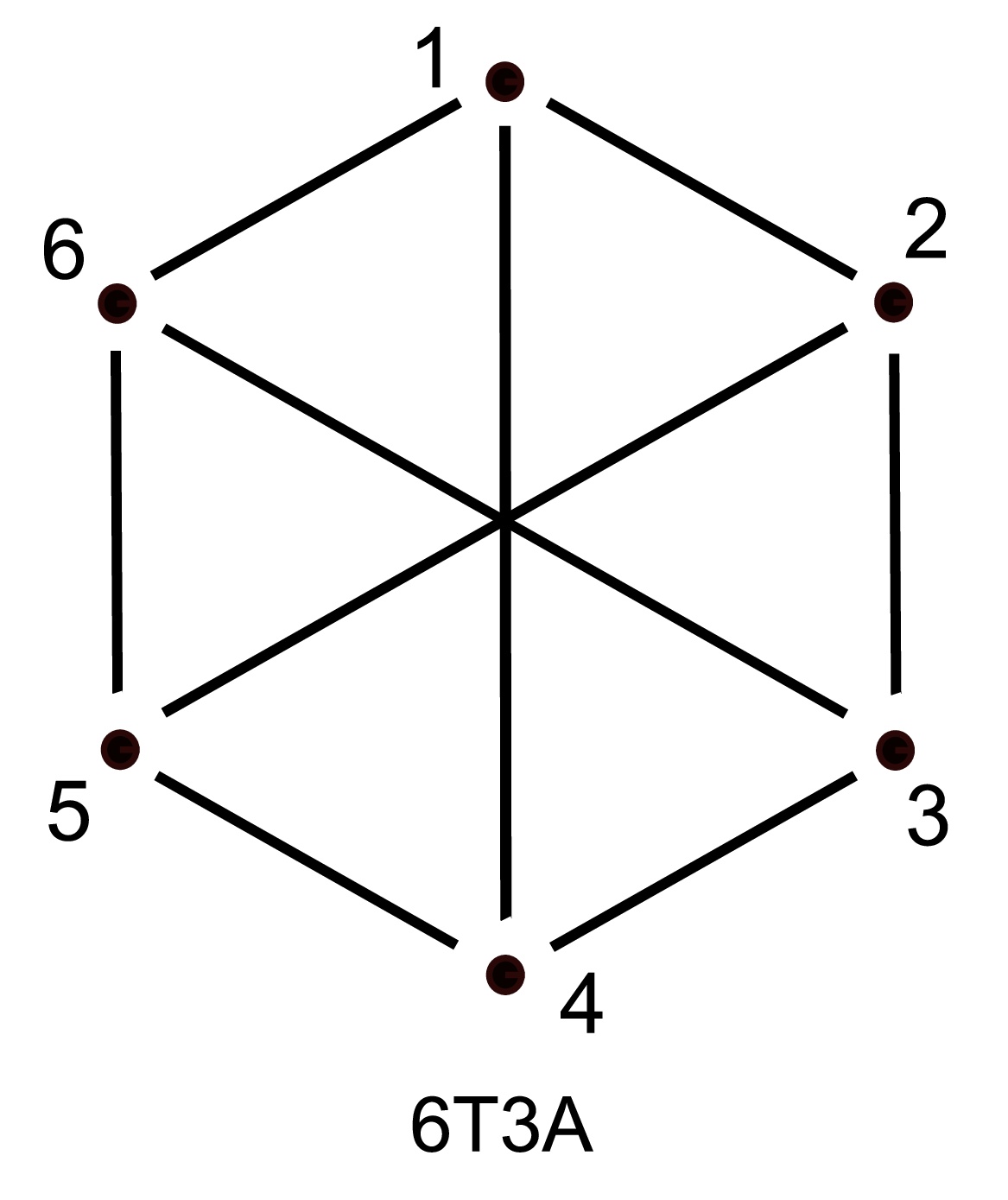}     &
			\includegraphics[angle=0, width=0.14\textwidth]{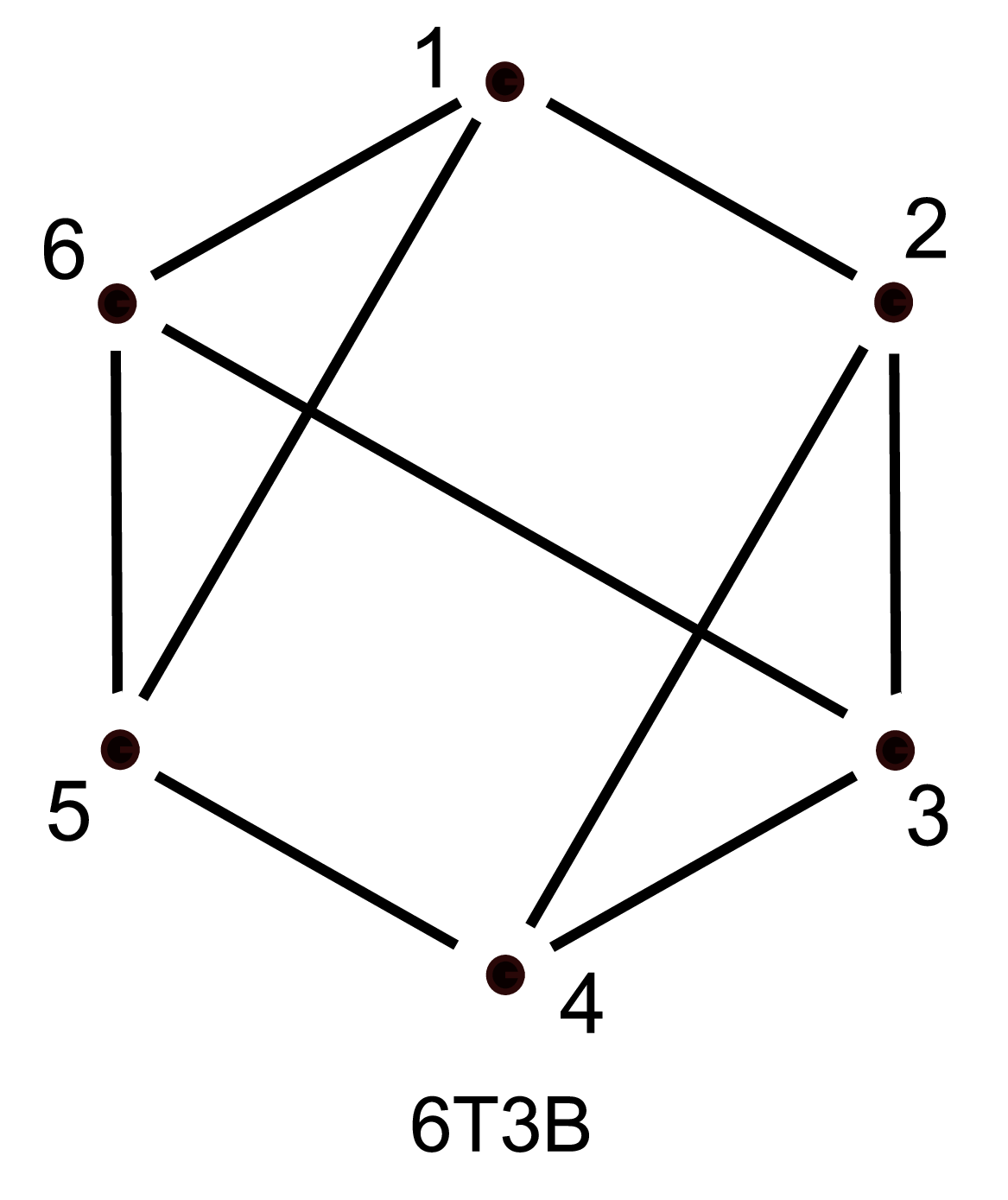}     &
			\includegraphics[angle=0, width=0.14\textwidth]{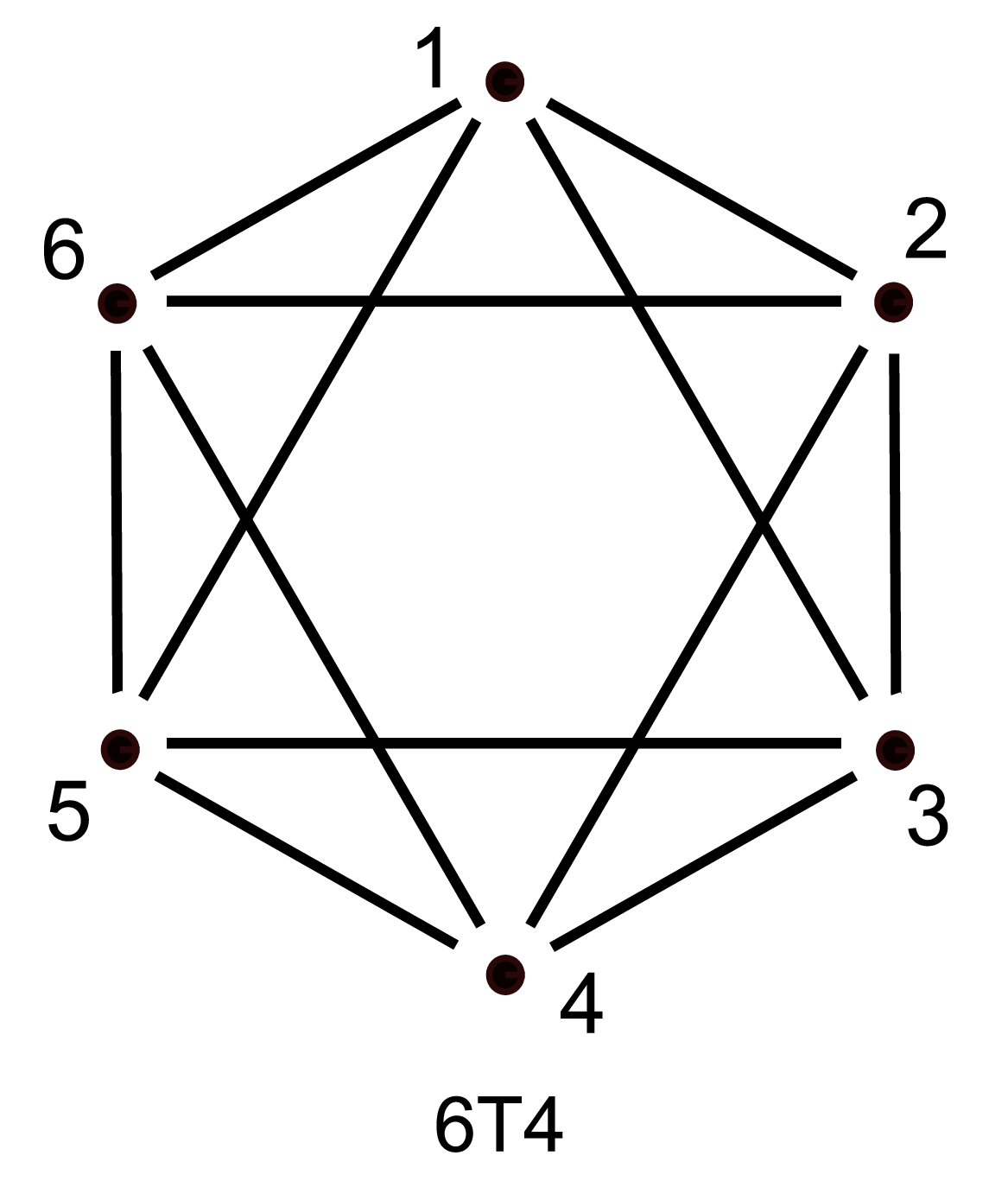}     &
			\includegraphics[angle=0, width=0.14\textwidth]{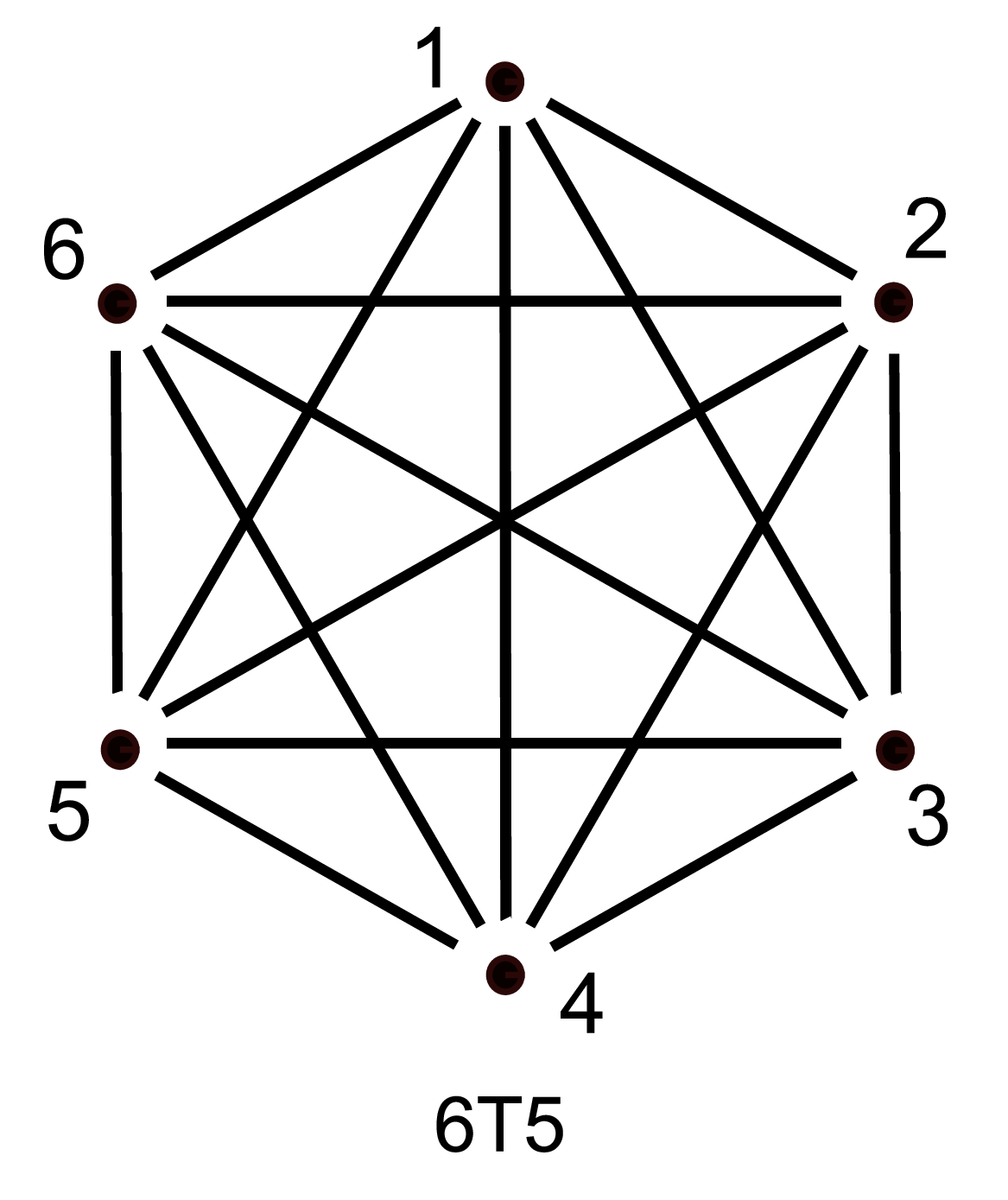}     \\
%			\includegraphics[width=0.14\textwidth]{8TR2}     &
%			\includegraphics[width=0.14\textwidth]{8TR3}     &
%			\includegraphics[width=0.14\textwidth]{8TR4A}    &
%			\includegraphics[width=0.14\textwidth]{8TR4B}     &
%  		                                                                                      &
%		                                                                                      \\
%			\includegraphics[width=0.14\textwidth]{8TR5A}     &
%			\includegraphics[width=0.14\textwidth]{8TR5B}     &
%			\includegraphics[width=0.14\textwidth]{8TR6}     &
%			\includegraphics[width=0.14\textwidth]{8TR7}     &
%  		                                                                                      &
%                                                                                              \\
		\end{tabular}
		\caption{The various conceptual schemes studied for the 4T and 6T cases (top and bottom respectively) with the associated nomenclature. As explained in section \ref{part:perf_study}, we do not study schemes with intrinsically imbalanced photometric inputs other than the open ones, because of lower performance.
		%In the 8T case, only the internal baselines linked to the first telescope are drawned, for the sake of clarity and because the 8T schemes are geometrically symetric.
		}
		\label{fig:schemes}
\end{figure*}
%%%%%%%%%%%%%%%%%%%%%%%%%%%%%%%%%%%%%%%%%%%%%%%%%%%%%%%
%
%

Theoretically, it is possible to cophase an array of $N$ telescopes by measuring only $N-1$ baselines. However because of the noisy measurements and of the varying observing conditions during a night, some baselines can deliver information of poor quality, so that it is beneficial to have some redundancy with additional baselines. It is then possible to retrieve the phase on a baseline in several different ways, ensuring a better fringe tracking stability. The drawback is that when the number of measured baselines increases, each one is less sensitive because the flux of the telescopes is divided between more baselines. The sensitivity of the fringe sensor then depends on a competition between the information redundancy and the sensitivity of the individual baselines. The aim of this section is to determine the most efficient schemes with respect to their intrinsic performance and operationnal advantages.

Several on-going projects will work with 4 (GRAVITY, MATISSE) to 6 (VSI, MIRC) telescopes. Therefore we focus on these 2 cases, assuming that all telescopes are identical. We consider the following schemes illustrated in Fig.~\ref{fig:schemes}:
\begin{itemize}
\item The \textbf{open schemes} are made up of the minimal number of baselines, that is $N-1$ and are noted $N$TO. In this case the interferometric outputs are intrisically imbalanced in flux, in order to have baselines with equivalent performance. For instance, in the 4TO case, we do not distribute $50\%$ of the intensity of the telescope 2 onto baselines $\{12\}$ and $\{23\}$, but $\sim 40\%$ and $\sim 60\%$ respectively (see Appendix~\ref{app:openschemes} for the details of this optimization).
\item In the \textbf{redundant schemes}, the flux of each telescope is equally divided between the same number $R$ of baselines. When $R$\,=\,2 the schemes are more precisely called circular. The nomenclature to designate them in the following is $N$T$R$, possibly with an additionnal letter when there are several possibilities for the same value of $R$.
\end{itemize}

		%%%%%%%%%%%%%%%%%%%%%%%%%%%%%%%%%%%%%%%%%%%%%%%%%%%%%%%
		%%%%%%%%%%%%%%%%%%%%%%%%%%%%%%%%%%%%%%%%%%%%%%%%%%%%%%%
		\subsection{Study of the combination schemes}             \label{part:perf_study}

We have decided to compare the various schemes on the base of three considerations: their intrinsic performances, their ability to provide the individual beam photometries without dedicated outputs, and their robustness to unpredictable and rapidly varying observing conditions.

\subsubsection{Performance study}

The principle of our analysis is similar to the one led for the GRAVITY fringe tracker by \citet{houairi_2008}. It consists in computing the vector of the optimal optical path estimators $x$ used to drive the delay lines, from the noisy and possibly redundant phase information $\phi$. These quantities are linked by the interaction matrix $\textbf{M}$ which is known:
\begin{equation}
   \phi = \textbf{M} \, x
\end{equation}
With redundant schemes, the system is overdetermined so that we use a $\chi^2$ minimization procedure to compute the control matrix $\textbf{W}$ and then $x$:
\begin{equation}
   x = \textbf{W} \, \phi
\end{equation}

Since $\phi$ is noisy, we have to take into account the error on the measurement when computing $\textbf{W}$, in order to reduce the impact of the noisiest baselines and prevent the solution from diverging. The quantity of interest is finally the error $\sigma_{ij}$ on the corrected differential pistons calculated for each baseline $\{ij\}$ with respect to a reference noise $\sigma_0$, which corresponds to the error of a simple two-telescope interferometer. The expression of $\sigma_0$ is derived from \citet{shao_1988} or, in a more general form, from \citet{tatulli_2010a}. It depends on the considered noise regime, so that the detector and photon noise regimes can be independently studied:
\begin{eqnarray}
	 \sigma_0^{det}   &= \displaystyle\frac{A}{K \, V}\\
	 \sigma_0^{\rm phot} &= \displaystyle\frac{B}{\sqrt{K} \, V}
\end{eqnarray}
where $A$ and $B$ are proportionality factors depending only on the fringe coding, so that this study is independent on the phase and the group delay estimators used. Results for the different schemes are therefore perfectly comparable within the same regime. Note that the above expressions also agree with our previous results concerning the group delay (Eqs.~\ref{eq:GDnoisemodel1} and \ref{eq:GDnoisemodel2}).

To analyze the behavior of the different schemes in realistic conditions, we consider the following three cases:
\begin{itemize}
	\item \textbf{Ideal case}: all the baselines are strictly equivalent in terms of flux and visibility.
	\item \textbf{Resolved source case}: one baseline of the array is highly resolving the source (cases e.g.\ of an asymmetric source or of a very long baseline). To study this case, we set the fringe visibility to 0.1 on one particular baseline, and to 1 on the others.
	\item \textbf{Low flux case}: the flux of one telescope is set to one tenth of the others, to simulate a quick variations of flux (e.g.\ scintillation) or a technical problem.
\end{itemize}
The results for these three cases are presented in Tab.\ \ref{tab:idealcase} to \ref{tab:lowflux},  showing the relative error $\epsilon_{ij} = \sigma_{ij}/\sigma_0$ on the corrected piston for the various baselines $\{ij\}$.

In the ideal case (Tab.~\ref{tab:idealcase}), the redundancy slightly degrades the performance in the detector noise regime (because the signal is coded on a larger number of pixels) but does not impact the performance in the photon noise regime. The differences are at maximum of the order of $30\%$ between the various schemes. Open and circular schemes provide similar performance. However, in the open schemes, the flux is not divided equally between the various baselines to reach an optimal SNR (see Appendix~\ref{app:openschemes}). Although the baselines at both ends of the array receive roughly $40\%$ more photons than the others, they are affected by a photometric imbalance, leading to a fringe contrast loss of roughly $10\%$ (i.e.\ an SNR loss around 20\%): this points to the fact that the input photons are not optimally used. On the other hand the schemes with more baselines benefit of some redundancy. These facts explain why open schemes are slightly less sensitive in the photon noise regime than redudant -- and balanced -- ones. A similar conclusion concerning open schemes was already reached by \citet{houairi_2008} in the 4T case.

In the case where a baseline resolves the target (Tab.~\ref{tab:ressource}), the benefit of redundancy clearly appears. Indeed, whereas the measurement error on the resolving baseline strongly increases with open schemes, the performance degradations are well contained with the redundant ones. There is still a significant improvement between $R$\,=\,2 and $3$, but only limited differences between more redundant schemes.

When a telescope has a reduced flux (Tab.~\ref{tab:lowflux}), the overall results do not significantly vary between the various schemes. Having a minimal redundancy ($R$\,=\,2) appears optimal in the detector noise regime, since more baselines induce a larger overall read-out noise. In the photon noise regime, redundant schemes have very close performances and are slightly more efficient than the open ones. Hence circular scheme should be favored with respect to open ones and the use of more redundant schemes is not essential from the performance point of view.

Taking into account the relatively close performance between the redundant concepts and regarding their instrumental complexity (number of baselines to be coded, optical transmission, etc.), schemes with $R$\,=\,2 or 3 should be favored.

%
%
%%%%%%%%%%%%%%%%%%%%%%%%%%%%%%%%%%%%%%%%%%%%%%%%%%%%%%%
\begin{table}[t]
	\begin{center}
		\begin{tabular}{ccc}
\hline
\hline
             & Detector noise & Photon noise \\
Scheme & $\epsilon_{ij}$ &  $\epsilon_{ij}$ \\
\hline
4TO & 1.6 & 1.3\\
4T2 &  1.7 & 1.2 \\
4T3 &  2.1 & 1.2 \\
\hline
6TO & 1.8 & 1.4  \\
6T2 &  1.8 &  1.3 \\
6T3A &  2.2 &  1.3 \\
6T3B &  2.2 - 2.3 & 1.3 \\
6T4 &  2.6 & 1.3 \\
6T5 &  2.9 & 1.3 \\
%\hline
%4TO & 1.62 & 1.31\\
%4T2 &  1.73 & 1.22 \\
%4T3 &  2.12 & 1.22 \\
%\hline
%6TO & 1.81 & 1.36  \\
%6T2 &  1.83 &  1.29 \\
%6T3A &  2.23 &  1.29 \\
%6T3B &  2.19 - 2.32 & 1.26  - 1.34 \\
%6T4 &  2.58 & 1.29 \\
%6T5 &  2.89 & 1.29 \\
%\hline
%%8TO & 1.81 & 1.36 \\
%8T2 & 1.87 &  1.32 \\
%8T3 &  2.27 - 2.3  &   1.3 0- 1.33  \\
%8T4A &  2.64  & 1.32\\
%8T4B &  2.60 -2.69 & 1.30 - 1.34\\
%8T5A &  2.88 - 2.98  & 1.29 - 1.33\\
%8T5B &  2.93 - 2.98  & 1.31 - 1.33\\
%8T6 &  3.24 & 1.32 \\
%8T7 &   &  \\
\hline
		\end{tabular}
	\end{center}
	\caption{Results of the performance study in the ideal case, where all the baselines are equivalent and noted $\{ij\}$.}
	\label{tab:idealcase}
\end{table}
%%%%%%%%%%%%%%%%%%%%%%%%%%%%%%%%%%%%%%%%%%%%%%%%%%%%%%%
%
%
%
%
%%%%%%%%%%%%%%%%%%%%%%%%%%%%%%%%%%%%%%%%%%%%%%%%%%%%%%%
\begin{table}[t]
	\begin{center}
		\begin{tabular}{ccccc}
\hline
\hline
             & \multicolumn{2}{c}{Detector noise} & \multicolumn{2}{c}{Photon noise} \\
Scheme & $\epsilon_{12}$ &  $\epsilon_{ij}$ &   $ \epsilon_{12}$ & $ \epsilon_{ij}$ \\
\hline
4TO & 16.2 & 1.6 & 13.1 & 1.3 \\
4T2 &  3.4  &  2.0 &  2.4  &  1.4  \\
4T3 &  3.0  &  $2.1 - 2.4$ & 1.7  &  $1.2 - 1.4$ \\
\hline
6TO & 18.1 & 1.8 & 13.6 & 1.4 \\
6T2 &  4.3  &  2.0   &  3.1  &  1.4 \\
6T3A &  3.3  &  $2.3 - 2.4$  & 1.9  &  $1.3 - 1.4$ \\
6T3B &  3.2 - 3.6  &  $2.2 - 2.5$ &  1.8 - 2.1  &  $1.3 - 1.4$ \\
6T4 &  3.3  &  $2.6 - 2.8$  &  1.7  &  $1.3 - 1.4$ \\
6T5 &  3.5  &  $2.9 - 3.1$ &  1.6  &  $1.3 - 1.4$  \\
%\hline
%%8TO & 18.1 & 1.81 & 13.6  & 1.36 \\
%8T2 & 5.1 & 2.0 & 3.6 & 1.4 \\
%8T3 & 3.6 & 2.3 - 2.5 & 2.0 & 1.2 - 1.4 \\
%8T4A & 3.5 &2.7 - 2.8 & 1.8 & 1.3 - 1.4\\
%8T4B & 3.4 & 2.6 - 2.9 & 1.7 & 1.3 - 1.5 \\
%8T5A & 3.7 & 2.9 - 3.2 & 1.6 & 1.3 - 1.4\\
%8T5B & 3.7 & 2.9 - 3.2 & 1.6 & 1.3 - 1.4\\
%8T6 & 3.8 & 3.2 - 3.4 & 1.5 & 1.3 - 1.4\\
%8T7 &&&&  \\
\hline
		\end{tabular}
	\end{center}
	\caption{Results of the performance study when the baseline $\{12\}$ resolves the source. The other baselines are noted $\{ij\}$ and are roughly equivalent.}
	\label{tab:ressource}
\end{table}
%%%%%%%%%%%%%%%%%%%%%%%%%%%%%%%%%%%%%%%%%%%%%%%%%%%%%%%
%
%

%
%
%%%%%%%%%%%%%%%%%%%%%%%%%%%%%%%%%%%%%%%%%%%%%%%%%%%%%%%
\begin{table}[t]
	\begin{center}
		\begin{tabular}{ccccc}
\hline
\hline
             & \multicolumn{2}{c}{Detector noise} & \multicolumn{2}{c}{Photon noise} \\
Scheme & $\epsilon_{1j}$ &  $\epsilon_{ij}$ &   $ \epsilon_{1j}$ & $ \epsilon_{ij}$ \\
\hline
4TO & 5.1 & 1.6   & 3.1 & 1.3 \\
4T2 &  4.7  &  1.9  &  2.5  &  1.4  \\
4T3 &  5.6  &  2.4   &  2.5  &  1.4  \\
\hline
6TO & 5.7 & 1.8 & 3.2 & 1.4 \\
6T2 & 4.8  &  1.9 &  2.6 &  1.4 \\
6T3A &  5.8  &  2.2 & 2.5  &  1.4 \\
6T3B &  $5.7 - 5.8$  &  $2.2 - 2.5$  &  $2.5 - 2.6$  &  $1.2 - 1.4$\\
6T4 &  6.6  &  $2.7-2.9$ &  2.5  &  $1.3-1.4$ \\
6T5 &  7.3  &  3.1  &  2.5 &  1.4  \\
%\hline
%4TO & 5.12 & 1.62   & 3.06 & 1.31 \\
%4T2 &  4.66  &  1.95  &  2.50  &  1.36  \\
%4T3 &  5.64  &  2.41   &  2.47  &  1.37  \\
%\hline
%6TO & 5.71 & 1.81 & 3.18 & 1.36 \\
%6T2 & 4.82  &  1.96 &  2.63  &  1.37 \\
%6T3A &  5.78  &  2.42 & 2.53  &  1.39 \\
%6T3B &  $5.7 - 5.8$  &  $2.2 - 2.5$  &  $2.51 - 2.58$  &  $1.27 - 1.44$\\
%6T4 &  6.58  &  $2.72-2.93$ &  2.51  &  $1.35-1.42$ \\
%6T5 &  7.34  &  3.14  &  2.50  &  1.39  \\
%\hline
%%8TO & 5.8 & 1.8 & 3.2 & 1.4 \\
%8T2 & 5.0 &2.0 & 2.7 &1.4 \\
%8T3 & 5.8 & 2.5 & 2.5 & 1.4 \\
%8T4A & 6.6 & 2.8 & 2.5 &1.4 \\
%8T4B & 6.6 & 2.6 - 3.0 & 2.5 & 1.4 \\
%8T5A & 7.1 & 3.1 - 3.3 & 2.5 & 1.4 \\
%8T5B & 7.4 & 3.1 & 2.5 & 1.4 \\
%8T6 & 8.1 & 3.5 & 2.5 & 1.4 \\
%8T7 &&&&  \\
\hline
		\end{tabular}
	\end{center}
	\caption{Results of the performance study in the flux drop-out case. The pupil 1 has a low flux and the related baselines are noted $\{1j\}$. The unaffected baselines are noted $\{ij\}$.}
	\label{tab:lowflux}
\end{table}
%%%%%%%%%%%%%%%%%%%%%%%%%%%%%%%%%%%%%%%%%%%%%%%%%%%%%%%
%
%

\subsubsection{Extracting the photometry}
The knowledge of the photometry is theoretically not mandatory to measure the fringe phase. However, a real-time photometric monitoring is very useful during operation: it provides an additional diagnosis in case of flux-related issues and it allows the image quality to be optimized in all beams simultaneously (otherwise the only way to optimise the flux of each telescope is to optimise them sequentially). Moreover, the knowledge of the photometries allow the fringe visibility to be computed in real-time, revealing possible technical issues (or even astrophysical ``issues'' such as unknown binaries).

Some of the schemes that we study allow the instantaneous photometry to be extracted on each pupil without the need of dedicated photometric outputs. We found that, in the context of pairwise combinations, the photometry can be recovered from the fringe signal itself for every pupil that is part of a closed (sub-)array constituted of an odd number of pupils. Otherwise the system linking the fringe signals to the photometries is degenerated. Thus, the 4T2, 6T2 and 6T3A schemes cannot extract the photometry since they only contain rings of 4 and/or 6 telescopes, whereas the 4T3 and 6T3B can, since there are triangular sub-arrays. This is summarized in Tab.~\ref{tab:photom}. Note that for arrays with an odd number of telescopes, circular schemes ($R$\,=\,2) always allow the photometry to be directly estimated.

%
%Monitoring the photometries is possible with other schemes but should pass by dedicated photometric outputs. Because of a less important flux in the interferometric outputs, the fringe sensor sensitivity shall be lowered. Assuming that $10\%$ of the input fluxes is required to well estimate the photometries, the phases should be roughly $23\%$ and $11\%$ less precise in detector and photon noise regimes respectively. Adding that in the noise values calculated previously, it clearly appears that in the case the photometries are required, a scheme naturally providing the photometries should be favored.  \\
%

\subsubsection{Robustness}

\begin{figure*}[t]
	\centering
	\begin{tabular}{cc}
      \includegraphics[angle=0, height=0.232\textheight]{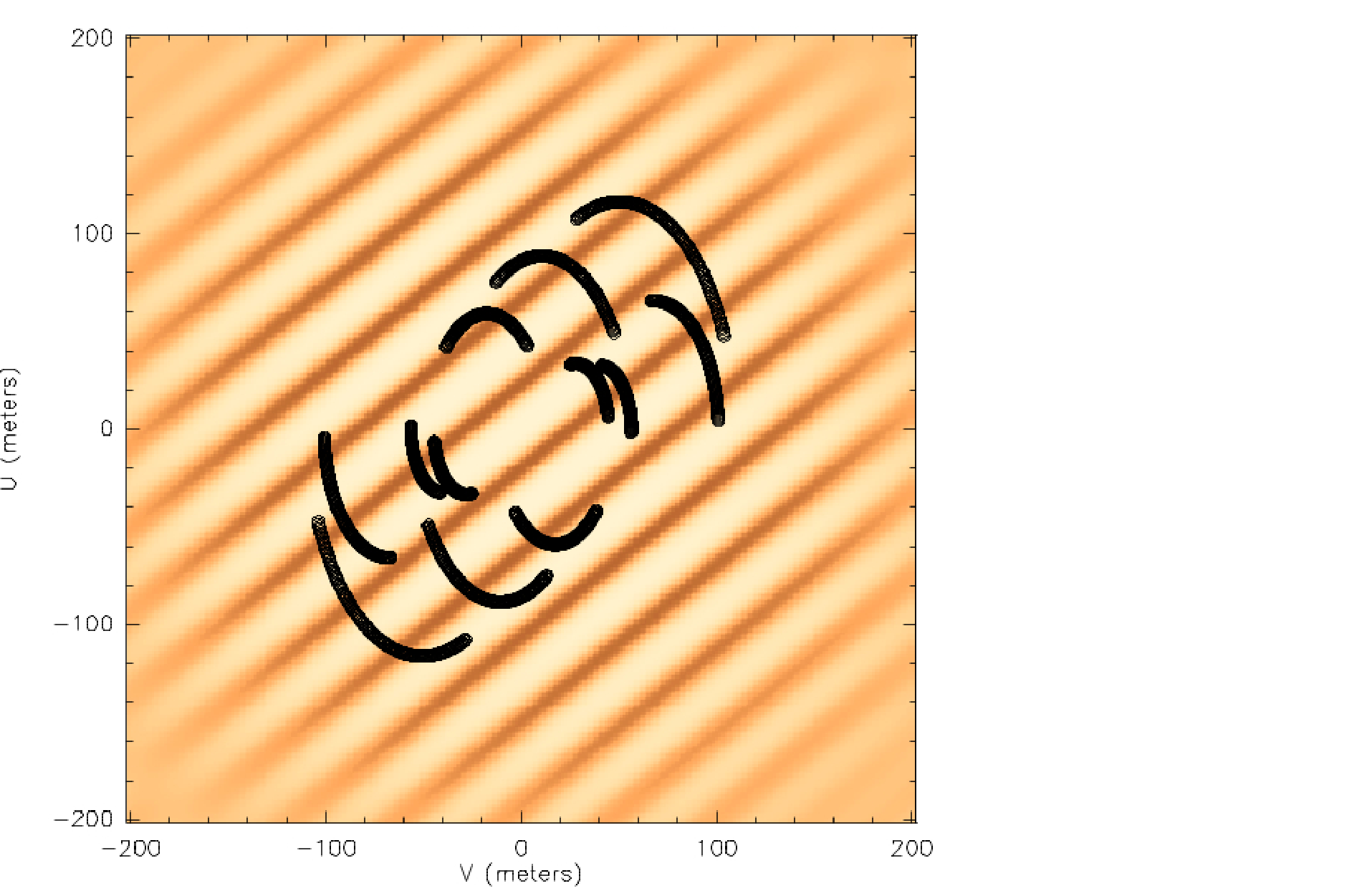} &
      \includegraphics[angle=0, height=0.25\textheight]{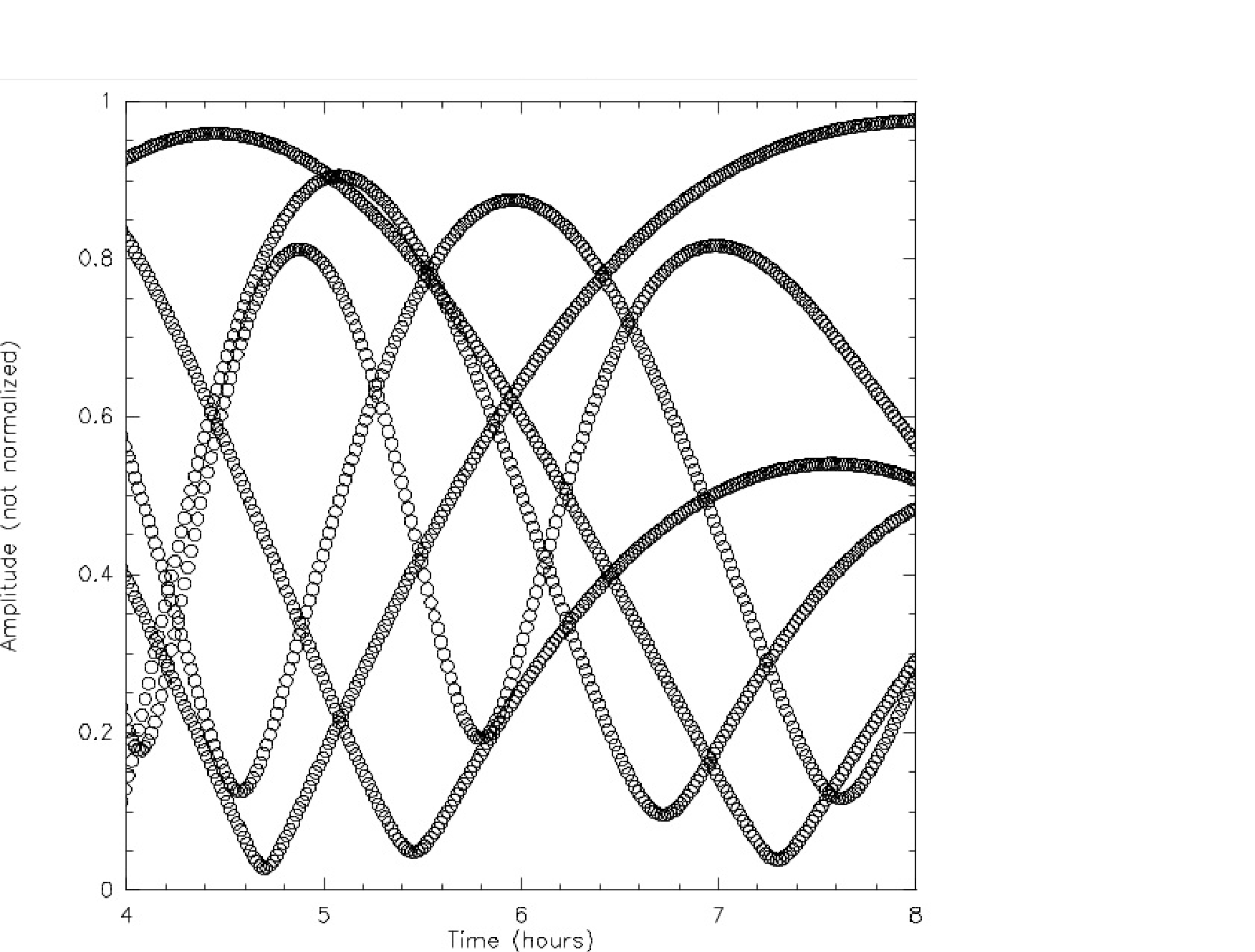}
	\end{tabular}
	\caption[Predicted fringe contrast when observing a binary]{ Predicted fringe contrast when observing a binary star with equal fluxes and a separation of about 10\,mas with the four UTs and a fringe sensor working in the H-band. The left panel shows the (u,v) tracks overlaid on the fringe contrast from the model. The right panel shows the fringe contrast versus time for 4h for each baseline. The figures have been made with the \texttt{aspro} public software from JMMC. \label{fig:contrast_on_resolved_binary}}
	\label{fig:aspro_vis}
\end{figure*}

When observing unknown asymmetrical sources, like well resolved binary stars, unpredictable baselines can exhibit very low visibilities, changing with a time scale of less than one hour (see Fig.~\ref{fig:aspro_vis} for an example). The fringe position may then become impossible to measure on some baselines, leading to a possible discontinuity in the array cophasing. The case of a resolved source previously studied (see Tab.~\ref{tab:ressource}) is an example of such a situation: when one baseline highly resolves the source, the comparison between the open schemes and the redundant ones clearly shows the benefit of having additional baselines. If we now assume that two baselines fully resolve the source, the schemes with $R\geq3$ provide better performances than open and circular schemes, and so on. In general, redundancy allows bootstrapping to be performed and therefore the tracking stability to be increased along an observation night, so that schemes with a high number of baselines are favored.

%
%
%%%%%%%%%%%%%%%%%%%%%%%%%%%%%%%%%%%%%%%%%%%%%%%%%%%%%%
\begin{table*}[t]
\centering
 \begin{tabular}{cccccccccccccc}
  \hline
  \hline
  Scheme      && 4TO & 4T2 & 4T3 && 6TO & 6T2 & 6T3A & 6T3B & 6T4 & 6T5 \\
  \hline
  Photometries\,? && no & no  & yes && no & no  & no  & yes  &yes  & yes   \\
  \hline
%  \hline
%  Scheme      & 8T2 & 8T3 & 8T4A & 8T4B & 8T5A & 8T5B & 8T6 & 8T7 & \\
%  \hline
%  Photometries? & no & yes & no & yes & yes & yes & yes & yes &  \\
%  \hline
 \end{tabular}
  \caption{Ability of the combination schemes to provide the inputs photometries without dedicated outputs. The schematic representation of the schemes can be found in Fig.~\ref{fig:schemes}.}
 \label{tab:photom}
\end{table*}
%%%%%%%%%%%%%%%%%%%%%%%%%%%%%%%%%%%%%%%%%%%%%%%%%%%%%%
%
%
%
%
%%%%%%%%%%%%%%%%%%%%%%%%%%%%%%%%%%%%%%%%%%%%%%%%%%%%%%%
%%%%%%%%%%%%%%%%%%%%%%%%%%%%%%%%%%%%%%%%%%%%%%%%%%%%%%%
%
%

%Such a strategy also maximizes the cophasing stability in case of frequent flux drop-outs (due to low Strehls or telescopes chopping) since unaffected baselines are kept locked and thus only the extinguished beams have to be cophased back after extinction.

		%%%%%%%%%%%%%%%%%%%%%%%%%%%%%%%%%%%%%%%%%%%%%%%%%%%%%%%
		%%%%%%%%%%%%%%%%%%%%%%%%%%%%%%%%%%%%%%%%%%%%%%%%%%%%%%%
		\subsection{Choice of the combination schemes}

The various schemes studied here provide similar performances in an ideal situation. When considering more realistic conditions, the benefit of the redundancy clearly appears, by improving the tracking robustness in various observing conditions. Additionally, among all the schemes, some provide the input fluxes in real time without the need of dedicated outputs, which is extremely useful for the state machine. We conclude that the best compromises between robustness and sensitivity are the 4T3 and 6T3B schemes.
% In the 6T case, more baselines slightly decrease the performance and dramticaaly increase the implementation difficulties. 
Because of their similar performance and their easier practical implementation, we also consider that the circular schemes 4T2 and 6T2 are suitable, if monitoring the photometric fluxes is not required. In the 4T case, these conclusions are in agreement with the results of \citet{houairi_2008} for the dedicated fringe tracker of GRAVITY. The results in the 6T case are also in agreement with the choices made for CHAMP \citep{bergerDh_2006} at the CHARA array with a 6T2 configuration, even though we favor a scheme with more baselines for robustness purposes.

Despite the fact we study only 2 cases (4 and 6 pupils), it appears to be the trend that, in the context of pairwise combinations with an even number of telescopes, an optimal fringe sensor should measure either $N$ or $3N/2$ baselines ($R$\,=\,2 or 3 respectively) depending on the need for photometries. With an odd number of telescopes, circular schemes should be optimal thanks to their capability to directly monitor the photometry.

%We have now determine the optimal way to measure the fringe position (parts~\ref{part:phase} and~\ref{part:groupdelay}) and to ensure the highest sensitivity and operationnal efficiency of the fringe sensor (this part). Melting all these results we are going to estimate and compare the performances of  the 4-telescope conept based on the 4T3 scheme providing the phase and group delay on each of the 6 baselines thanks to a static ABCD fringe coding, dispersed over 5 spectral channels.

			%%%%%%%%%%%%%%%%%%%%%%%%%%%%%%%%%%%%%%%%%%%%%%%%%%%%%%%
			%%%%%%%%%%%%%%%%%%%%%%%%%%%%%%%%%%%%%%%%%%%%%%%%%%%%%%%
			%%%%%%%%%%%%%%%%%%%%%%%%%%%%%%%%%%%%%%%%%%%%%%%%%%%%%%%
			\section{Estimated performance of the chosen concepts}  \label{part:finalperf}

Now that the optimal fringe sensing concepts have been identified, we study their on-sky performance within the VLTI infrastructure. To this aim, we have developed a dedicated software simulation tool called \textit{Sim2GFT} (2GFT standing for the ``2nd Generation Fringe Tracker'' of the VLTI). This simulator, consisting in a set of IDL routines, aims at performing realistic simulations of future observations with the 2GFT fringe sensor and to evaluate its performance in terms of residual piston jitter after closed-loop control. In the rest of this section, we assume that single-mode fibers are used to filter the input wavefronts, following \citet{tatulli_2010a}.

    \subsection{The Sim2GFT simulator}

Sim2GFT is largely based on the GENIEsim software \citep{absil_2006}, and therefore follows the same architecture and philosophy. The simulations are taking into account all major contributors to the final performance, from the atmosphere and the telescopes down to the fringe sensor and delay lines. The signal-to-noise ratio on the phase measurement in the fringe sensor is mainly driven by the amount of coherent and incoherent photons (including the atmospheric and instrumental thermal emission), and by the way they are distributed on the detector. In order to properly estimate the amount of coherent and incoherent photons, all the VLTI and 2GFT subsystems are described by their influence on the intensity, piston, and wavefront quality of the light beams collected by each telescope. The estimated instrumental visibility within the fringe sensor takes into account the visibility loss due to piston jitter, atmospheric refraction, intensity mismatch between the beams due to atmospheric turbulence (scintillation), and longitudinal dispersion in the delay lines. In the case of piston jitter, a semi-empirical law based on on-sky FINITO data is used to include both the effect of atmospheric piston and vibration-induced piston. Another key element in the simulation is the coupling of the light beams into single-mode fibers, which we estimate by separating the contribution of tip-tilt (through the overlap integral between an offset Airy pattern and the fiber mode) and higher order aberrations (through the estimated Strehl ratio---without tip-tilt---that acts as a multiplicative factor).

The operation of 2GFT is closely related to the detector read-out scheme. Assuming a HAWAII-2RG focal plane array, we consider that the ABCD outputs of all baselines are spread on a single detector line, and that the spectral dispersion is performed on five contiguous detector lines. The detector is read line by line, with a read-out time that depends on the particular arrangement of the ABCD outputs on the lines (it amounts to $201\,\mu$s for our design). Deriving a reliable estimation of the phase and group delay requires the five spectral channel to be used\footnote{To perform a phase delay estimation with the ABCD scheme, one spectral channel is theoretically sufficient. However, for a better robustness to dispersion effects, we assume that the information from all five spectral channels is needed and will be used in practice.}. However, it must be noted that the phase and group delay estimations can be updated each time a new detector line is read, although it will be partly redundant with the previous estimation---this corresponds to the sliding-window estimation already in use at the Keck fringe tracker~\citep{colavita_2010b}.

The closed-loop behaviour of the fringe tracker is simulated by feeding back the fringe sensor phase delay measurements to the VLTI delay lines, using a simple PID as a controller. Group delay measurements are not explicitly used in our simulations, although in practice they will be used to make sure that fringe tracking is performed on the appropriate (white-light) fringe. The closed-loop simulation relies on a frequency-domain description of the input disturbance (by its power spectral density) and of the subsystems (by their transfer function). The repetition frequency of the loop and the controller gain are optimised as a function of the input photon flux and atmospheric piston to produce the smallest possible piston residual at the output of the closed loop. In order to ensure a stable fringe tracking, we require the sensing noise to be smaller than 100~nm RMS for 90\% of the measurements on any individual baseline, which would correspond to an ${\rm SNR} > 4$ on the fringes in K band for 90\% of the measurements.

In the following sections, we describe the estimated performance for fringe sensing and fringe tracking of the 4T3 redundant concept with ABCD encoding on five spectral channels over the K band (from 1.9 to 2.4~$\mu$m). The estimations are based on an expected K-band transmission of 3\% for the whole VLTI/2GFT instrument.

    \subsection{Fringe sensing performance}

%
%%%%%%%%%%%%%%%%%%%%%%%%%%
\begin{figure}[t]
   \centering
   \includegraphics[angle=0, width=0.48\textwidth]{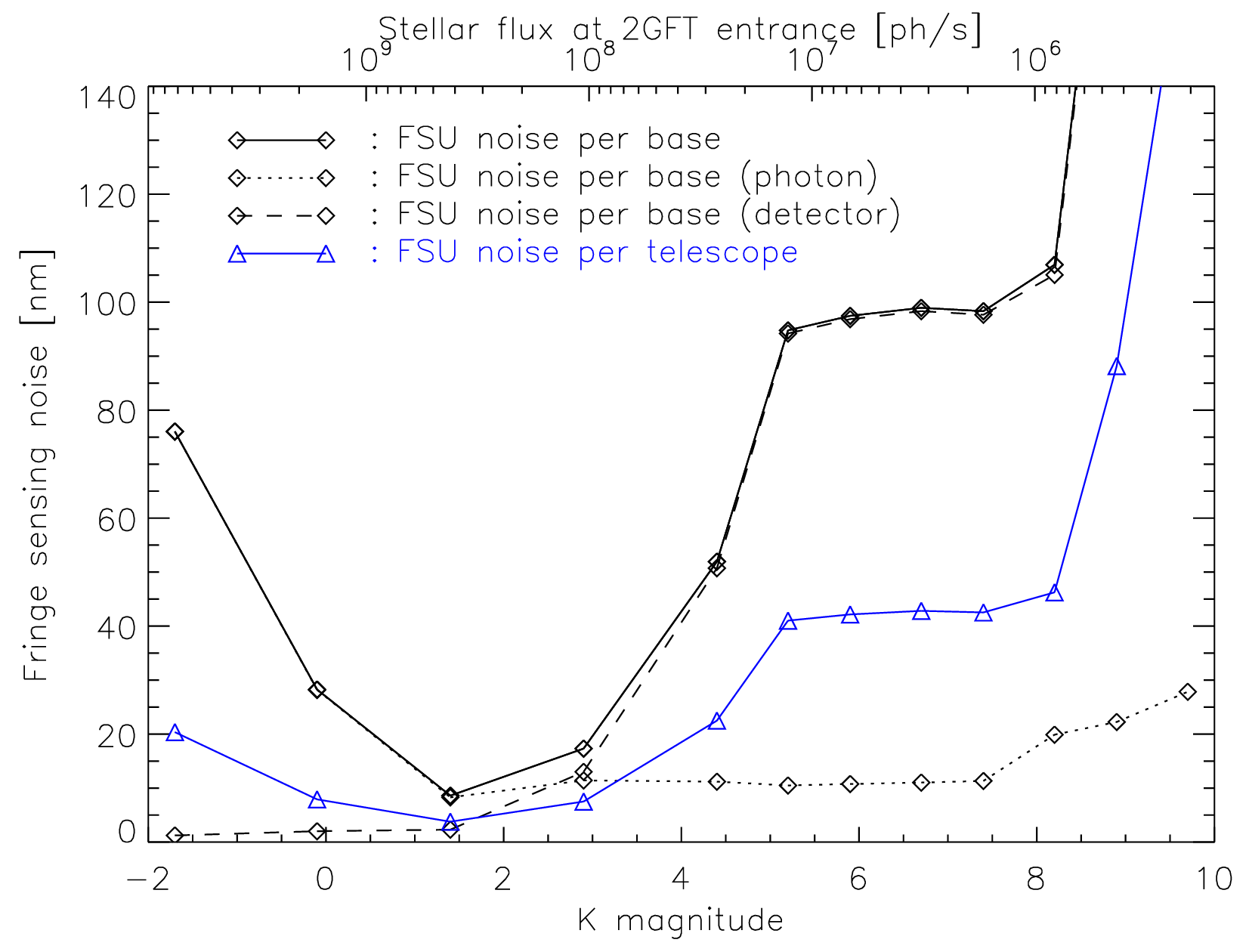}
   \caption{Fringe sensing noise plotted at percentile 0.9 (i.e., the noise is actually smaller than the plotted curves for 90\% of the occurrences) as a function of the target's K magnitude (or of the stellar flux in photons per second at the entrance of 2GFT) in the 4T3-ABCD case, assuming a K0\,III target and using the A0-G1-K0-I1 quadruplet of ATs at the VLTI. The fringe tracking loop is operated at its maximum frequency as long as the fringe sensing noise per baseline remains $<100$\,nm RMS for 90\% of the measurements on any individual baseline. The closed-loop repetition frequency is reduced to maintain this level of performance otherwise (this happens for $K>5$ in the present case, as also shown in Fig.~\ref{fig:closeloop}), until this level cannot be reached any more (beyond $K=7.5$ in the present case). Note that the increase in sensing noise for bright targets is due to the stellar photosphere being resolved, which reduces the available coherent flux.}
   \label{fig:fsunoise}
\end{figure}
%%%%%%%%%%%%%%%%%%%%%%%%%%
%
%
%%%%%%%%%%%%%%%%%%%%%%%%%%
\begin{figure*}[t]
  \centering
  \begin{tabular}{cc}
    \includegraphics[angle=0, width=0.48\textwidth]{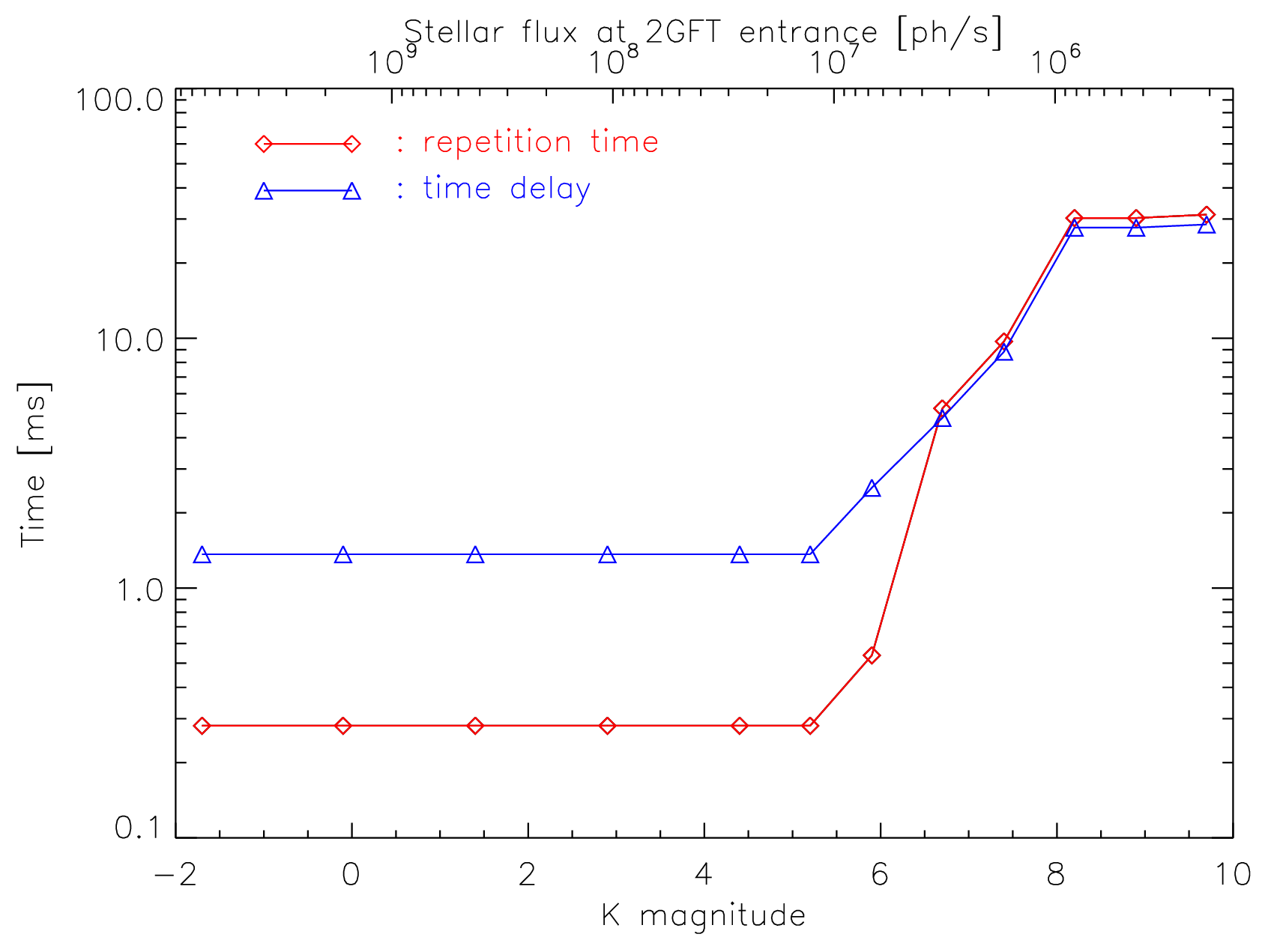} &
    \includegraphics[angle=0, width=0.48\textwidth]{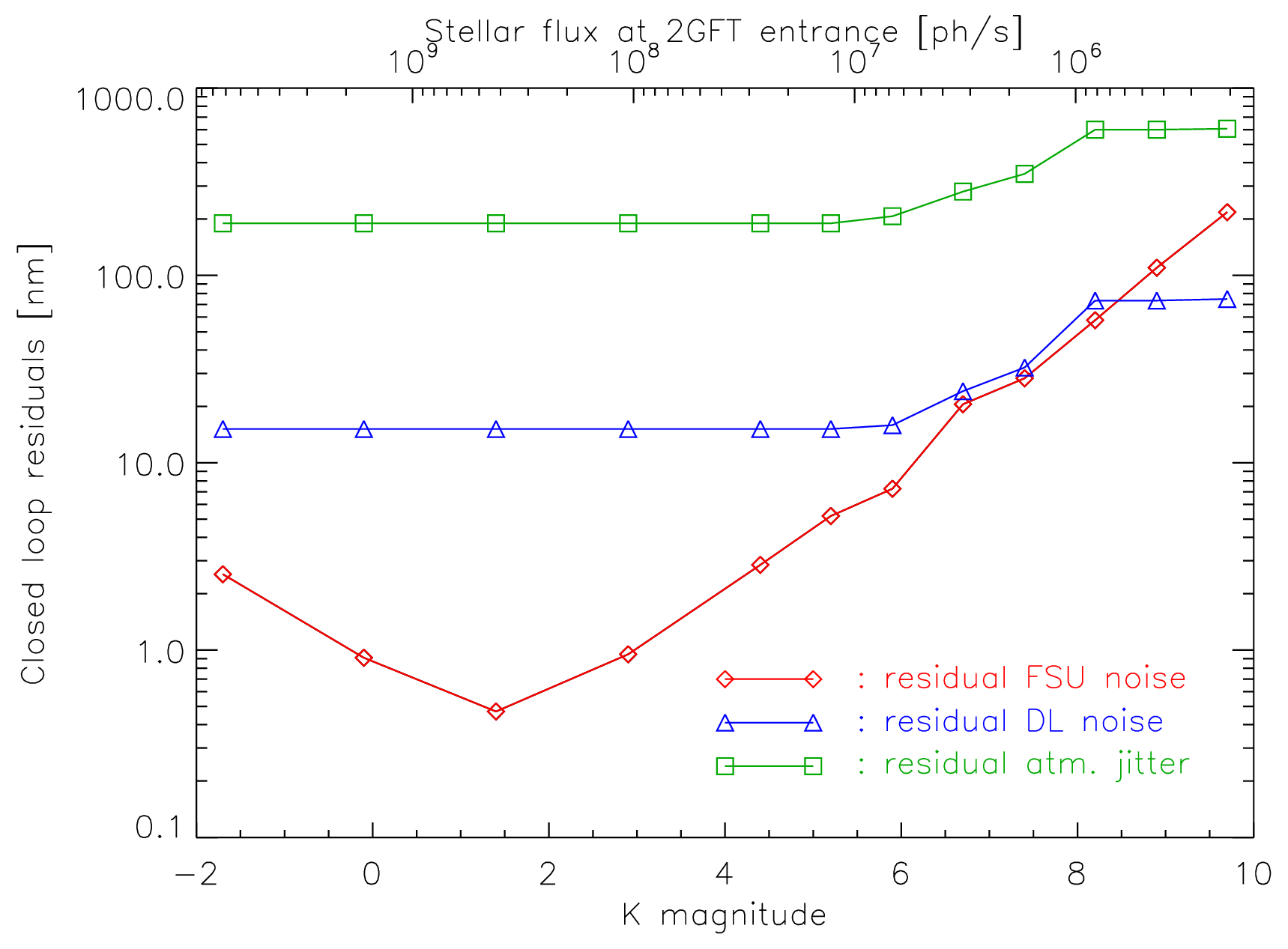}
  \end{tabular}
  \caption{\textit{Left}: Closed-loop repetition time and time delay in the loop as a function of target $K$ magnitude. For stars fainter than $K=6$, the loop repetition time is increased (i.e., its frequency decreased) to ensure a sufficient SNR on the detected fringes in each individual measurement (until the specified SNR cannot be reached any more whatever the integration time). \textit{Right}: Noise residuals at the output of the closed loop, for the three main contributors: fringe sensing (FSU), delay line (DL) and atmospheric noises.}
  \label{fig:closeloop}
\end{figure*}
%%%%%%%%%%%%%%%%%%%%%%%%%%
%

End-to-end simulations of VLTI/2GFT have been performed using the 1.8-m Auxiliary Telescopes (ATs) for a K0\,III star located at various distances ranging from about 10\,pc to 2\,kpc, in standard atmospheric conditions: seeing $\varepsilon=0.85"$, coherence time $\tau_0=3$\,ms, outer scale ${\cal L}_{\rm out}=25$\,m, and sky temperature $T_{\rm sky}=285$\,K. The target star is assumed to be located close to zenith. For each magnitude, the closed-loop repetition frequency has been chosen as high as possible within hardware limitations ($<4$\,kHz), while keeping the average fringe sensing noise smaller than 100~nm RMS on all measured baselines.

Fig.~\ref{fig:fsunoise} illustrates the sensing noise per baseline as a function of stellar magnitude (black diamonds). The respective contributions of photon noise and detector noise are represented by dotted and dashed lines. On the bright-side end of the plot, photon noise dominates the noise budget. The increase in photon noise from $K=1.5$ to $K=-2$ is due to the star being (strongly) resolved, which reduces the available coherent flux. Detector noise becomes larger than photon noise around $K=3$, and the fringe sensing noise reaches its allowed limit ($<100$\,nm RMS for 90\% of the measurements) around $K=5$. For fainter magnitudes, Sim2GFT makes sure that the fringe sensing noise remains at the same level by reducing the closed-loop repetition frequency (i.e., increasing the integration time on the fringe sensing detector). This is possible only until magnitude $K=7.5$ in the present case, where a phase sensing noise of 100\,nm per baseline cannot be reached any more for any integration time, because of the strong fringe blurring that appears at long DITs. The points plotted in the figure at $K>7.5$ do not comply with our requirements any more, and have been computed for the repetition frequency that minimizes the fringe sensing noise ($\sim 33$\,Hz in the considered cases).

Also represented in Fig.~\ref{fig:fsunoise} is the fringe sensing noise per telescope, which results from the optimized estimation of individual telescope pistons from all measured baselines, as explained in Section~\ref{part:perf_study}. The fringe sensing noise per telescope is significantly smaller than the measurement noise on each individual baseline because the estimation of the former is based on the information collected by multiple baselines.

The same kind of performance study has been carried out in the case of the Unit Telescopes, showing a similar general behaviour as in the case of ATs. The only differences are:
\begin{itemize}
\item the magnitude where stable closed-loop fringe tracking becomes impossible, which is now around $K=9.5$,
\item the decrease in the coupling efficiency for stars fainter than $V=10$, which is due to the reduced performance of the MACAO adaptive optics system.
\end{itemize}
The latter effect, which is almost nonexistent in the case of ATs (equipped with STRAP for tip-tilt control), speeds up the drop of closed-loop performance at faint magnitudes. The maximum loop repetition frequency ($\sim 4$\,kHz) can actually be maintained until $K\simeq 8.5$ in the case of UTs. The presence of telescope vibrations in the case of UTs is taken into account in a semi-empirical way in our simulations, through an estimation of the visibility loss due to vibration-induced piston jitter, so that the SNR in the fringe sensing process is estimated in a realistic way. However, let us note that telescope vibrations are expected to strongly affect the residual piston jitter at the output of the closed loop (an effect not simulated in Sim2GFT), so that the results presented in right-hand side plot of Fig.~\ref{fig:closeloop} (in the case of ATs) would be significantly degraded in the case of UTs.

    \subsection{Fringe tracking performance}

Fig.~\ref{fig:closeloop} shows the characteristic times of the closed loop and the noise residuals at the output of the fringe tracking loop. The left-hand side plot shows that for magnitudes brighter than $K=5$, the loop can be operated at its maximum repetition frequency (3.6\,kHz in this case). For fainter targets, the repetition time is gradually increased to keep a sufficient SNR on each individual fringe measurement. The sudden increase in repetition time around $K=6$ is due to a modification in the loop behavior: for bright stars, only one spectral channel is read for each repetition time and the information at other wavelengths is taken from previous repetition times, while for fainter stars all spectral channels are read during each repetition time (the main goal of this being to keep the time delay\footnote{The time delay of the loop is defined as the amount of time between the middle of the overall integration time used for a phase estimation (i.e., including the contribution of all spectral channels), and the moment when the detector read-out sequence is completed for the considered spectral channel.} in the loop reasonably short even at low repetition frequencies). The time delay is longer than the repetition time in the bright target case, because only one spectral channel is read per repetition time, while the phase estimation uses the phase information from all five spectral channels.

The left-hand side plot of Fig.~\ref{fig:closeloop} can be used to derive a limiting magnitude for the chosen fringe sensing concept. One just needs to define a repetition time threshold above which fringe tracking becomes inefficient. Here, we assume a maximum allowed repetition time of 10\,ms (i.e., minimum frequency of 100\,Hz),\footnote{For an integration time of 10\,ms on the fringe sensor, the estimated loss of visibility due to piston jitter in standard atmospheric conditions is only 5\% in the case of ATs, while it amounts to 28\% in the case of UTs (an effect mainly due to vibrations). Operating at lower frequencies would become impractical in the case of UTs, but could be considered in the case of ATs (especially in good atmospheric conditions).} which gives a limiting magnitude of $K=7.5$ on the ATs. In the case of UTs, the limiting magnitude amounts to $K=9.5$. In both cases, this coincidentally corresponds to the magnitude where maintaining a phase measurement error below 100\,nm is not possible, which indicates that a DIT of 10\,ms is actually a sound choice to define limiting magnitudes in closed-loop fringe tracking operation. Note that a limiting magnitude of $K=7.5$ in closed-loop fringe tracking with 90\% locking ratio under standard atmospheric conditions corresponds quite well to what has been demonstrated on-sky with the PRIMA fringe sensor unit on the ATs \citep{sahlmann_2009}.

The right-hand side of Fig.~\ref{fig:closeloop} shows the noise residuals at the output of the fringe tracking loop, computed per telescope. Note that the fringe sensing noise residual at the output of the loop is much smaller than the actual fringe sensing noise (evaluated at the detection level), due to closed-loop filtering. Also note that the fringe sensing noise is always much smaller than the atmospheric noise under typical atmospheric conditions, a behavior directly related to the constraint imposed on the phase sensing noise per baseline in each repetition time ($<100$\,nm RMS for 90\% of the measurements). For these reasons, the fringe sensing noise does not significantly affect the residual noise level at the output of the fringe tracking loop. The influence of the fringe sensor on the residual piston noise comes rather from its intrinsic sensitivity, which determines the maximum repetition frequency that can be reached for a given coherent flux.

We have also performed simulations in various atmospheric conditions, ranging from bad (seeing $\epsilon_0=1.1"$ and coherence time $\tau_0=2$\,ms) to excellent ($\epsilon_0=0.5"$ and $\tau_0=10$\,ms). The influence of atmospheric conditions on the fringe tracking performance is mainly twofold: on one hand it determines the input atmospheric noise that needs to be corrected, and on the other hand it affects the amount of available coherent photons since it determines the injection efficiency into single-mode fibres. Our simulations have shown that the limiting magnitude increases by about 2~magnitudes between bad and excellent conditions. For instance, if one defines the limiting magnitude at 100\,Hz, it varies between $K=6.2$ and $K=8.5$ depending on the conditions. These limiting magnitudes do not mean however that fringes cannot be detected at fainter magnitudes. We estimate that the ultimate limit for fringe detection (fringes detected for 50\% of the measurements at an SNR of 4, using a DIT of 25\,ms) should be around $K=9.5$ for ATs used in good atmospheric conditions.

			%%%%%%%%%%%%%%%%%%%%%%%%%%%%%%%%%%%%%%%%%%%%%%%%%%%%%%%
			%%%%%%%%%%%%%%%%%%%%%%%%%%%%%%%%%%%%%%%%%%%%%%%%%%%%%%%
			%%%%%%%%%%%%%%%%%%%%%%%%%%%%%%%%%%%%%%%%%%%%%%%%%%%%%%%
			\section{Conclusions and perspectives}

We determined the optimal 4- and 6-telescopes fringe tracker concepts. We showed that for realistic atmospheric conditions, the measurements of the various phase states (e.g., ABCD) that are needed to derive the fringe phase should better be done simultaneously in order to limit the influence of external disturbances (piston, scintillation, vibrations, etc.) on the measurement precision. Furthermore, spectrally dispersing the fringes allows the group delay to be evaluated with one set of contemporaneous data, which (like for the phase measurement) minimizes the influence of disturbances. We also showed that this method is more robust to longitudinal dispersion effects. Therefore, we concluded that the optimal way to measure the fringe position (phase and group delay) is to perform a static ABCD fringe coding, dispersed over about five spectral channels.

We also demonstrated that the co-axial pairwise combination schemes with a moderate redundancy provide the best compromise between sensitivity and robust operations. They are less sensitive to varying observing conditions, and some schemes also allow the photometries to be directly extracted from the fringe signal, which is useful for the state machine. We finally favored the 4T3 and 6T3B schemes for 4- and 6-telescope operations respectively.

Merging these results, we have simulated the expected performance of the 4-telescope concept. For an efficient fringe tracking, with fringes locked at least 90\% of the time, we expect limiting magnitudes of 7.5 and 9.5 at K band with ATs and UTs respectively. These performances are close to those of single baseline fringe trackers currently in operation. Another important result is that the fringe tracker ultimate performances are not limited by the fringe sensing measurement errors, but rather by the time delay between the measurement of the piston and its correction by the delay lines.

Finally, in the coming years, a new generation of infrared detectors should be available. By providing very high acquisition frequencies and an extremely low read-out noise at the limit of photon-counting, multi-axial schemes should be reconsidered as a possible solution for fringe-traciking, as they would not be limited by the large amount of pixels needed to encode the interferometric signal.

\begin{acknowledgements}
The authors are grateful to the referee, whose careful and thorough review of the text and theoretical formalism helped them improve the papers clarity and quality considerably.
\end{acknowledgements}
%%%%%%%%%%%%%%%%%%%%%%%%%%%%%%%%%%%%%%%%%%%%%%%%%%%%%%
%%                 REFERENCES                       %%
%%%%%%%%%%%%%%%%%%%%%%%%%%%%%%%%%%%%%%%%%%%%%%%%%%%%%%

\bibliographystyle{aa}
%\bibliography{biblio}
\bibliography{biblio}

\begin{thebibliography}{23}
\expandafter\ifx\csname natexlab\endcsname\relax\def\natexlab#1{#1}\fi

\bibitem[{{Absil} {et~al.}(2006){Absil}, {Di Folco}, {M\'erand}, {Augereau},
  {Coud\'e du Foresto}, {Aufdenberg}, {Kervella}, {Ridgway}, {Berger}, {ten
  Brummelaar}, {Sturmann}, {Strumann}, {Turner}, \& {McAlister}}]{absil_2006}
{Absil}, O., {Di Folco}, E., {M\'erand}, A., {et~al.} 2006, \aap, 452, 237

\bibitem[{{Berger} {et~al.}(2006){Berger}, {Monnier}, {Millan-Gabet}, {ten
  Brummelaar}, {Muirhead}, {Pedretti}, \& {Thureau}}]{bergerDh_2006}
{Berger}, D.~H., {Monnier}, J.~D., {Millan-Gabet}, R., {et~al.} 2006, in Proc.
  of SPIE, Vol. 6268

\bibitem[{{Colavita}(2010)}]{colavita_2010a}
{Colavita}, M.~M. 2010, \pasp, 122, 712

\bibitem[{{Colavita} {et~al.}(2010){Colavita}, {Booth}, {Garcia-Gathright},
  {Vasisht}, {Johnson}, \& {Summers}}]{colavita_2010b}
{Colavita}, M.~M., {Booth}, A.~J., {Garcia-Gathright}, J.~I., {et~al.} 2010,
  \pasp, 122, 795

\bibitem[{{Colavita} {et~al.}(1999){Colavita}, {Wallace}, {Hines}, {Gursel},
  {Malbet}, {Palmer}, {Pan}, {Shao}, {Yu}, {Boden}, {Dumont}, {Gubler},
  {Koresko}, {Kulkarni}, {Lane}, {Mobley}, \& {van Belle}}]{colavita_1999a}
{Colavita}, M.~M., {Wallace}, J.~K., {Hines}, B.~E., {et~al.} 1999, \apj, 510,
  505

\bibitem[{{Di Lieto} {et~al.}(2008){Di Lieto}, {Haguenauer}, {Sahlmann}, \&
  {Vasisht}}]{dilieto_2008}
{Di Lieto}, N., {Haguenauer}, P., {Sahlmann}, J., \& {Vasisht}, G. 2008, in
  Proc. of SPIE, Vol. 7013

\bibitem[{{Fried}(1966)}]{fried_1966a}
{Fried}, D.~L. 1966, Journal of the Optical Society of America (1917-1983), 56,
  1372

\bibitem[{{Gai} {et~al.}(2003){Gai}, {Corcione}, {Lattanzi}, {Bauvir},
  {Bonino}, {Gardiol}, {Gennai}, {Loreggia}, {Massone}, \&
  {Menardi}}]{gai_2003b}
{Gai}, M., {Corcione}, L., {Lattanzi}, M.~G., {et~al.} 2003, Memorie della
  Societa Astronomica Italiana, 74, 472

\bibitem[{{Gai} {et~al.}(2004){Gai}, {Menardi}, {Cesare}, {Bauvir}, {Bonino},
  {Corcione}, {Dimmler}, {Massone}, {Reynaud}, \& {Wallander}}]{gai_2004}
{Gai}, M., {Menardi}, S., {Cesare}, S., {et~al.} 2004, in Proc. of SPIE, ed.
  {W.~A.~Traub}, Vol. 5491, 528--+

\bibitem[{{Gillessen} {et~al.}(2010){Gillessen}, {Eisenhauer}, {Perrin},
  {Brandner}, {Straubmeier}, {Perraut}, {Amorim}, {Sch{\"o}ller},
  {Araujo-Hauck}, {Bartko}, {Baumeister}, {Berger}, {Carvas}, {Cassaing},
  {Chapron}, {Choquet}, {Clenet}, {Collin}, {Eckart}, {Fedou}, {Fischer},
  {Gendron}, {Genzel}, {Gitton}, {Gonte}, {Gr{\"a}ter}, {Haguenauer}, {Haug},
  {Haubois}, {Henning}, {Hippler}, {Hofmann}, {Jocou}, {Kellner}, {Kervella},
  {Klein}, {Kudryavtseva}, {Lacour}, {Lapeyrere}, {Laun}, {Lena}, {Lenzen},
  {Lima}, {Moratschke}, {Moch}, {Moulin}, {Naranjo}, {Neumann}, {Nolot},
  {Paumard}, {Pfuhl}, {Rabien}, {Ramos}, {Rees}, {Rohloff}, {Rouan}, {Rousset},
  {Sevin}, {Thiel}, {Wagner}, {Wiest}, {Yazici}, \& {Ziegler}}]{gillessen_2010}
{Gillessen}, S., {Eisenhauer}, F., {Perrin}, G., {et~al.} 2010, in Proc. of
  SPIE, Vol. 7734

\bibitem[{{Houairi} {et~al.}(2008){Houairi}, {Cassaing}, {Perrin},
  {Eisenhauer}, {Brandner}, {Straubmeier}, \& {Gillessen}}]{houairi_2008}
{Houairi}, K., {Cassaing}, F., {Perrin}, G., {et~al.} 2008, in Proc. of SPIE,
  Vol. 7013

\bibitem[{{Jurgenson} {et~al.}(2008){Jurgenson}, {Santoro}, {Baron}, {McCord},
  {Block}, {Buscher}, {Haniff}, {Young}, {Coleman}, \&
  {Creech-Eakman}}]{jurgenson_2008}
{Jurgenson}, C.~A., {Santoro}, F.~G., {Baron}, F., {et~al.} 2008, in Proc. of
  SPIE, Vol. 7013

\bibitem[{{Le Bouquin} {et~al.}(2009){Le Bouquin}, {Abuter}, {Haguenauer},
  {Bauvir}, {Popovic}, \& {Pozna}}]{lebouquin_2009}
{Le Bouquin}, J.-B., {Abuter}, R., {Haguenauer}, P., {et~al.} 2009, {\aa}p,
  493, 747

\bibitem[{{Lopez} {et~al.}(2008){Lopez}, {Antonelli}, {Wolf}, {Lagarde},
  {Jaffe}, {Navarro}, {Graser}, {Petrov}, {Weigelt}, {Bresson}, {Hofmann},
  {Beckman}, {Henning}, {Laun}, {Leinert}, {Kraus}, {Robbe-Dubois}, {Vakili},
  {Richichi}, {Abraham}, {Augereau}, {Behrend}, {Berio}, {Berruyer},
  {Chesneau}, {Clausse}, {Connot}, {Demyk}, {Danchi}, {Dugu{\'e}}, {Finger},
  {Flament}, {Glazenborg}, {Hannenburg}, {Heininger}, {Hugues}, {Hron},
  {Jankov}, {Kerschbaum}, {Kroes}, {Linz}, {Lizon}, {Mathias}, {Mathar},
  {Matter}, {Menut}, {Meisenheimer}, {Millour}, {Nardetto}, {Neumann},
  {Nussbaum}, {Niedzielski}, {Mosoni}, {Olofsson}, {Rabbia}, {Ratzka}, {Rigal},
  {Roussel}, {Schertl}, {Schmider}, {Stecklum}, {Thiebaut}, {Vannier}, {Valat},
  {Wagner}, \& {Waters}}]{lopez_2008}
{Lopez}, B., {Antonelli}, P., {Wolf}, S., {et~al.} 2008, in Proc. of SPIE, Vol.
  7013

\bibitem[{{Malbet} {et~al.}(2008){Malbet}, {Buscher}, {Weigelt}, {Garcia},
  {Gai}, {Lorenzetti}, {Surdej}, {Hron}, {Neuh{\"a}user}, {Kern}, {Jocou},
  {Berger}, {Absil}, {Beckmann}, {Corcione}, {Duvert}, {Filho}, {Labeye}, {Le
  Coarer}, {Li Causi}, {Lima}, {Perraut}, {Tatulli}, {Thi{\'e}baut}, {Young},
  {Zins}, {Amorim}, {Aringer}, {Beckert}, {Benisty}, {Bonfils}, {Cabral},
  {Chelli}, {Chesneau}, {Chiavassa}, {Corradi}, {De Becker}, {Delboulb{\'e}},
  {Duch''ne}, {Forveille}, {Haniff}, {Herwats}, {Hofmann}, {Le Bouquin},
  {Ligori}, {Loreggia}, {Marconi}, {Moitinho}, {Nisini}, {Petrucci},
  {Rebordao}, {Speziali}, {Testi}, \& {Vitali}}]{malbet_2008}
{Malbet}, F., {Buscher}, D., {Weigelt}, G., {et~al.} 2008, in Proc. of SPIE,
  Vol. 7013

\bibitem[{{Meisner} \& {Le Poole}(2003)}]{meisner_2003}
{Meisner}, J.~A. \& {Le Poole}, R.~S. 2003, in Proc. of SPIE, ed. W.~A.
  {Traub}, Vol. 4838, 609--624

\bibitem[{{Monnier} {et~al.}(2004){Monnier}, {Berger}, {Millan-Gabet}, \& {ten
  Brummelaar}}]{monnier_2004}
{Monnier}, J.~D., {Berger}, J., {Millan-Gabet}, R., \& {ten Brummelaar}, T.~A.
  2004, in Proc. of SPIE, ed. {W.~A.~Traub}, Vol. 5491, 1370--+

\bibitem[{Papoulis(1984)}]{papoulis_1984}
Papoulis, A. 1984, {Probability, Random Variables and Stochastic Processes}
  (McGraw-Hill)

\bibitem[{{Pedretti} {et~al.}(2004){Pedretti}, {Thureau}, {Wilson}, {Traub},
  {Monnier}, {Ragland}, {Carleton}, {Millan-Gabet}, {Schloerb}, {Brewer},
  {Berger}, \& {Lacasse}}]{pedretti_2004}
{Pedretti}, E., {Thureau}, N.~D., {Wilson}, E., {et~al.} 2004, in Proc. of
  SPIE, ed. W.~A. {Traub}, Vol. 5491, 540--+

\bibitem[{{Sahlmann} {et~al.}(2009){Sahlmann}, {M{\'e}nardi}, {Abuter},
  {Accardo}, {Mottini}, \& {Delplancke}}]{sahlmann_2009}
{Sahlmann}, J., {M{\'e}nardi}, S., {Abuter}, R., {et~al.} 2009, \aap, 507, 1739

\bibitem[{{Shao} {et~al.}(1988){Shao}, {Colavita}, {Hines}, {Staelin}, \&
  {Hutter}}]{shao_1988}
{Shao}, M., {Colavita}, M.~M., {Hines}, B.~E., {Staelin}, D.~H., \& {Hutter},
  D.~J. 1988, \aap, 193, 357

\bibitem[{{Tatulli} {et~al.}(2010){Tatulli}, {Blind}, {Berger}, {Chelli}, \&
  {Malbet}}]{tatulli_2010a}
{Tatulli}, E., {Blind}, N., {Berger}, J.~P., {Chelli}, A., \& {Malbet}, F.
  2010, \aap, 524, A65+

\bibitem[{{Wilson} {et~al.}(2004){Wilson}, {Pedretti}, {Bregman}, {Mah}, \&
  {Traub}}]{wilson_2004}
{Wilson}, E., {Pedretti}, E., {Bregman}, J., {Mah}, R.~W., \& {Traub}, W.~A.
  2004, in Proc. of SPIE, ed. W.~A. {Traub}, Vol. 5491, 1507--+

\end{thebibliography}
\vspace{10cm}

%%%%%%%%%%%%%%%%%%%%%%%%%%%%%%%%%%%%%%%%%%%%%%%%%%%%%%
%%                  APPENDIX                       %%
%%%%%%%%%%%%%%%%%%%%%%%%%%%%%%%%%%%%%%%%%%%%%%%%%%%%%%
\newpage
\appendix

			%%%%%%%%%%%%%%%%%%%%%%%%%%%%%%%%%%%%%%%%%%%%%%%%%%%%%%%%%%%%%%%%%%%%%
			%%%%%%%%%%%%%%%%%%%%%%%%%%%%%%%%%%%%%%%%%%%%%%%%%%%%%%%%%%%%%%%%%%%%%
			\section{Phase error: detection and delay noises expressions}                    \label{app:atmnoise}

Considering an ABCD fringe coding \citep{colavita_1999a}, the phase is extracted as follows. First we have the 4 ABCD measurements in quadrature:
\begin{equation}
\left\{
	\begin{array}{lll}
	 A &\varpropto V \cos(\phi)  &\\
	 B &\varpropto V \cos(\phi + \pi/2) &= - V \sin(\phi) \\
	 C &\varpropto V \cos(\phi + \pi) &= - V \cos(\phi) \\
	 D &\varpropto V \cos(\phi + 3\pi/2) &= V \sin(\phi)
	\end{array}
\right.
\end{equation}
where $V$ and $\phi$ are the fringe contrast and phase respectively. We extract the real and imaginary part of the complex fringe signal:
\begin{equation} \label{eq:ABCDdef}
\left\{
	\begin{array}{ll}
	 A-C &\varpropto V \cos(\phi) \\
	 D-B &\varpropto V \sin(\phi)
	\end{array}
\right.
\end{equation}
and finally we estimate the phase through its cotangent:
\begin{equation}
\mathrm{tan}(\hat{\phi}) = \frac{D-B}{A-C}
\end{equation}

We are interested here by the statistical error on the phase measurement, which depends on three sources of noises: detector noise, photon noise and delay noises. Since these noises are statistically independent, the variance on the phase measurement $\sigma_\phi^2$ is simply the quadratic sum of these three noises:
\begin{equation}
 	\sigma_\phi^2 = \sigma_{det}^2 + \sigma_{phot}^2 + \sigma_{del}^2
\end{equation}

	\subsection{Detection noises}
The  detector and photon noises terms ($\sigma_{det}^2$ and $\sigma_{phot}^2$ respectively) are derived from \citet{shao_1988} for the ABCD fringe coding, and for sake of simplicity we put them together into the so-called signal detection noise $\sigma_{sig}^2$:
\begin{eqnarray}
	\sigma_{sig}^2 &= \sigma_{det}^2 + \sigma_{phot}^2 \\
 	\sigma^2_{det} &=  2\,\displaystyle\frac{4\,\sigma_e^2} {V^2 \, K^2} \\
	\sigma^2_{phot} &= 2\,\displaystyle\frac{K} {V^2 \, K^2}
\end{eqnarray}
where $K$ is the number of photo-events collected during the exposure and $\sigma^2_e$ is the detector read-out noise.

	\subsection{Delay noise}
The delay noise is due to the delay between the various measurements needed to estimate the phase and therefore only concern a temporal phase estimator. Because of instrumental or atmospheric disturbances (e.g.\ fluctuation of the differential piston or scintillation) the phase estimation can highly biased. Since \citet{fried_1966a} has shown that atmospheric piston and scintillation are uncorrelated, we can study both effects independently:

\begin{equation}
	\sigma_{del}^2 = \sigma_{pist}^2 + \sigma_{sci}^2
\end{equation}

\subsubsection{Piston noise: $\sigma_{pist}$}
We note here $\phi_p(t)$ the piston term introduced by the atmosphere at a moment $t$ and consider that each (A,C) and (B,D) measurement last half the total integration time $t_0$. Taking the point in the middle of the interval $t_0$ as the reference, the interferometric signal writes:
\begin{eqnarray}
	& A - C &\varpropto V\,\cos (\phi+\phi_p(t - t_0/4))  \\
	& B - D &\varpropto V\,\sin (\phi+\phi_p(t + t_0/4))
\end{eqnarray}
We note $\dphi_p = \phi_p(t+t_0/4) - \phi_p(t - t_0/4) $ the piston fluctuation between both measurements:
\begin{eqnarray}
	& A - C &\varpropto V\,\cos (\phi-\dphi_p/2)  \\
	& B - D &\varpropto V\,\sin (\phi +\dphi_p/2)
\end{eqnarray}
$\dphi_p$ being unknown, the phase estimator $\tilde{\phi}$ is:
\begin{eqnarray}
	\mathrm{tan}\,\tilde{\phi} = \frac{B-D}{A-C} = \frac{\sin(\phi + \dphi_p/2)}{\cos(\phi - \dphi_p/2)}
\end{eqnarray}
As soon as $\dphi_p$ is non null, the phase measurement is biased. If we consider the statistic variations of the piston, this bias can be considered as an additional noise. We now calculate the standard deviation of this phase measurement linked to the piston variations between 2 exposures separated by a time $t_0/2$. The standard deviation of the piston for this time will be noted $\sigma(\dphi_p, t_0/2)$. Assuming that the piston variations are small ($\sigma(\dphi_p, t_0/2) \ll 1$\,rad) and using the second order expansion formula of \citet{papoulis_1984}, the measured phase variance writes as:
\begin{equation}
 \sigma^2(\tilde{\phi}) = \left. \left( \frac{\partial \tilde{\phi}}{\partial \dphi_p} \right) \right|^2_{\langle \dphi_p \rangle}
                          \sigma^2(\dphi_p, t_0/2)
\end{equation}
where $\langle\dphi_p\rangle$ is the mean piston variation during $t_0/2$. One shows then that:
\begin{equation}
	\frac{\partial \tilde{\phi}}{\partial \dphi_p} = \frac{1}{2}\,\frac{\cos(2\,\phi)}{\cos^2(\phi-\dphi_p/2)}\, \left(1 + \left( \frac{\sin(\phi + \dphi_p/2)}{\cos(\phi - \dphi_p/2)}\right)^2\right)^{-1}
\end{equation}
Assuming that $\langle \dphi_p \rangle =0$, we obtain the scintillation noise:
\begin{equation}
	\sigma^2(\tilde{\phi}) = \frac{1}{4} \, \cos^2(2\,\phi)\, \sigma^2(\dphi_p, t_0/2)
\end{equation}
This result depends on the mean phase position. Assuming that $\phi$ is uniformly distributed over $[0, 2\pi]$, one finally obtains:
\begin{eqnarray}
\label{eq:sig2pist}
%\langle \sigma(\tilde{\phi}) \rangle_\phi &= 0.5 \sigma(\phi_p, t_0/2) \\
%\langle \sigma(\tilde{\phi})^2 \rangle_\phi &= 0.37 \sigma(\phi_p, t_0/2)
\sigma(\tilde{\phi})^2 &= 0.125 \, \sigma^2(\delta\phi_p, t_0/2)
\end{eqnarray}
This deviation is evaluated here by means of VLTI/FINITO data, and results are presented in Table~\ref{tab:piston_noise} for typical integration times from 2 to 8\,ms for ATs and from 1 to 4\,ms for UTs.

\begin{table}
 \begin{center}
  \begin{tabular}{cccc}
&   \multicolumn{3}{c}{\textbf{ATs}} \\
\hline
$t_0$ [ms]   & 2 & 4 & 8 \\
 E & $\lambda/$114 & $\lambda/$72 & $\lambda/$43\\
 E & $\lambda/$103 & $\lambda/$62 & $\lambda/$36 \\
 G & $\lambda/$90 & $\lambda/$57 & $\lambda/$34 \\
 G &  $\lambda/$91 & $\lambda/$60 & $\lambda/$34 \\
 M & $\lambda/$86 & $\lambda/$53 & $\lambda/$31 \\
 M & $\lambda/$81 & $\lambda/$51 & $\lambda/$29 \\
 B & $\lambda/$20 & $\lambda/$13 & $\lambda/$8 \\
 B & $\lambda/$29 & $\lambda/$19 & $\lambda/$12 \\
\\
&   \multicolumn{3}{c}{\textbf{UTs}} \\
\hline
$t_0$ [ms]   & 1 & 2 & 4 \\
 G & $\lambda/$32 & $\lambda/$20 & $\lambda/$12 \\
 M & $\lambda/$22 & $\lambda/$13 & $\lambda/$8 \\
 M & $\lambda/$23 & $\lambda/$19 & $\lambda/$10 \\
 B & $\lambda/$20 & $\lambda/$13 & $\lambda/$8 \\
  \end{tabular}
  \caption{Piston noise calculated with different sets of data on VLTI telescopes in H-band. The noise is written respectivily to the wavelength, for 3 different integration times. The values correspond to the worst case ($\sigma(\tilde{\phi}) = \sigma(\phi_p, t_0/2)$). Atmospheric conditions are: Excellent (E), Good (G), Medium (M), Bad (B). The corresponding observing conditions can be found in Tab.\ \ref{tab:atm_conditions}.}
 \label{tab:piston_noise}
 \end{center}
\end{table}

\subsubsection{Scintillation noise: $\sigma_{sci}$}
The influence of scintillation (i.e., photometric variations) between (A,C) and (B,D) measurements is to induce fringe contrast fluctuations, which can bias the phase measurement. This effect will be studied in the same manner than in the previous section. Considering an ideal interferogram, the real and imaginary parts of the coherent signal write:
\begin{eqnarray}
  &A-C &\varpropto V_{sci}(t-t_0/4) \, \cos \phi   \\
  &B-D &\varpropto V_{sci}(t+t_0/4) \, \sin \phi
\end{eqnarray}
where $V_{sci}$ is the contrast attenuation term due to the photometric imbalance between the two beams $I_1$ and $I_2$:
\begin{equation}
  V_{sci} = \frac{2 \sqrt{I_1 \, I_2}}{I_1 + I_2}
\end{equation}
Noting the flux variation $\delta_i = I_i(t+t_0/4) - I_i(t-t_0/4)$, the phase estimator writes:
\begin{equation}
   \tan\,\tilde{\phi} = \frac{B-D}{A-C} = \alpha \, \tan \phi
\end{equation}
where:
\begin{equation}
 \alpha = \sqrt{\frac{I_1 + \delta_1/2}{I_1 - \delta_1/2}\,\frac{I_2+\delta_2/2}{I_2 - \delta_2/2}} \times\frac{I_1 + I_2 - \delta_1/2 - \delta_2/2}{I_1 + I_2 + \delta_1/2 + \delta_2/2}
\end{equation}
Simplifying the first and second terms by $I_1 I_2$ and $I_1 + I_2$ respectively :
\begin{eqnarray}
 	\alpha		&=& \sqrt{\frac{(1 + x_1/2)\,(1 + x_2/2)}{(1 - x_1/2)\,(1 - x_2/2)}}\times\frac{1 - y_1/2 - y_2/2}{1 + y_1/2 + y_2/2} \\
	\mathrm{with} \nonumber \\
	x_i &=& \delta_i/I_i\\
	y_i &=& \delta_i/(I_1+I_2)
\end{eqnarray}
If the flux varies between both quadratures, $\alpha \neq 1$ and the phase estimation is biased. If we consider the statistic variations of the both photometries, this bias can be considered as an additional noise. We therefore calculate the measured photometric variance functions of the variance of the relative photometries $\sigma^2 (x_i, t_0/2)$ between two exposures distant of $t_0/2$. We assume that the two pupils are sufficiently distant to be considered as uncorrelated, which is the case if the baseline is longer than the atmospheric outer scale (typically 20\,m). Since the atmosphere follows the same statistics on both, it implies $\langle x_1 \rangle = \langle x_2 \rangle = \langle x \rangle$ and $\sigma(x_1, t_0/2) = \sigma(x_2, t_0/2) = \sigma(x, t_0/2)$:
\begin{eqnarray}
   \sigma^2(\tilde{\phi}) &=& \left. \left( \frac{\partial \tilde{\phi}}{\partial x_1} \right) \right|^2_{\langle x_1 \rangle, \langle x_2 \rangle}
                          \sigma^2(x_1, t_0/2)
                         + \left. \left( \frac{\partial \tilde{\phi}}{\partial x_2} \right) \right|^2_{\langle x_1 \rangle, \langle x_2 \rangle}
                          \sigma^2(x_2, t_0/2) \nonumber \\
                          &=& 2 \left. \left( \frac{\partial \tilde{\phi}}{\partial x_i} \right) \right|^2_{\langle x_1 \rangle, \langle x_2 \rangle}
                          \sigma^2(x, t_0/2)
\end{eqnarray}
with:
\begin{eqnarray}
\label{eq:partial}
    \left. \left( \frac{\partial \tilde{\phi}}{\partial x_i} \right) \right|_{\langle x_1 \rangle, \langle x_2 \rangle} &=& \frac{\partial\, ( \alpha  \tan \phi)}{\partial x_i} \, \frac{1}{1 + (\alpha \tan \phi)^2}
\end{eqnarray}
In order to obtain an analytical expression of this quantity, we assume the flux variaitons to be small: $\delta_i \ll I_i$. We do a first order expansion of $\alpha$ and only conserve the terms of the first order:
\begin{eqnarray}
	\alpha &\sim & (1 + x_1/4)^2\,(1 + x_2/4)^2\,(1 - y_1/2 - y_2/2)^2 \\
			   &\sim & (1 + x_1/2)\,(1 + x_2/2)\,(1 - y_1 - y_2) \\
		       &\sim &(1 + x_1/2 + x_2/2)(1 - y_1 - y_2) \\
		       &\sim & 1\, +\, x_1/2\, +\, x_2/2\, -\, y_1\, -\, y_2
\end{eqnarray}
We expand the latter expression to simplify it:
\begin{eqnarray}
	\alpha &\sim &  1\, +\, \frac{\delta_1}{2 I_1}\, +\,\frac{\delta_2}{2 I_2}\, -\frac{\delta_1 + \delta_2}{I_1 + I_2}\\
%				&\sim &  1\, +\, \frac{\delta_1 I_2 (I_1 + I_2) + \delta_2 I_1 (I_1+I_2) - 2(\delta_1 + \delta_2)I_1 I_2}{2(I_1 + I_2)I_1 I_2}\\
				&\sim &  1\, +\, \frac{\delta_1 I_2^2 + \delta_2 I_1^2 - (\delta_1 + \delta_2)I_1 I_2}{2(I_1 + I_2)I_1 I_2}\\
				&\sim &  1\, +\, \frac{I_2 - I_1}{I_2 + I_1}\,\frac{\delta_1 I_2 - \delta_2 I_1}{2 I_1 I_2}\\
\mathrm{And \: finally:} \nonumber \\
\alpha &\sim &  1\, +\, \frac{1}{2}\,\frac{I_2 - I_1}{I_2 + I_1}\,(x_1 - x_2)
\end{eqnarray}
Therefore, considering $\alpha \sim 1$ in the second term of the expression \ref{eq:partial}:
\begin{equation}
     \left. \left( \frac{\partial \tilde{\phi}}{\partial x_i} \right) \right|_{\langle x_1 \rangle, \langle x_2 \rangle} = \frac{1}{2} \frac{I_2 - I_1}{I_2 + I_1} \cos\phi \, \sin\phi
\end{equation}
Noting that: 
\begin{equation}
	\left(\frac{I_2 - I_1}{I_2 + I_1}\right)^2 = 1 - 4\frac{I_1 I_2}{(I_1 + I_2)^2} = 1 - V_{sci}^2
\end{equation}
we finally show the variance of the phase measurement due to the photometric noise is:
\begin{equation}
  \sigma^2(\tilde{\phi}) = \frac{1}{2} \, (\sin \phi \: \cos \phi)^2 (1 - V_{sci}^2\left(\langle I_1 \rangle, \langle I_2 \rangle\right))\,\, \sigma^2(x, t_0/2)
\end{equation}
Note that the result depends on the mean value of the scintillating visibility $V_{sci}$. Hence a perfectly balanced system should present a null photometric noise. This is an unrealistic effect due to our {\it symetric} modeling of the photometric variation with a step. In practice, the quick variations of photometries (i.e. during the integration) induce a noise even for a perfectly symetric combiner. To obtain a more realistic value, we can consider a (worst) case with a mean imbalance between fluxes of a factor of 10, so that $V_{sci} \sim 0.57$ and $1 - V_{sci}^2 \sim 0.67$.

If we finally average this result over every realisation of $\phi$ (still assuming its statistics to be uniform between $0$ and $2\pi$):
\begin{eqnarray}
\label{eq:phot-noise}
%   \langle \sigma(\tilde{\phi}) \rangle_\phi &= 0.32 \displaystyle\frac{\sigma(x, t_0/2)}{\langle x \rangle} \\
%   \langle \sigma(\tilde{\phi})^2 \rangle_\phi &= 0.125 \displaystyle\frac{\sigma(x, t_0/2)}{\langle x \rangle}
   \sigma^2(\tilde{\phi}) &=& 0.04\,\sigma^2(x, t_0/2)
\end{eqnarray}
Similarly to the piston noise, we present in Table~\ref{tab:photom_noise} the results obtained from ESO data on ATs and UTs, for different integration times.

%What we can see from these results (tab. \ref{tab:photom_noise}) for ATs and UTs is that even in bad conditions and long integration time the photometric noise should be much more less than the pure phase precision that we can expect in such conditions. Hence the photometric noise seems to be quite negligible beside photon and detector noise whatever the considered atmospheric conditions.
%Moreover if the fringes are supposed to be locked near a local maximum, $\phi \sim 0$ and as a result the photometric noise tends to zero.\\
%To conclude the photometric noise should be almost null in a fringe tracking regime, and quite negligible in bad tracking conditions.

\begin{table}[t]
 \begin{center}
  \begin{tabular}{cccc}
&   \multicolumn{3}{c}{\textbf{ATs}} \\
   \hline
$t_0$ [ms]   & 2 & 4 & 8 \\
 G & $\lambda/$499 & $\lambda/$369 & $\lambda/$290 \\
 M & $\lambda/$549 & $\lambda/$301 & $\lambda/$163 \\
 M &  $\lambda/$298 & $\lambda/$196 & $\lambda/$130 \\
 M & $\lambda/$400 & $\lambda/$277 & $\lambda/$192 \\
 B & $\lambda/$101 & $\lambda/$67 & $\lambda/$37 \\
\\
& \multicolumn{3}{c}{\textbf{UTs}} \\
 \hline
$t_0$ [ms]  & 1 & 2 & 4 \\
 G & $\lambda/$162 & $\lambda/$122 & $\lambda/$59 \\
 M & $\lambda/$107 & $\lambda/$76 & $\lambda/$35 \\
 B & $\lambda/$101 & $\lambda/$52 & $\lambda/$21
  \end{tabular}
  \caption{The photometric noise written respectivily to the wavelength in H-band, for 3 differents integration times. The values correspond to the worst case as defined in Eq. \ref{eq:phot-noise}. Atmospheric conditions are: Exceptionnal (E), Good (G), Medium (M), Bad (B). The corresponding observing conditions can be found in Tab.\ \ref{tab:atm_conditions}.}
 \label{tab:photom_noise}
 \end{center}
\end{table}
			\section{Theoretical dynamic range for the group delay estimation with dispersed fringes}                    \label{app:DRdisp}

We analyze here the case of a dispersed estimator for the group delay, similar to what is implemented on PRIMA, PTI or KI. We remind that the coherence envelope $E(x)$ corresponds to the Fourier transform modulus of the coherent signal:
\begin{equation}
	   E(x) \varpropto \left|\,  \int_0^\infty I(\lambda) V(\lambda) e^{i 2 \pi x_{GD} /\lambda}\, e^{-i 2 \pi x /\lambda} \,d\lambda \,\right|
\end{equation}
where $x$ is the OPD, $x_{GD}$ the position of the envelope center, and $I(\lambda)$ and $V(\lambda)$ the source intensity and visibility, both depending of the wavelength $\lambda$. We consider a spectral band centered around $\lambda_0$ and of width $\Delta \lambda$, so that the coherence length $L_c$ of the wide-band interferogram is $L_c = \lambda_0^2/\Delta \lambda$. The fringes are dispersed over $N_\lambda$ spectral channels of equal width $\delta \lambda = \Delta \lambda/N_\lambda$. In term of wavenumber, the wide- and narrow-band widths write $\Delta \sigma = 1/L_c$ and $\delta \sigma = \Delta \sigma/N_\lambda$.
%: the coherence length of broad-band interferogram on each channel is therefore roughly equal to $\lambda_0^2/(\Delta\lambda/N_\lambda = N_\lambda L_c$.

For sake of simplicity we consider here an ideal case, that is all the considered quantities are achromatic, in particular the source flux $I$ and visibility $V$ do not depend on the wavelength. We assume we dispose of a fringe coding (ABCD for instance) allowing the complex fringe signal $Z_k$ to be computed in each channel $k$, this latter being defined as:
\begin{equation}
	Z_k = I_k V_k e^{i 2\pi \sigma_k x_{GD}} = I V e^{i 2\pi \sigma_k x_{GD}}
\end{equation}
where $\sigma_k = 1/\lambda_k$ is the effective wavenumber on each spectral channel. The discrete Fourier transform of this coherent signal is then:
\begin{eqnarray}
	&\mathcal{F}(x) &= \sum_{k=1}^{N_\lambda} Z_k \, e^{-i \, 2\pi \sigma_k x} \nonumber \\
	&						&=\sum_{k=1}^{N_\lambda} I\,V\, e^{-i \, 2\pi \sigma_k (x-x_{GD})}
\end{eqnarray}
and we finally compute the squared coherence envelope:
\begin{eqnarray}
	&E^2(x) &\varpropto |\mathcal{F}(x)|^2 \\
	&				&= \mathcal{F}(x) \, \mathcal{F}^*(x) \\
	&				&=  I^2\,V^2\, \sum_{k=1}^{N_\lambda} \sum_{l=1}^{N_\lambda} e^{-i \, 2\pi (\sigma_k - \sigma_l) (x-x_{GD})}
\end{eqnarray}
where $\mathcal{F}^*$ is the complex conjugate of $\mathcal{F}$. Since each spectral channel has the same width, $\sigma_k - \sigma_l = \delta \sigma \, (k-l)$ and we finally get:
\begin{equation}
	E^2(x) \varpropto  I^2\,V^2\, \sum_{k=1}^{N_\lambda} \sum_{l=1}^{N_\lambda} e^{-i \, 2\pi \,\delta \sigma \, (x-x_{GD})\,(k-l)}
\end{equation}
The group delay is obtained when this quantity is maximum, that is when all the phasors in the double summation are in phase. In the present simple case, it is obvious it happens when $x=x_{GD}$, which leads to:
\begin{equation}
\forall (k,l), \quad	e^{-i \, 2\pi \, \delta \sigma \, (x-x_{GD})} = 1
\end{equation}
And solving this equation finally gives:
\begin{equation}
	x = x_{GD} \quad [1/\delta \sigma]
\end{equation}
where $[ \,]$ is the modulo symbol. In other word, by dispersing the fringes, we find the group delay with an ambiguity equal to $1/\delta \sigma$. From the definition of $\delta\sigma$, it finally corresponds to an ambiguity (or a dynamic range) equal to $N_\lambda \, \displaystyle\frac{\lambda^2}{\Delta \lambda}$.

			%%%%%%%%%%%%%%%%%%%%%%%%%%%%%%%%%%%%%%%%%%%%%%%%%%%%%%%%%%%%%%%%%%%%%
			%%%%%%%%%%%%%%%%%%%%%%%%%%%%%%%%%%%%%%%%%%%%%%%%%%%%%%%%%%%%%%%%%%%%%
			\section{Noise propagation on pairwise combination schemes}            \label{app:optschemecalc}

The study conducted in Section \ref{part:optschemes} aims at comparing various co-axial pairwise combination schemes looking at the phase and group delay measurements precision in various configurations. This study is based on  analytical descriptions of measurement noises. We describe here various points which have been necessary to carry out this study but which are not essential for the comprehension of the results.

\subsection{Reference noise}

Thanks to \citet{shao_1988}, \citet{tatulli_2010a} and our study (Eqs.~\ref{eq:GDnoisemodel1} and \ref{eq:GDnoisemodel2}), we know the analytical expression of the phase and group delay noises, in detector and photon noise regimes and for co-axial pairwise combinations. They express as:
\begin{eqnarray}
\label{eq:refdet}
&\sigma_0^{det} &= \displaystyle\frac{A}{K V} 	\\
\label{eq:refphot}
&\sigma_0^{phot} &= \displaystyle\frac{B}{\sqrt{K} V}
\end{eqnarray}
$K$ and $V$ being the number of photo-events and the fringe visibility. $A$ and $B$ are proportionality factors depending on the fringe coding, which have no influence in the following. These expressions correspond to the noise for a two-telescope (one baseline) instrument and are considered as noise references in the following.

\subsection{Individual baseline noise}

When we consider an interferometric array with more than 2 telescopes, the flux of each telescope is distributed between several different baselines, increasing the noise on each baselines. We consider two cases here: the open and redundant schemes.

\subsubsection{Redundant schemes}

The most simple cases are the redundant schemes in which the flux of each pupil is divided between the same number $R$ of baselines. Compared to a two-telescope instrument, the total flux $K$ on each baseline is divided by $R$, so that the measurement noise is:
\begin{eqnarray}
&\sigma^{det}	 &= \displaystyle A \frac{R}{K V} \nonumber \\
&						&= R\,	\sigma_0^{det}		\label{eq:refdet2} \\
&\sigma^{phot} &= \displaystyle B \frac{\sqrt{R}}{\sqrt{K} V}	\nonumber \\
&						&= \sqrt{R}\, \sigma_0^{phot} \label{eq:refphot2}
\end{eqnarray}
We are therefore able to compare the different schemes on detector and photon noise regimes on the base of a reference noise.

\subsubsection{Open schemes} \label{app:openschemes}

The open schemes use the minimal number of baselines enabling the array to be cophased, that is $N-1$ baselines. In this case the array is not symmetric, so that splitting the flux of intermediate pupils into equal parts (i.e. taking $50\%$ of their flux for each baseline) implies unequal performances for the different baselines. In this study we want the open schemes to have intrinsically equivalent baselines, that is with the same SNR on the fringe position measurements. To do so, we have to consider intrinsically imbalanced photometric inputs for each baselines and we evaluate the optimal fraction of the flux to inject in the different baselines.

% In this case we cannot consider the flux of each pupil is equally distributed between each baseline. For instance in the 4TO case, if we take half of the flux of pupil 2 to combine with all the flux of pupil 1, there will be a strong photometric imbalance resulting into a loss of contrast, so that baselines $\{12\}$ and $\{34\}$ will be less sensitive than the $\{23\}$.

Considering two identical telescopes $i$ and $j$, we combine their light by taking a fraction $\delta_i$ and $\delta_j$ of the incoming fluxes on each telescope respectively. In this case, the total flux available on the baseline is:
\begin{equation}
	K' = K (\delta_i + \delta_j)/2
\end{equation}
and the fringe contrast $V$ is possibly reduced because of the photometric imbalance:
\begin{equation}
V' = V \, \frac{2 \sqrt{\delta_i \delta_j}}{\delta_i + \delta_j}
\end{equation}
Now considering the noise expressions in Eq. \ref{eq:refdet} and \ref{eq:refphot}, we can easily write the measurement noises in this case, still as a function of our reference noises:
\begin{eqnarray}
\label{eq:asymdet}
&\sigma_{ij}^{det}	&=     \displaystyle\frac{1}{\sqrt{\delta_i \delta_j}} \, \sigma_0^{det} \\
\label{eq:asymphot}
&\sigma_{ij}^{phot}	&= \displaystyle\sqrt{\frac{\delta_i+\delta_j}{2 \delta_i \delta_j}} \, \sigma_0^{phot}
\end{eqnarray}

\begin{figure}
	\begin{center}
	\includegraphics[width=.2\textwidth]{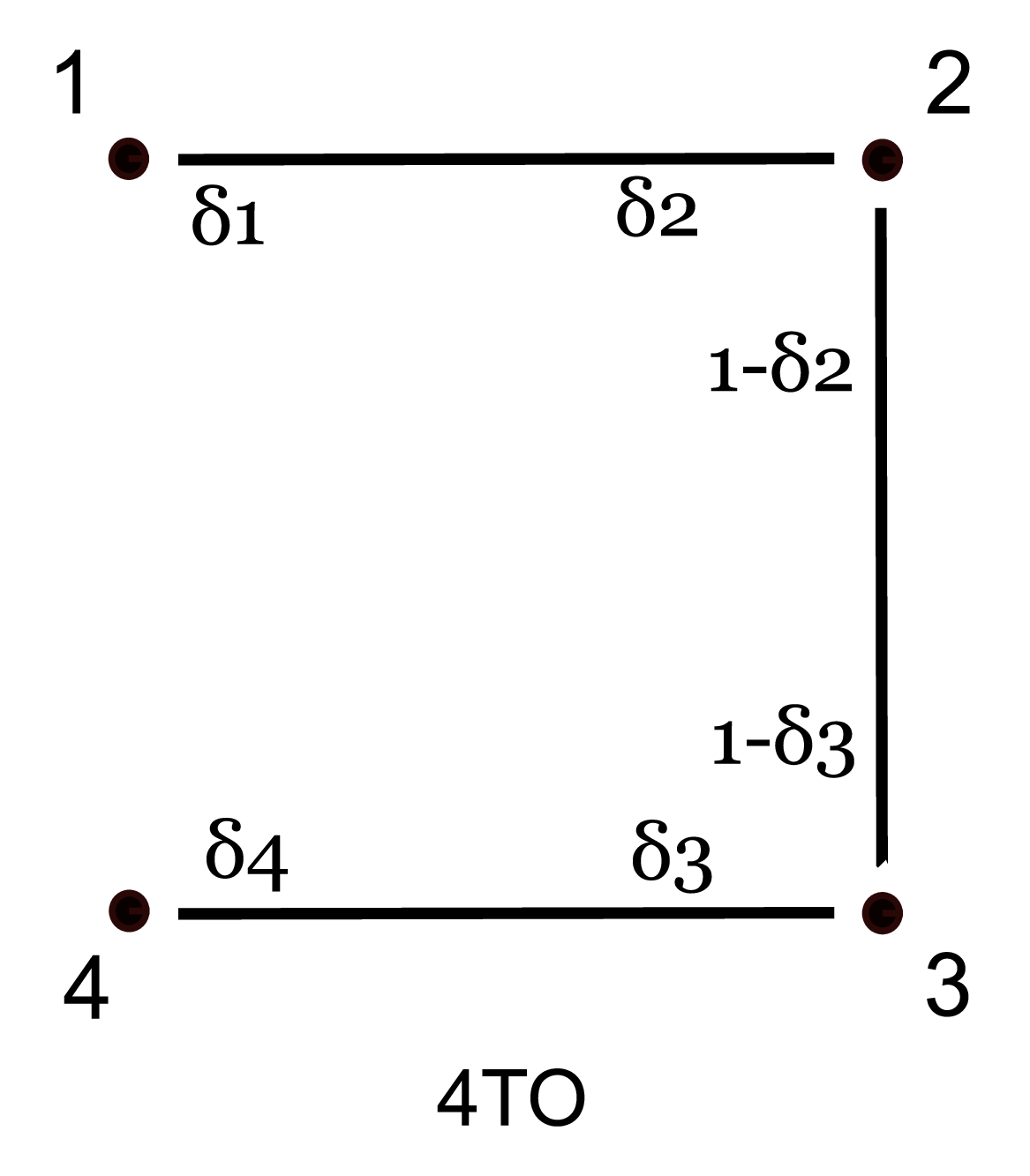}
	\includegraphics[width=.2\textwidth]{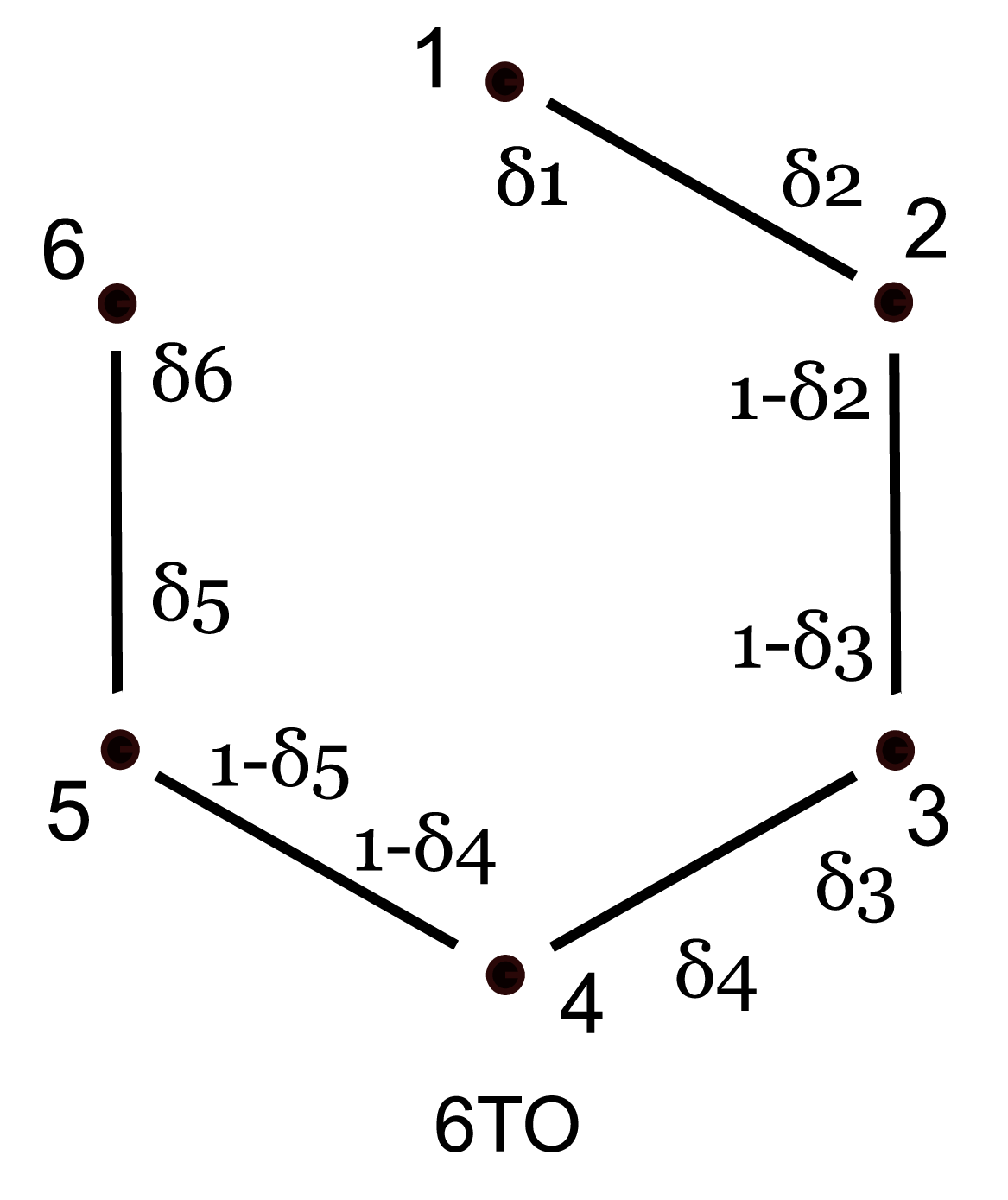}
	\end{center}
	\caption{Open schemes we consider in the 4 and 6 telescopes cases. The nomenclature for the flux split ratio $\delta_i$ are represented on the figures. \label{fig:openschemes}}
\end{figure}

The open schemes with 4 and 6 telescopes are presented on Fig. \ref{fig:openschemes}, with the associated nomenclature in term of splitting ratio $\delta_i$. We determine in the following their values.

\paragraph{4TO case}

For symmetry reasons we consider $\delta_1 = \delta_4 = 1$ and $\delta_2 = \delta_3 = \delta$, and therefore the measurement errors on the 3 baselines write:
\begin{center}
\begin{tabular}{ll}
%\hline \hline
%Detector noise regime &	Photon noise regime \\
%\hline
$   \sigma_{12}^{det} = \sigma_{34}^{det}  = \frac{1}{\sqrt{\delta}}\sigma_{0}^{det}	$ & $   \sigma_{12}^{phot} = \sigma_{34}^{phot} = \sqrt{\frac{1+\delta}{2\delta}} \sigma_{0}^{phot}$\\
\\
$   \sigma_{23}^{det} = \frac{1}{1-\delta}	\sigma_{0}^{det}$ & $   \sigma_{23}^{phot} = \frac{1}{\sqrt{1-\delta}} \sigma_{0}^{phot}$	 \\
%\hline
\end{tabular}
\end{center}
Our goal is to have equivalent baselines, i.e., we want $\sigma_{ij}$ to be equal on the three baselines. Solving this system in detector and photon noise regimes leads to:
\begin{center}
\begin{tabular}{ll}
%\hline \hline
%%Detector noise regime	&	Photon noise regime\\
%\hline
$\delta^{det} = 0.38$ & $\delta^{phot} = 0.42$ \\
&\\
$   \sigma_{ij}^{det} = 1.62 \sigma_{0}^{det}$ & $ \sigma_{ij}^{phot} = 1.31 \sigma_{0}^{phot}$ \\
%\hline
\end{tabular}
\end{center}

\paragraph{6TO case}

For symmetry reasons we have $\delta_1 = \delta_6 = 1$, $\delta_2 = \delta_5$ and $\delta_3 = \delta_4$. The measurement errors on the 5 baselines write:
\begin{center}
\begin{tabular}{ll}
%\hline \hline
%Detector noise regime &	Photon noise regime \\
%\hline
$   \sigma_{12}^{det}= \sigma_{56}^{det}= \frac{1}{\sqrt{\delta_2}} \sigma_{0}^{det}	$ &
$   \sigma_{12}^{phot}= \sigma_{56}^{phot}= \sqrt{\frac{1+\delta_2}{2\delta_2}} \sigma_{0}^{phot}$ \\
\\
$   \sigma_{23}^{det}= \sigma_{45}^{det}= \frac{1}{\sqrt{(1-\delta_2)(1-\delta_3)}}	 \sigma_{0}^{det}$ &
$   \sigma_{23}^{phot}= \sigma_{45}^{phot}$\\
& \qquad\,\,$= \sqrt{\frac{2-\delta_2-\delta_3}{2(1-\delta_2)(1-\delta_3)}} \sigma_{0}^{phot}$ \\
\\
$   \sigma_{34}^{det}= \frac{1}{\delta_3} \sigma_{0}^{det}$	& $   \sigma_{34}^{phot}= \frac{1}{\sqrt{\delta_3}} \sigma_{0}^{phot}$\\
%\hline
\end{tabular}
\end{center}
In the same way than previously, we estimate the optimal value of the different $\delta_i$:
\begin{center}
\begin{tabular}{ll}
%\hline \hline
%Detector noise regime	&	Photon noise regime\\
%\hline
$\delta_{2}^{det} = 0.31$ & $\delta_{2}^{phot} = 0.37$ \\
$\delta_{3}^{det} = 0.55$ & $\delta_{3}^{phot} = 0.54$ \\
&\\
$   \sigma_{ij}^{det} = 1.81 \sigma_{0}^{det}$ & $   \sigma_{ij}^{phot} = 1.36 \sigma_{0}^{phot}$ \\
%\hline
\end{tabular}
\end{center}

\vskip12pt
For the 4TO and 6TO cases we note the different values of $\delta_i$ are close in detector and photon noise regimes, so that such schemes are practically possible. In both regime we consider the same values: $\delta=0.40$ in the 4T case; $\delta_2 = 0.34$ and $\delta_3 = 0.54$ in the 6T case.

\subsection{Estimating the individual fringe position and final measurement noise}

We have to estimate $N-1$ differential pistons in order to cophase the interferometric array. In practice we measure $B$ differential pistons (noted $\tilde{\phi}$), with $B > N-1$ for redundant schemes, and $B = N-1$ for the open ones. Noting $x$ the vector of the $N-1$ optical path estimators used to drive the delay lines, the equation system linking $\tilde{\phi}$ and $x$ is:
\begin{equation}
 \tilde{\phi} = {\bf M} x
\end{equation}
where ${\bf M}$ is the so-called interaction matrix, which is known. We now need to inverse this system by computing the control matrix ${\bf W}$ :
\begin{equation} \label{eq_x_result}
  \hat{x} = {\bf W} \tilde{\phi}
\end{equation}
For the redundant schemes, ${\bf M}$ is rectangular and we compute ${\bf W}$ on the base of a singular value decomposition of ${\bf M}$. We therefore solve the system in the sens of a least square minimization, i.e. we minimize the quantity:
\begin{equation}
\chi^2 = |\tilde{\phi} - {\bf M} \hat{x}|^2
\end{equation}
However the measurements $\tilde{\phi}$ are noisy and we have to weight them to minimize the influence of the noisiest baselines. Considering that the measurements have gaussian statistics and are statistically independent, the $\chi^2$ writes:
\begin{equation}
\chi^2 = \left| \frac{\tilde{\phi} - {\bf M} \hat{x}}{\sigma} \right|^2
\end{equation}
where $\sigma$ is the vector of the error on the measurement $\tilde{\phi}$, given by eq. \ref{eq:refdet2} and \ref{eq:refphot2} depending on the noise regime. We modify in consequence the differential phase vector $\tilde{\phi}$ and the interaction matrix ${\bf M}$ as follow:
\begin{eqnarray}
&  {\bf M}_{ij}		& \rightarrow {\bf M}_{ij} / \sigma_i, \quad  j \in [1, N-1], i \in [1, B]	\\
&  \tilde{\phi}_i 	& \rightarrow \tilde{\phi}_i / \sigma_i				\label{def_phi_modif}
\end{eqnarray}

\subsection{Statistical error on the estimated differential pistons}

To compare the various schemes, we are interested by the error on the differential piston $x_{ij} = x_i - x_j$, which corresponds to the error on the correction applied to the delay lines:
\begin{equation}
%&\phi_{ij} &= \hat{x}_i - \hat{x}_j							\\
x_i = \sum_{k=1}^{B} {\bf W}_{ik} \tilde{\phi}_k
\end{equation}
Given the definition of $\tilde{\phi}_k$ (Eq.~\ref{def_phi_modif}), the statistical error on these terms is $\sigma(\tilde{\phi}_k) = 1$. We finally get the quadratic error $\sigma^2_{ij}$ on the corrected differential piston:
\begin{equation}
\sigma^2_{ij} = \sum_{k=1}^B ({\bf W}_{ik} - {\bf W}_{jk})^2
\end{equation}

\end{document}